\UseRawInputEncoding
\pdfoutput=1
\documentclass[manuscript]{aastex63}

\shorttitle{Extended HNCO, SiO, and HC$_{3}$N emission in 43 southern star-forming regions}
\shortauthors{He et al.}

\begin{document}
\title{Extended HNCO, SiO, and HC$_{3}$N emission in 43 southern star-forming regions}


\author[0000-0002-8760-8988]{Yu-Xin He}
\affiliation{Xinjiang Astronomical Observatory, Chinese Academy of Sciences, Urumqi 830011, PR China}
\affiliation{Key Laboratory of Radio Astronomy, Chinese Academy of Sciences, Urumqi 830011, PR China}
\affiliation{Chinese Academy of Sciences South America Center for Astronomy, National Astronomical Observatories, CAS, Beijing 100012, China}
\affiliation{Departmento de Astronom\'{i}a, Facultad de Ciencias F\'{i}sicas y Matem\'{a}ticas, Universidad de Concepci\'{o}n, Concepci\'{o}n, Chile}

\author{Christian Henkel}
\affiliation{Max-Planck-Institut f{\"u}r Radioastronomie, Auf dem H{\"u}gel 69, 53121 Bonn, Germany}
\affiliation{Xinjiang Astronomical Observatory, Chinese Academy of Sciences, Urumqi 830011, PR China}
\affiliation{Astron. Dept., King Abdulaziz University, P.O. Box 80203, 21589 Jeddah, Saudi Arabia}

\author{Jian-Jun Zhou}
\affiliation{Xinjiang Astronomical Observatory, Chinese Academy of Sciences, Urumqi 830011, PR China}
\affiliation{Key Laboratory of Radio Astronomy, Chinese Academy of Sciences, Urumqi 830011, PR China}

\author{Jarken~Esimbek}
\affiliation{Xinjiang Astronomical Observatory, Chinese Academy of Sciences, Urumqi 830011, PR China}
\affiliation{Key Laboratory of Radio Astronomy, Chinese Academy of Sciences, Urumqi 830011, PR China}

\author{Amelia M.\ Stutz}
\affiliation{Departmento de Astronom\'{i}a, Facultad de Ciencias F\'{i}sicas y Matem\'{a}ticas, Universidad de Concepci\'{o}n, Concepci\'{o}n, Chile}
\affiliation{Max-Planck-Institute for Astronomy, K\"onigstuhl 17, 69117 Heidelberg, Germany}

\author{Hong-Li Liu}
\affiliation{Departmento de Astronom\'{i}a, Facultad de Ciencias F\'{i}sicas y Matem\'{a}ticas, Universidad de Concepci\'{o}n, Concepci\'{o}n, Chile}

\author{Wei-Guang Ji}
\affiliation{Xinjiang Astronomical Observatory, Chinese Academy of Sciences, Urumqi 830011, PR China}
\affiliation{Key Laboratory of Radio Astronomy, Chinese Academy of Sciences, Urumqi 830011, PR China}

\author{Da-Lei Li}
\affiliation{Xinjiang Astronomical Observatory, Chinese Academy of Sciences, Urumqi 830011, PR China}
\affiliation{Key Laboratory of Radio Astronomy, Chinese Academy of Sciences, Urumqi 830011, PR China}

\author{Gang Wu}
\affiliation{Xinjiang Astronomical Observatory, Chinese Academy of Sciences, Urumqi 830011, PR China}
\affiliation{Key Laboratory of Radio Astronomy, Chinese Academy of Sciences, Urumqi 830011, PR China}

\author{Xin-Di Tang}
\affiliation{Xinjiang Astronomical Observatory, Chinese Academy of Sciences, Urumqi 830011, PR China}
\affiliation{Key Laboratory of Radio Astronomy, Chinese Academy of Sciences, Urumqi 830011, PR China}

\author{Toktarkhan Komesh}
\affiliation{Xinjiang Astronomical Observatory, Chinese Academy of Sciences, Urumqi 830011, PR China}
\affiliation{Department of Solid State Physics and Nonlinear Physics, Faculty of Physics and Technology, Al-Farabi Kazakh National \\~~University, Almaty, 050040, Kazakhstan}
\affiliation{University of the Chinese Academy of Sciences, Beijing 100080, People's Republic of China}

\author{Serikbek Sailanbek}
\affiliation{Xinjiang Astronomical Observatory, Chinese Academy of Sciences, Urumqi 830011, PR China}
\affiliation{Department of Solid State Physics and Nonlinear Physics, Faculty of Physics and Technology, Al-Farabi Kazakh National \\~~University, Almaty, 050040, Kazakhstan}
\affiliation{University of the Chinese Academy of Sciences, Beijing 100080, People's Republic of China}

\begin{abstract}

  We have selected 43 southern massive star-forming regions to study the spatial distribution of HNCO 4$_{04}$--3$_{03}$, SiO 2--1 and HC$_{3}$N 10--9 line emission and to investigate their spatial association with the dust emission.
  The morphology of HNCO 4$_{04}$--3$_{03}$ and HC$_{3}$N 10--9 agrees well with the dust emission.
  HC$_{3}$N 10--9 tends to originate from more compact regions than HNCO 4$_{04}$--3$_{03}$ and SiO 2--1.
  We divided our sources into three groups: those in the Central Molecular Zone (CMZ), those associated with bubbles (Bubble), and the remaining sources, which are termed 'normal star forming regions' (NMSFR). These three groups, subdivided into three different categories with respect to line widths, integrated intensities, and column densities, hint at the presence of different physical and chemical processes.
  We find that the dust temperature $T_{\rm d}$, and the abundance ratios of $N_{\rm HNCO}/N_{\rm SiO}$ and $N_{\rm HNCO}/N_{\rm HC3N}$ show a decreasing trend towards the central dense regions of CMZ sources, while $N_{\rm HC3N}/N_{\rm SiO}$ moves into the opposite direction.
  Moreover, a better agreement is found between $T_{\rm d}$ and $N_{\rm HC3N}/N_{\rm SiO}$ in Bubble and NMSFR category sources.
  Both outflow and inflow activities have been found in eight of the sixteen bubble and NMSFR sources.
  The low outflow detection rate indicates that in these sources the SiO 2--1 line wing emission is either below our sensitivity limit or that the bulk of the SiO emission may be produced by the expansion of an H{\sc\,ii} region or supernova remnant, which has pushed molecular gas away forming a shock and yielding SiO.

\end{abstract}
\keywords{astrochemistry  ---  stars: formation  ---  ISM: abundances  ---  ISM: clouds  ¨C--  ISM: molecules  ---  radio lines: ISM}

\section{Introduction} \label{sec:intro}
Shocks are a ubiquitous phenomenon in the interstellar medium (ISM) of galaxies. They may be driven by supernova explosions, the pressure of photoionized gas, stellar winds, and collisions between fast-moving clumps of interstellar gas, where the fluid-dynamical disturbances proceed at a velocity that exceeds the local sound speed because of the presence of large pressure gradients \citep{1993ARA&A..31..373D}. The processes of birth, evolution and death of stars are always associated with shocks \citep{2015sf2a.conf...87G}. Therefore, a systematic study of shocks in regions of star formation is of great importance for our general understanding of the physical and chemical boundary conditions of star formation.

Isocyanic acid (HNCO), which was detected for the first time in the Galactic radio source Sgr B2 by \citet{1971BAAS....3Q.388S}, is a ubiquitous molecule. Based on the morphology of the emission, the HNCO abundance with respect to H$_{2}$ and the relation to SiO emission, several authors tested the hypothesis that HNCO could be a good tracer of interstellar shocks \citep[e.g.][]{2000A&A...361.1079Z,2005ApJ...618..259M,2006NewA...11..594M}. Afterwards, \citet{2010A&A...516A..98R} tested this hypothesis by observing several transitions of HNCO towards a well-known shocked Galactic source, L1157--mm, and proposed that HNCO is a good shock tracer and the gas phase abundance of HNCO is achieved by both grain mantle evaporation through shock waves and by neutral-neutral reactions in the gas phase involving CN and O$_{2}$.

The silicon monoxide (SiO) molecule is an excellent tracer of molecular gas processed by the action of high-velocity ($\sim$20 -- 50 km s$^{-1}$) shocks in regions of star formation \citep{1992A&A...254..315M,1997A&A...321..293S,1999A&A...343..585C,2008A&A...482..809G,2008A&A...490..695G,2016ApJ...822...85L}. \citet{1992A&A...254..315M} reported that this molecule is enhanced by large factors (in some cases by $>$ 10$^{6}$) towards molecular outflows, which strongly suggests that grain disruption by shocks is the major mechanism for releasing SiO into the gas phase \citep{1992A&A...254..315M}. Furthermore, SiO emission can also trace Photon Dominated Regions (PDRs) with $\sim$10 -- 20 km s$^{-1}$ shocks \citep[e.g.][]{2001A&A...372..291S}. In the Galactic center region, the gas is characterized by large line widths, indicating a high degree of turbulence. There, SiO is widespread  (see e.g. the surveys in the J = 1--0 transition by \citealt{1997ApJ...482L..45M} and \citealt{2013MNRAS.433..221J}, and in the J = 2--1 transition by \citealt{2012MNRAS.419.2961J}), which was interpreted as evidence for large scale or ubiquitous (fast) shocks. \citet{1998A&A...334..646H} observed 33 targets in the Galactic center region and found that at least some SiO is detected in all cloud cores, mostly associated with cool gas. They proposed that this was due to shocks caused by local turbulence and/or cloud-cloud collisions. In the Galactic disc, quasithermal SiO emission is tightly correlated with high temperature regions \citep[e.g.][]{1989ApJ...343..201Z}. In addition, observations indicate that the properties of molecular peaks in the Galactic center region are markedly different from the Galactic disk, where thermal SiO emission is confined to very small regions in the vicinity of outflows associated with star formation \citep{1982ApJ...252L..29D,1998A&A...334..646H}.

Interstellar cyanoacetylene (HC$_{3}$N) was first detected in Sgr B2 by \citet{1971ApJ...163L..35T}. It is an excellent dense gas tracer \citep{1976ApJ...205...82M,1977ApJ...216..738M,1991JKAS...24..217C,1996ApJ...460..343B}. Both chemical models and observations indicate that the abundance of HC$_{3}$N is enhanced towards hot cores (through the gas-phase reaction C$_{2}$H$_{2}$ + CN $\rightarrow$ HC$_{3}$N + H, see \citealt{1999ASIC..540...97V} and \citealt{2009MNRAS.394..221C} for details).

Shocks are often related to outflowing gas. The above mentioned tracers, however, are not ideal when trying to study the entire velocity field encompassing both quiescent, outflowing {\it and} inflowing gas. For this purpose the HCO$^+$ $J$ = 1--0 line has been chosen. With its skewed line profiles in case of outflow or inflow, it provides the required signature \citep{2007ApJ...669L..37W,2015MNRAS.450.1926H,2016MNRAS.461.2288H,2019MNRAS.488.4638L}.

The two shock tracers HNCO and SiO are chemically different. HNCO is thought to arise from dust ice mantle sublimation under soft shocks \citep{2005ApJ...618..259M}, while SiO can be significantly enhanced in the gas phase by strong shocks due to a partial/total evaporation of the SiO present in both the grain mantles and the grain cores \citep[e.g.][]{2005ApJ...627L.121J,2008A&A...482..549J,1997A&A...321..293S,2008A&A...490..695G}. Moreover, HNCO and HC$_{3}$N are easily dissociated by UV radiation and are hot core tracers \citep{2008ApJ...678..245M,2014A&A...562A...3M,1998A&A...329.1097R}, while SiO is more robust against UV radiation. In past years, some HNCO or SiO molecular line surveys of massive galactic dense clumps have already been published \citep{1998A&AS..132..211H,2000A&A...361.1079Z,2013A&A...555A..18L,2016A&A...586A.149C}. However, most of the line surveys of massive Galactic dense clumps performed so far are based on single-pointing observations in which case the spatial distribution of HNCO and SiO emissions cannot be explored. Furthermore, beside sources in the Galactic center region, the large-scale SiO J = 2--1 line emission has been measured only in a few regions of the Galactic disk, such as the filamentary infrared dark cloud (IRDC) G035.39$-$00.33 \citep{2010MNRAS.406..187J} and the three IRDCs G028.37$+$00.07, G034.43$+$00.24, and G034.77$-$00.55 \citep{2018MNRAS.474.3760C}, with the IRDC G034.77$-$00.55 being further studied in detail by using ALMA data \citep{2019ApJ...881L..42C}. Therefore, it is clear that high-sensitivity mapping observations of HNCO, SiO and HC$_{3}$N are needed to characterise the spatial distributions of these species and their relation to physical properties of dense clumps. This paper presents a systematic study of a sample of 43 massive dense clumps. The spectral-line mapping data used here were taken from the Millimetre Astronomy Legacy Team 90 GHz (MALT90) survey \citep{2011ApJS..197...25F,2013PASA...30...57J}. After describing the sample selection and observations in Sect.~\ref{sec:data}, the observational results are presented in Sect.~\ref{sec:dust_tem_mol_col_density}. \textbf{In Sects.~\ref{sec:indiv} and \ref{sec:discussion},} we analyze and discuss the results and summarize the paper in Sect.~\ref{sec:conclusions}.

\section{Data} \label{sec:data}
\subsection{Observations} \label{sec:obs}
In the following, we will study a sample of 43 massive southern dense clumps using data taken from the Millimetre Astronomy Legacy Team 90\,GHz (MALT90) Survey, the APEX Telescope Large Area Survey of the Galaxy (ATLASGAL), and the Herschel infrared Galactic Plane Survey (Hi-GAL).

MALT90 is a large international project aimed at characterizing the physical conditions, chemical, and evolutionary state of over 2,000 high-mass dense cores in the southern sky at 90\,GHz with the ATNF Mopra 22 m telescope \citep{2011ApJS..197...25F,2013PASA...30...38F,2013PASA...30...57J}. With the full 8 GHz bandwidth of MOPS and the OTF mapping mode of Mopra, 3 arcmin $\times$ 3 arcmin maps can be efficiently obtained simultaneously in 16 molecular lines. The angular resolution and typical rms noise of this project are about 38$^{\prime\prime}$ and $T$$_{\rm A}^{\ast}$ = 0.25 K per 0.11 km s$^{-1}$ channel, respectively (for the main beam efficiency, see Sect.~\ref{sec:sampl}). The data cubes of HNCO 4$_{04}$--3$_{03}$, SiO 2--1, HC$_{3}$N 10--9, and HCO$^{+}$ 1--0, which will be used in the following analysis, were downloaded from the MALT90 online archive \footnote{http://atoa.atnf.csiro.au/MALT90}.

ATLASGAL is the first systematic survey of the inner Galactic plane in the sub-millimetre aimed at studying continuum emission from the highest density regions of dust at 345 GHz with the APEX 12 m telescope \citep{2009A&A...504..415S,2013A&A...549A..45C}. The angular resolution of the APEX telescope at this frequency is 19$^{\prime\prime}$$\!\!$.2, and the r.m.s. is 50-70 mJy beam$^{-1}$ in the final 870 $\mu$m survey images.

Hi-GAL is an unbiased photometric survey of the inner Galactic plane at 70 and 160 $\mu$m with PACS \citep{2010A&A...518L...2P}, and at 250, 350, and 500 $\mu$m with SPIRE \citep{2010A&A...518L...3G} on board of the Herschel satellite, aimed to catalogue star-forming regions and studying cold structures across the interstellar medium \citep{2010PASP..122..314M}. While we will not use the 70 $\mu$m band, which may trace a high temperature but low mass dust component, the corresponding angular resolutions for the longer wavelength bands are 12$^{\prime\prime}$, 18$^{\prime\prime}$, 25$^{\prime\prime}$, and 37$^{\prime\prime}$, respectively. The PACS and SPIRE photometric maps used here, processed by \citet{2016A&A...591A.149M}, were taken from the online archive \footnote{http://tools.asdc.asi.it/HiGAL.jsp}.

\subsection{Sample selection} \label{sec:sampl}
The source sample of the present paper was selected from the MALT90 observations applying the following criteria: (i) at the peak of each HNCO 4$_{04}$--3$_{03}$, SiO 2--1 and HC$_{3}$N 10--9 integrated intensity emission map, the averaged intensity over a circular area with a diameter roughly equal to the Mopra beamwidth (38$^{\prime\prime}$, see Sect.~\ref{sec:obs}) must lie above the detection threshold of $3\sigma$. (ii) The HNCO 4$_{04}$--3$_{03}$ line presents relatively simple profiles, which means no more than two velocity components, along the line of sight towards the peak position of its integrated intensity emission map.

Using these criteria, all those 43 sources with extended molecular line emissions have been selected for our sample. In the following these sources were divided into three subsamples: 27 sources are part of the Central Molecular Zone (CMZ) of the Galaxy, 10 sources are associated with expanding bubbles of ionized gas (Bubble), and 6 are "normal" star forming (NMSFR) clouds with no obvious star-formation and interaction signatures from the surrounding Spitzer 8$\mu$m continuum emissions \citep{2003PASP..115..953B,2009PASP..121..213C}. Moreover, 4 Bubble sources (G322.159+00.635, G327.293$-$00.579, G345.004$-$00.224  and G350.101+00.083) and 3 NMSFR sources (G329.030$-$00.202, G335.586$-$00.289 and G348.754$-$00.941) in our sample are identified as infrared dark clouds (IRDCs) by analysing Spitzer data in \citet{2009A&A...505..405P}.

In the following, the statistical results including only 10 Bubble and even less NMSFR clouds would be not fully sufficient to provide reliable statistical results, so that we will also compare our findings with additional measurements taken from the literature in Sect.~\ref{sec:discussion}. The basic information is summarized in Tables~\ref{tab:source_parameters_1} and \ref{tab:source_parameters_2}. Spectral line data were reduced using CLASS (Continuum and Line Analysis Single-Disk Software), a part of the GILDAS (Grenoble Image and Line Data Analysis Software) software \footnote{GILDAS is a radio astronomy software developed by IRAM. See {\tt http://www.iram.fr/IRAMFR/GILDAS}/.}. After baseline subtraction, the final spectra were converted to units of main beam brightness temperature ($T_{\rm mb}$) according to the usual expression $T_{\rm mb} = T_{\rm a}^* / \eta_{\rm mb}$, where $\eta_{\rm mb}$ is the frequency-dependent beam efficiency. The main beam efficiency of Mopra at 86 GHz is 0.49, and at 110 GHz it is 0.44 \citep{2005PASA...22...62L}. Unless otherwise specified, we use $T_{\rm mb}$ throughout the paper. ATLASGAL and Hi-GAL data were also analysed towards our 43 southern massive dense clumps (see Sect.~\ref{sec:dust_tem_mol_col_density}).

\section{Dust temperature and molecular column density} \label{sec:dust_tem_mol_col_density}
\subsection{Dust temperature} \label{sec:dust_tem}
Combining the \emph{Herschel} Hi-GAL observations at 160, 250, 350, 500 $\mu$m, and the APEX ATLASGAL observations at 870 $\mu$m, we obtained dust temperature maps pixel by pixel from a single temperature gray-body dust emission model which is given by

\begin{equation}
I_\nu(T_d,N_{\rm H_2}) \, = \, B_\nu(T_d)\left(1-e^{{\rm R}\times \mu m_{\rm H} N_{\rm H_2}\kappa_\nu}\right),
\label{eq-Idust}
\end{equation}

where $I_\nu$ is the observed intensity, $B_\nu(T_d)$ is the Planck function at a dust temperature $T_d$, $N_{\rm H_2}$ is the column density of molecular hydrogen, R is the gas to dust mass ratio (assumed to be 100), $\mu$ = 2.8 is the mean molecular weight of the interstellar medium \citep{2008A&A...487..993K}, and $\kappa_\nu$ is the dust absorption coefficient. The $\kappa_\nu$ curve is approximated by a power-law of $\kappa_{\nu} = \kappa_{0}(\nu/\nu_{0})^{\beta}$, which depends on frequency ($\nu$ $<$ 1THz) and spectral index $\beta$ \citep{1983QJRAS..24..267H}. \citet{2015ApJ...815..130G} tested 14 IRAS sources selected from the 1.1 mm Bolocam Galactic Plane Survey \citep{2010ApJS..188..123R,2013ApJS..208...14G} and the ATLASGAL catalog, and found that $\overline \beta$\,=\,1.6 is in agreement with the absorption coefficient law of silicate-graphite grains, with 3$\times$10$^{4}$ years of coagulation, and without ice coatings analyzed by \citet{2011A&A...532A..43O}. This choice of $\beta$ is also consistent with previous studies \citep[see e.g.][]{2011A&A...535A.128B}. The sources in this paper can also be considered as a subsample of the targets analysed by \citet{2015ApJ...815..130G}, who determined dust temperatures for all molecular clumps from the MALT90 survey using the aforementioned absorption coefficient law presented by \citet{2011A&A...532A..43O}. However, \citet{2015ApJ...815..130G} just provide average and peak dust temperature for each source in their work, so we calculate dust temperature maps here. We follow their method without fitting $\beta$ and directly use $\kappa_{\nu}$=32.9, 13.9, 7.7, 4.3 and 1.7 cm$^{2}$ g$^{-1}$ values for the 160, 250, 350, 500, and 870 $\mu$m data, which were extracted from \citet{2011A&A...532A..43O}. Before performing Spectral Energy Distribution (SED) fitting, we made use of the routine \footnote{https://github.com/esoPanda/FTbg/.} developed by \citet{2015MNRAS.450.4043W} to remove emission from the background and foreground. Next, all previously mentioned images are convolved to the SPIRE 500 $\mu$m data resolution (i.e., to 37$^{\prime\prime}$) using the kernels provided by \citet{2011PASP..123.1218A}, and then re-gridded to the same pixelization ($\sim$12$^{\prime\prime}$).

The line optical thickness ($\tau_{\nu}$) of HNCO 4$_{04}$--3$_{03}$ could not be derived through fitting the hyperfine structure, because the hyperfine structure is not spectrally resolved. With limited bandwith, we also could not derive $\tau_{\nu}$ of HNCO 4$_{04}$--3$_{03}$, SiO 2--1 and HC$_{3}$N 10--9 by comparing the intensities of two different isotopologues of the same species. Therefore, we calculate the value of optical depth using equation \ref{eq02}. Assuming local thermodynamic equilibrium (LTE) conditions, all levels are populated according to the same excitation temperature ($T_{\rm ex}$). In addition, we performed checks on the mean volume densities of the H$_{\rm 2}$ molecule, which is given by $n_{\rm H_2} =  \frac{N_{\rm H_2}}{\theta \times d}$, combining column density ($N_{\rm H_2}$, see Sect.~\ref{sec:mol_col_density_and_frac_abundance}) with source size ($\theta$, see Sect.~\ref{sec:indiv}), and find that about 75\% of the sources are denser than 1.2$\times$10$^{4}$ cm$^{-3}$, which indicates that the gas and dust are thermally coupled (see \citealt{2019MNRAS.483.5355M} for details). Distances ($d$) for CMZ sources are assumed to be 8.15 kpc \citep{2019ApJ...885..131R}, and are taken from SIMBAD\footnote{http://simbad.u-strasbg.fr/simbad/} for the remaining sources. Therefore we assumed that $T_{\rm ex}$ is equal to the temperature of the dust, $T_{\rm d}$. The optical thickness of the line emission ($\tau$) can then be derived from the equation \citep{2011ApJ...738...11H}:

\begin{equation}
T_{\rm MB} \, = \, \frac{h\nu}{k_{\rm B}}\left[F(T_{\rm ex})-F(T_{\rm bg})\right]\left(1-e^{-\tau_{\nu}} \right) \, ,
\label{eq02}
\end{equation}

where $T_{\rm MB}$ is the main-beam brightness temperature, h is the Planck constant, k$_{\rm B}$ is the Boltzmann constant, $\nu$ is the transition frequency, $T_{\rm ex}$ is the line excitation temperature, $T_{\rm bg}$ is the background temperature, and $F(T_{\rm ex})$ is the (average) photon occupation number, which is given by $F(T) = 1/(e^{h\nu/k_{\rm B}T}-1)$. With $T_{\rm ex}$ from $T_{\rm d}$, and $T_{\rm mb}$ and $T_{\rm bg}$ being known from the observations, the optical depths $\tau_\nu$ can also be determined.

\subsection{Column densities and fractional abundances}  \label{sec:mol_col_density_and_frac_abundance}

The beam-averaged column densities of the three molecules (the size of the beam is 38$^{\prime\prime}$), $N_{\rm MOL}$, can be derived from \citet{1991ApJ...374..540G}:

\begin{equation}
N_{\rm MOL} \, = \, \frac{8\pi \nu^3}{c^3}\frac{Q_{\rm rot}}{g_{\rm u} A_{\rm ul}}\frac{\exp(E_{_{l}}/k_{\rm B}T_{\rm ex})}{[1-\exp(-h\nu/k_{\rm B}T_{\rm ex})]}\int \tau_{\rm \nu} \,dv~,
\label{eqn-den-colum-2}
\end{equation}

where $g_{\rm u}$ is the statistical weight of the upper level, $A_{\rm ul}$ is the Einstein coefficient for spontaneous emission, $E_{_{l}}$ is the energy of the lower state, and $Q_{\rm rot}$ is the partition function. All the above parameters of HNCO 4$_{04}$--3$_{03}$, SiO 2--1 and HC$_{3}$N 10--9 are presented in detail by \citet{2012ApJ...756...60S}. The calculated $\tau_{\nu}$ and column densities of these three molecules are listed in Table~\ref{tab:source_parameters_1}, column 10 and Table~\ref{tab:source_parameters_2}, column 7, respectively.

The hydrogen column density, $N_{\rm H_{\rm 2}}$,  can be derived from the ATLASGAL dust continuum maps smoothed to the MALT90 resolution of 38$^{\prime\prime}$ \citep{2008A&A...487..993K}:

\begin{equation}
N_{\rm H_{\rm 2}} \, = \, \frac{S_{\rm{\nu}} \, \rm{R}}{B_{\rm \nu}(T_{\rm d}) \, \Omega \, \kappa_{\rm{\nu}} \, \mu \, m_{\rm H}} \,,
\label{eqn-h2-colum}
\end{equation}

where S$_{\rm \nu}$ is the 870 $\rm \mu$m continuum flux, $\rm \Omega$ is the beam solid angle (which has been smoothed to the MALT90 resolution of 38$^{\prime\prime}$), and ${\rm \kappa_{\nu}}$ is the dust-absorption coefficient (taken as 1.7 cm$^{2}$ g$^{-1}$, see Sect.~\ref{sec:dust_tem}).

The fractional abundances of the molecules were calculated by dividing the beam-averaged molecular column density by the hydrogen column density, $x_{\rm MOL}$\,=\,$N_{\rm MOL}$/$N_{\rm H_{\rm 2}}$. The calculated fractional abundances of HNCO, SiO and HC$_{3}$N are listed in Table~\ref{tab:source_parameters_2}, column 9.

\section{Individual target analysis} \label{sec:indiv}
Figs.~\ref{fig:Figure_A1}--\ref{fig:Figure_A22} show the diagrams of the HNCO, SiO and HC$_{3}$N 3 $\times$ 3 arcmin$^2$ integrated intensity maps superposed on the ATLASGAL 870 $\mu$m emission images and dust temperature maps, the beam averaged spectra of HNCO 4$_{04}$--3$_{03}$, SiO 2--1 and HC$_{3}$N 10--9 at the position marked by yellow crosses on the maps, and the trends of normalized abundance ratios of $N_{\rm HNCO}/N_{\rm SiO}$, $N_{\rm HNCO}/N_{\rm HC3N}$ and $N_{\rm HC3N}/N_{\rm SiO}$ along Galactic longitude and latitude passing through the yellow cross for the 43 southern sources, respectively. The HNCO lines, and a part of the strong SiO and HC$_{3}$N lines, which show single Gaussian profiles, were fitted with a single Gaussian using CLASS. Sources showing more than one velocity were fitted using two Gaussian components. For instance, CMZ sources G359.977$-$00.072, G359.895$-$00.069, G359 .868$-$00.085, G359.453$-$00.112, G000.067$-$00.077, G000.104$-$00.080, G000.106$-$00.001, G000.908$+$00.116, G001.655$-$00.062 and G001.883$-$00.062, Bubble sources G345.004$-$00.224, G351.443$+$00.659 and G351.775$-$00.537, and NMSFR sources G329.030$-$00.202 and G335.586$-$00.289 showing more than one velocity component were fitted with two Gaussian components (the two Gaussian components and total fits are shown with red and green lines in Figs.~\ref{fig:Figure_A1}--\ref{fig:Figure_A22}, respectively). The Gaussian fitting velocity components with full-width half-maximum (FWHM) linewidth ($\Delta$V) of HNCO 4$_{04}$--3$_{03}$, SiO 2--1 and HC$_{3}$N 10--9 emission are presented in Table~\ref{tab:source_parameters_1}. The integrated line intensity ($\int {T_{\rm MB}} dv$\,) is derived by integrating the area underneath the lines over the velocity range indicated in Table~\ref{tab:source_parameters_2}, column 6. The details of the individual 43 sources are described in Appendix~\ref{sec:Individual_targets_analysis}. In most CMZ sources, we find that the abundance ratios of $N_{\rm HNCO}/N_{\rm SiO}$ and $N_{\rm HNCO}/N_{\rm HC3N}$ show a decreasing trend towards the yellow cross, which is similar to the dust temperature, while $N_{\rm HC3N}/N_{\rm SiO}$ moves into the opposite direction. In Bubble and NMSFR category sources, dust temperature and $N_{\rm HC3N}/N_{\rm SiO}$ roughly follow a similar trend towards the center, where stars are forming (see Sect.~\ref{sec:outflow_inflow}).

\begin{figure*}
  \centering
	\includegraphics[width = 0.7\linewidth]{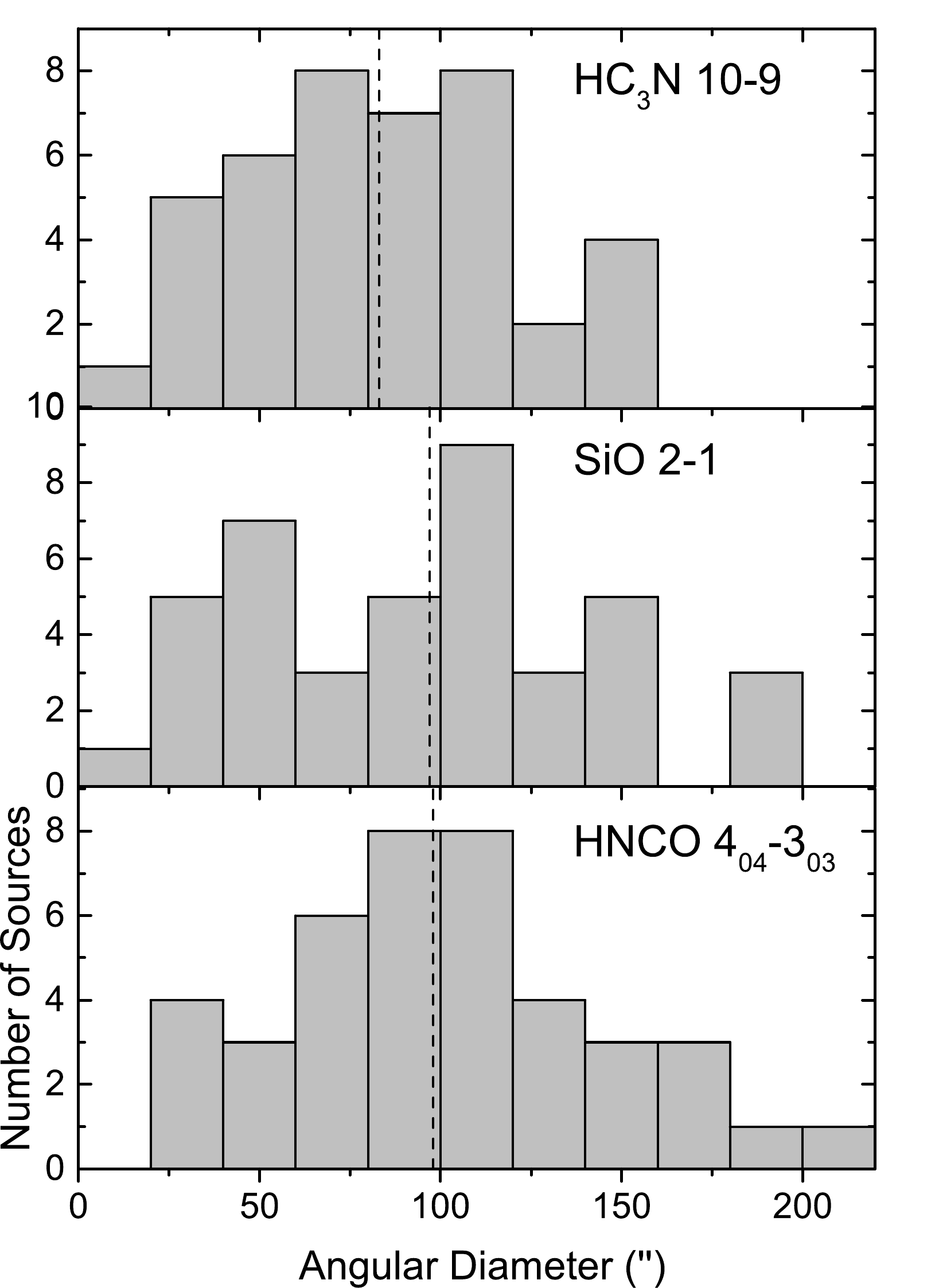}
    \caption{The beam deconvolved angular diameter distributions for HNCO 4$_{04}$--3$_{03}$, SiO 2--1 and HC$_{3}$N 10--9. The median values of each molecule are indicated by dashed black vertical lines.}
    \label{fig:Figure_Angular_diameter_distribution}
\end{figure*}

We characterize the size of HNCO 4$_{04}$--3$_{03}$, SiO 2--1 and HC$_{3}$N 10--9 emission using a beam deconvolved angular diameter of a circle with the same area as the half-peak intensity, which is given by $\theta = 2(\frac{A}{\pi}-\frac{\theta_{beam}^2}{4})^{1/2}$, where A is the area within the half-peak intensity, and $\theta_{beam}$ is the FWHM beam size. The angular diameters of HNCO range from 25$^{\prime\prime}$ to 200$^{\prime\prime}$, with a mean value of 102$^{\prime\prime}$$\pm$45$^{\prime\prime}$, the error being the standard deviation of an individual measurement. Nearly all the angular diameters of HNCO are well above the beam size. The angular diameters of SiO range from 14$^{\prime\prime}$ to 196$^{\prime\prime}$, with a mean value of 97$^{\prime\prime}$$\pm$48$^{\prime\prime}$. Molecular lines of HNCO 4$_{04}$--3$_{03}$ and SiO 2--1 show similar sizes for most sources. The angular diameters of HC$_{3}$N range from 18$^{\prime\prime}$ to 151$^{\prime\prime}$, with a mean value of 81$^{\prime\prime}$$\pm$37$^{\prime\prime}$. Nearly half of the angular diameters of HC$_{3}$N are smaller than the telescope beam size. The average angular diameters of NMSFR, Bubble and CMZ sources are 63$^{\prime\prime}$$\pm$23$^{\prime\prime}$, 61$^{\prime\prime}$$\pm$24$^{\prime\prime}$ and 124$^{\prime\prime}$$\pm$38$^{\prime\prime}$ for HNCO, 43$^{\prime\prime}$$\pm$7$^{\prime\prime}$, 51$^{\prime\prime}$$\pm$20$^{\prime\prime}$ and 122$^{\prime\prime}$$\pm$37$^{\prime\prime}$ for SiO, and 39$^{\prime\prime}$$\pm$10$^{\prime\prime}$, 48$^{\prime\prime}$$\pm$24$^{\prime\prime}$ and 100$^{\prime\prime}$$\pm$28$^{\prime\prime}$ for HC$_{3}$N, respectively. That HC$_{3}$N turns out to be more compact than HNCO and SiO cannot be an effect of relatively weak HC$_{3}$N lines which are not detected in the outskirts of the clouds due to the limited sensitivity of our data. To the contrary, the line tending to be weakest in our study is SiO, thus suggesting that HC$_{3}$N really originates from a smaller volume than HNCO and SiO. Besides, these three species show more spatially extended emission towards the CMZ sources as compared with the other two categories. The distributions of the angular diameters for HNCO 4$_{04}$--3$_{03}$, SiO 2--1 and HC$_{3}$N 10--9 are shown in Fig.~\ref{fig:Figure_Angular_diameter_distribution}.

\section{Discussion} \label{sec:discussion}
\subsection{Previous studies} \label{sec:previ}
In order to investigate relations between properties of HNCO 4$_{04}$--3$_{03}$, SiO 2--1 and HC$_{3}$N 10--9 and to compare our results with previous studies, we introduce a number of comparisons with previous publications directly related to our data in this section. Moreover, we also discuss inflow and outflow activities in the Bubble and NMSFR sources (see Sect.~\ref{sec:ratios_inte_abundance}).

In recent years, statistical studies of molecular clumps in the Milky Way have mainly relied on samples associated with infrared dark clouds (IRDCs) in \citet{2014A&A...562A...3M}, \citet{2012ApJ...756...60S}, \citet{2008ApJ...678.1049S} and \citet{2010ApJ...714.1658S}. IRDCs are extinction features against the background mid-IR emission \citep[e.g.][]{2006ApJ...639..227S,1998ApJ...508..721C}, and clumps therein are cold (10-20 K) compared to clumps severely affected by ongoing massive star formation \citep{1998ApJ...508..721C}. However, observations show that clumps found in IRDCs can also be in various evolutionary stages or subject to H{\sc\,ii} region (bubble) feedback \citep{2012ApJ...756...60S}, which are similar to Bubble and NMSFR category sources defined by us in this study (see Sect.~\ref{sec:sampl}). Seven sources (4 in the Bubble category and 3 in the NMSFR category) in our sample have been identified as IRDCs by \citet{2009A&A...505..405P}. Besides, 18 out of 35 clumps of IRDCs studied by \citet{2014A&A...562A...3M} are located in the CMZ. All this suggests that IRDCs are the most natural sources for comparisons with our data. \citet{2008ApJ...678.1049S} and \citet{2010ApJ...714.1658S} assumed that $T_{\rm ex}$ is equal to the rotation temperatures of NH$_{3}$ and CH$_{3}$OH, and found clump kinetic temperatures in the range 10.3--20.8 K (15.5 K on average) and 13.4--24.5 K (17.3 K on average), respectively. In this work we determine (see Sect.~\ref{sec:dust_tem_mol_col_density}) only slightly higher dust temperatures in the range 18--33 K (20 K on average), so that the IRDCs provide a suitable sample with which our data can be compared. As most sources studied by \citet{2012ApJ...756...60S} show obvious signs of massive star-forming activities, they derived even higher dust temperatures in the range 17--52 K (average value 31 K). All of the above mentioned works have been carried out in a single pointing mode. Because peak positions tend to be strong, the following comparison is not dependent on the observational sensitivity.

To derive column densities, \citet{2014A&A...562A...3M} assumed that $T_{\rm ex}$ is equal to $E_{\rm u}$/k$_{\rm B}$ in the case of the linear molecules SiO and HC$_{3}$N, 6.3 K and 24.0K, and that $T_{\rm ex}$ = 2/3 $\times$ $E_{\rm u}$/k$_{\rm B}$, 7 K, in the case of HNCO, where $E_{\rm u}$ is the energy of the upper state. They derived HNCO column densities of 5.4$\times$10$^{12}$--7.9$\times$10$^{14}$ cm$^{-2}$. Their mean value, 2.2$\times$10$^{14}$ cm$^{-2}$, is about one order of magnitude higher than the value, 3.36$\times$10$^{13}$ cm$^{-2}$, obtained by \citet{2012ApJ...756...60S}. For HC$_{3}$N column densities, \citet{2014A&A...562A...3M} found a median value of 1.6$\times$10$^{14}$ cm$^{-2}$ towards 11 embedded clumps in two IRDCs, almost an order of magnitude higher than the highest value of 5.4$\times$10$^{13}$ cm$^{-2}$ found by \citet{2008ApJ...678.1049S}. Furthermore, \citet{2012ApJ...756...60S} determined an even lower median value of 4.77$\times$10$^{12}$ cm$^{-2}$ in 22 clumps. Because of these differences, we tested HNCO and HC$_{3}$N column density calculations using Eq.~\ref{eqn-den-colum-2} with different excitation temperature. We only find less than a factor of three differences between the column densities determined with 10 K and 30 K. So the significant difference is mainly caused by the spectral line intensity. By checking source positions in \citet{2014A&A...562A...3M}, where HNCO and HC$_{3}$N emissions are detected, we find almost all of them located in the CMZ. Consistent with our data analyzed in Sect.~\ref{sec:mol_line_properties_col_density}, higher spectral line intensities in the CMZ will result in one order of magnitude higher column densities than those in the other two categories.

\citet{2010ApJ...714.1658S} and \citet{2014A&A...562A...3M} derived values of 4.6$\times$10$^{12}$--3.8$\times$10$^{13}$ cm$^{-2}$ and 5.5$\times$10$^{12}$--4.8$\times$10$^{13}$ cm$^{-2}$, with average values of 1.5$\times$10$^{13}$ cm$^{-2}$ and 1.9$\times$10$^{13}$ cm$^{-2}$, for the column densities of SiO, respectively. \citet{2012ApJ...756...60S} have no CMZ sources, thus obtaining smaller SiO column densities of 1.36$\times$10$^{12}$--3.47$\times$10$^{13}$ cm$^{-2}$ with a median of 7.72$\times$10$^{12}$ cm$^{-2}$ in IRDC clumps. Moreover, there are two sources, G1.87$-$SMM 1 and G1.87$-$SMM 23, that have been studied by \citet{2014A&A...562A...3M} analysing the same MALT90 survey data also used by us. The molecular properties of the corresponding sources of G001.655$-$00.062 and G001.833$-$00.062 derived in this work are similar to the values they found.

Here we will also compare our results with those obtained by \citet{2017A&A...597A..11K} and \citet{2015A&A...579A.101A} in external galaxies in the following discussions. All of the above mentioned works have carried out their observation and analysis in a single pointing mode except \citet{2017A&A...597A..11K} in mapping the circumnuclear disk of the Seyfert galaxy NGC1068. By mapping analysis, we can more directly understand spatial variations in HNCO, SiO and HC$_{3}$N emission as already emphasized in Sect.~\ref{sec:indiv}. \citet{2017A&A...597A..11K} found that the mean ratios of $I_{\rm HNCO}/I_{\rm SiO}$ are 0.35$\pm$0.13 and 2.5$\pm$1.3 in the heavily shocked East Knot and mildly shocked West Knot of NGC1068, respectively. In nearby active galaxies, \citet{2015A&A...579A.101A} determined source-averaged column densities of simultaneously detected HNCO 4$_{04}$--3$_{03}$, SiO 2--1 and HC$_{3}$N 10--9 with signal-to-noise ratios S/N $>$ 3 in the starburst galaxies of M\,83, NGC\,253 \citep[see also][]{2006ApJS..164..450M}, and M\,82, the AGN and starburst galaxy NGC\,1068, and the ultra-luminous infrared galaxy (ULIRG) Arp220. The mean abundance ratios of $N_{\rm HNCO}/N_{\rm SiO}$, $N_{\rm SiO}/N_{\rm HC_{3}N}$, and $N_{\rm HNCO}/N_{\rm HC_{3}N}$ in starburst galaxies are 10.94, 0.18, and 1.73, respectively. The corresponding values for AGN and ULIRGs are 4.71, 0.45, and 2.13, and 2.86, 0.23, and 0.07, respectively.

For purposes of comparison with above mentioned works, the HNCO 4$_{04}$--3$_{03}$, SiO 2--1 and HC$_{3}$N 10--9 line properties have been extracted at the yellow crosses of all our 43 sources (see Figs.~\ref{fig:Figure_A1}--\ref{fig:Figure_A22} and Table~\ref{tab:source_parameters_1} and \ref{tab:source_parameters_2}).

\subsection{Molecular line properties and column densities} \label{sec:mol_line_properties_col_density}
\begin{figure*}
  \centering
	\includegraphics[width = 1.0\linewidth]{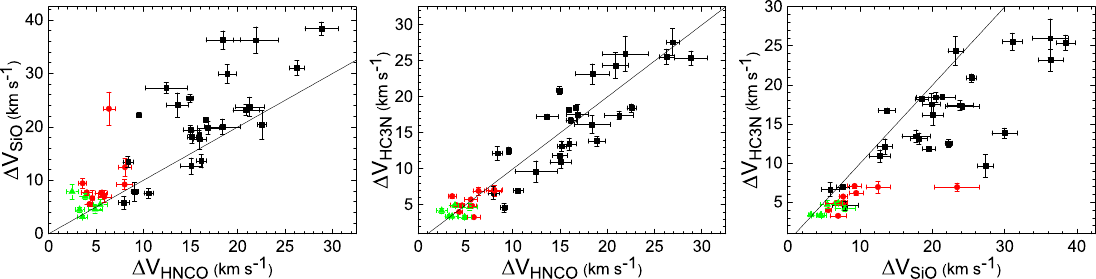}
    \caption{Comparison of the line width for HNCO and SiO (left), HNCO and HC$_{3}$N (middle), and SiO and HC$_{3}$N (right). NMSFR sources are plotted as green triangles. Sources associated with bubbles are plotted as red points, while sources in the CMZ are shown as black squares. The slopes of the black solid lines are 1.}
    \label{fig:Figure_24}
\end{figure*}

Fig.~\ref{fig:Figure_24} shows a comparison of line widths between HNCO and SiO, HNCO and HC$_{3}$N, as well as SiO and HC$_{3}$N. We find that SiO has larger line widths, which agrees with previous research of \citet{2000A&A...361.1079Z}. In addition, the line widths of HNCO and HC$_{3}$N are similar. The mean values of the HNCO line widths are 3.81$\pm$1.34, 5.71$\pm$1.63 and 18.19$\pm$7.46 km s$^{-1}$ for the NMSFR, Bubble, and CMZ samples, respectively. Separately for SiO and HC$_{3}$N, the corresponding values are 4.86$\pm$1.64, 8.80$\pm$2.08 and 23.83$\pm$9.11 km s$^{-1}$, and 3.87$\pm$0.57, 5.40$\pm$1.26 and 17.36$\pm$7.53 km s$^{-1}$. For NMSFR and CMZ sources, the SiO line widths are about 1.3 times larger than those of HNCO and HC$_{3}$N, while for sources in the Bubble category they are 1.4 to 1.6 times larger. SiO has the highest line widths, since it is tracing the strongest shocks.

Fig.~\ref{fig:Figure_25} shows a comparison of integrated intensities (in K km s$^{-1}$) averaged over a circular area positioned on the yellow cross of each map with a diameter corresponding to the $\sim$38$''$ Mopra beamwidth. We find a tight correlation between SiO and HC$_{3}$N (r = 0.89), and a linear fitting relationship:

\begin{equation}
I_{\rm HC_{3}N} = (1.15 \pm 0.09) \times I_{\rm SiO}+(1.08 \pm 2.45) \, .
\label{eq07}
\end{equation}

I$_{\rm HC_{3}N}$ and, to a lesser extent, I$_{\rm SiO}$ also show linear relationships with the integrated intensity of HNCO, with linear Pearson correlation coefficients of r = 0.81 and 0.66, respectively. The relationships are:

\begin{equation}
I_{\rm HC_{3}N} = (0.75 \pm 0.08) \times I_{\rm HNCO}+(6.00 \pm 2.92) \, .
\label{eq07}
\end{equation}

\begin{equation}
I_{\rm SiO} = (0.47 \pm 0.08) \times I_{\rm HNCO}+(8.49 \pm 2.89) \, .
\label{eq07}
\end{equation}

Applying equation~\ref{eq02}, Fig.~\ref{fig:Figure_26} plots optical depths. We find that the three molecular lines studied here are all optically thin ($\tau$ $\ll$ 1) in our star-forming regions, which agrees with the previous study of \citet{2014A&A...562A...3M}. For sources in the CMZ, the optical depths of the HNCO lines are similar to those of the HC$_{3}$N lines. The optical depths of the HC$_{3}$N and SiO lines appear to be higher relative to HNCO in NMSFR and Bubble sources.

Using the equations in Sect.~\ref{sec:mol_col_density_and_frac_abundance}, column densities and abundances of HNCO, SiO and HC$_{3}$N are derived and presented in columns 7 and 9 of Table~\ref{tab:source_parameters_2}. The column densities of HNCO, SiO and HC$_{3}$N for their whole sample to be compared with those by other groups, summarized in Sect.~\ref{sec:previ}, range from 1.91$\times$10$^{13}$--1.39$\times$10$^{15}$ cm$^{-2}$, 6.30$\times$10$^{12}$--1.83$\times$10$^{14}$ cm$^{-2}$ and 1.83$\times$10$^{13}$--4.50$\times$10$^{14}$ cm$^{-2}$, with mean values of 2.93$\times$10$^{14}$ cm$^{-2}$, 4.83$\times$10$^{13}$ cm$^{-2}$ and 1.19$\times$10$^{14}$ cm$^{-2}$, respectively. The corresponding median values are 1.98$\times$10$^{14}$ cm$^{-2}$, 3.00$\times$10$^{13}$ cm$^{-2}$, and 5.92$\times$10$^{13}$ cm$^{-2}$. The mean abundances of HNCO, SiO, and HC$_{3}$N are 3.13$\times$10$^{-9}$, 5.24$\times$10$^{-10}$, and 1.14$\times$10$^{-9}$, respectively. The corresponding median values are 3.44$\times$10$^{-9}$, 3.28$\times$10$^{-10}$, and 8.96$\times$10$^{-10}$. \citet{2012ApJ...756...60S} derived an SiO abundance comparable to that of ours for our whole sample.

The mean HNCO column density we find, 2.93$\times$10$^{14}$ cm$^{-2}$, is quite similar to those derived by \citet{2014A&A...562A...3M} towards their IRDC sources. Our median column density of 1.98$\times$10$^{14}$ cm$^{-2}$ is about five times higher than the median value found by \citet{2012ApJ...756...60S}. Because most sources of our work are part of the CMZ or associated with bubbles, the surrounding environment may cause this difference. The average HNCO column densities of NMSFR, Bubble and CMZ are 3.75$\times$10$^{13}$, 5.81$\times$10$^{13}$ and 4.37$\times$10$^{14}$ cm$^{-2}$, respectively. The corresponding median values are 3.97$\times$10$^{13}$, 5.60$\times$10$^{13}$ and 3.03$\times$10$^{14}$ cm$^{-2}$. We find that the average and median column densities of HNCO for each category increase as in the case of SiO monotonically from the NMSFR sources to the Bubble sources, to the CMZ sources. This suggests that shocks in the Bubble category are strong enough to evaporate a significant amount of HNCO from grain mantles without dissociating them in comparison with sources of the NMSFR category. HNCO in the CMZ is much more abundant than the other two species, SiO and HC$_{3}$N, and seems to react more sensitively to its extreme environment.

The mean values of SiO peak column densities of NMSFR, Bubble and CMZ clouds we derive are 9.58$\times$10$^{12}$ cm$^{-2}$, 2.88$\times$10$^{13}$ cm$^{-2}$, and 6.97$\times$10$^{13}$ cm$^{-2}$, respectively. The corresponding median values are 8.15$\times$10$^{12}$ cm$^{-2}$, 1.89$\times$10$^{13}$ cm$^{-2}$, and 4.76$\times$10$^{13}$ cm$^{-2}$. The NMSFR values are similar to those found by \citet{2010ApJ...714.1658S} for their sample of clumps within IRDCs, and by \citet{2014A&A...562A...3M} for their sample of massive clumps (see Sect.~\ref{sec:previ}).

For the HC$_{3}$N peak column densities of NMSFR, Bubble and CMZ sources we derive mean values of 3.49$\times$10$^{13}$, 5.81$\times$10$^{13}$, and 1.60$\times$10$^{14}$ cm$^{-2}$, and  median values of 3.08$\times$10$^{13}$, 5.09$\times$10$^{13}$, and 1.18$\times$10$^{14}$ cm$^{-2}$, respectively. Again column densities are rising from the NMSFR to the CMZ sources. The median HC$_{3}$N column density found by \citet{2014A&A...562A...3M} (11 out of their 12 detections are in the CMZ) is quite similar to our value for the CMZ clouds. Our average value is almost an order of magnitude higher than the highest value found by \citet{2008ApJ...678.1049S} towards massive clumps associated with IRDCs located in the Galactic disk. Our median column density in NMSFR (3.08$\times$10$^{13}$ cm$^{-2}$) exceeds the value found by \citet{2012ApJ...756...60S} by a factor of five. From the foregoing, we find that the CMZ contains large reservoirs of SiO, HNCO and HC$_{3}$N gas, with almost one order of magnitude higher column densities than in the disc, especially when being compared to the sources in the NMSFR category.

Fig.~\ref{fig:Figure_27} confirms that column densities of our three molecules increase from NMSFR, to Bubble, and on to sources with CMZ classification. A particularly strong correlation coefficient r = 0.88 is found for the column densities of SiO and HC$_{3}$N for the entire sample. However, the differences between the Pearson correlation coefficients are not that large, so that this is not a finding we will emphasize. Least squares fitting yields
\begin{equation}
log(N_{\rm HC_{3}N}) = (2.24 \pm 0.98)+(0.86 \pm 0.07) \times log(N_{\rm SiO}) \, .
\label{eq08}
\end{equation}

Moreover, a slightly weaker correlation is obtained between the column densities of HNCO and SiO (r = 0.71) for the whole sample, as well as between the column densities of HNCO and HC$_{3}$N (r = 0.79). Results given by linear fitting are
\begin{equation}
log(N_{\rm SiO}) = (5.33 \pm 1.26)+(0.58 \pm 0.09) \times log(N_{\rm HNCO}) \, ,
\label{eq08}
\end{equation}
and
\begin{equation}
log(N_{\rm HC_{3}N}) = (4.97 \pm 1.07)+(0.63 \pm 0.08) \times log(N_{\rm HNCO}) \, .
\label{eq08}
\end{equation}

The high correlations between SiO and HC$_{3}$N in integrated intensities and column densities above indicate that there is a close relationship during the process of their chemical evolution. This argument will contribute to determine the dominant chemical model through numerical simulations of star-forming regions, where large scale shocks are generated, e.g. due to external star forming activities or outflows from embedded young stellar objects.

\subsection{The ratios of integrated intensity and abundances} \label{sec:ratios_inte_abundance}
Fig.~\ref{fig:Figure_abundance} represents a comparison of the fractional abundances of HNCO, SiO and HC$_{3}$N. Two obvious clusters are seen between the CMZ sources and the other two categories, implying that the chemical properties of sources in the CMZ are different from sources far from the Galactic Center. A high correlation (right panel of Fig.~\ref{fig:Figure_abundance}) is obtained between the fractional abundances of SiO and HC$_{\rm 3}$N (r = 0.93) for the whole sample, \textbf{possibly because their parent species Si and C$_{2}$H$_{2}$ are both released from dust grains due to a sudden increase in temperature or to the erosion of dust grains via sputtering or grain-grain collisions in shocks.} The linear fitting result (red dashed line) is

\begin{equation}
log(x_{\rm HC_{3}N}) = (-0.94 \pm 0.50)+(0.86 \pm 0.05) \times log(x_{\rm SiO}) \, .
\label{eq04}
\end{equation}

The upper cluster only contains CMZ sources, while the lower cluster contains NMSFR and Bubble sources. Moreover, Fig.~\ref{fig:Figure_abundance} shows that HNCO is the most abundant molecule in the CMZ category, while the fractional abundance of SiO tends to be the lowest in the whole sample. For sources in the Bubble and NMSFR category, the fractional abundances of HNCO and HC$_{\rm 3}$N present a linear distribution with a slope of almost unity, while SiO is slightly less abundant..

In order to have a brief understanding of chemical properties of the three molecules HNCO, SiO and HC$_{3}$N, we computed integrated intensity and abundance ratios. ${I_{\rm HNCO}/I_{\rm SiO}}$, ${I_{\rm SiO}/I_{\rm HC_{3}N}}$, ${I_{\rm HNCO}/I_{\rm HC_{3}N}}$, ${N_{\rm HNCO}/N_{\rm SiO}}$, ${N_{\rm SiO}/N_{\rm HC_{3}N}}$, and ${N_{\rm HNCO}/N_{\rm HC_{3}N}}$ ratios are listed for the whole sample in Table~\ref{tab:source_ratios}. Statistics provided in Table~\ref{tab:source_statistics} include the mean, median and standard deviation (std), as well as minimum and maximum values of the sample.

\begin{figure*}
  \centering
	\includegraphics[width = 1.0\linewidth]{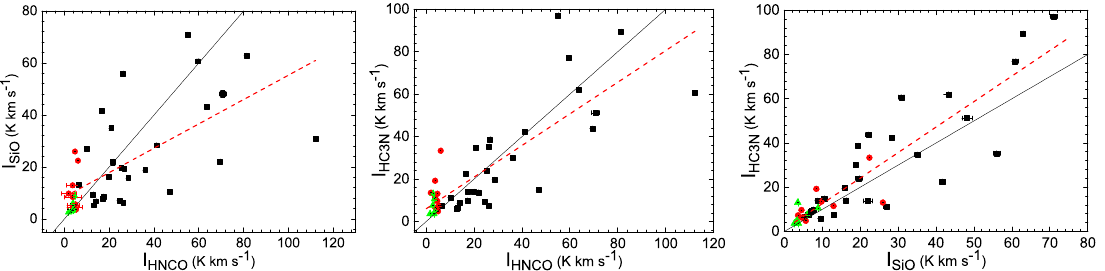}
    \caption{Same as Fig.~\ref{fig:Figure_24}, but for integrated intensity. The red dashed lines indicate the best-fit relation. The slopes of the black solid lines are 1.}
    \label{fig:Figure_25}
\end{figure*}

\begin{figure*}
  \centering
	\includegraphics[width = 1.0\linewidth]{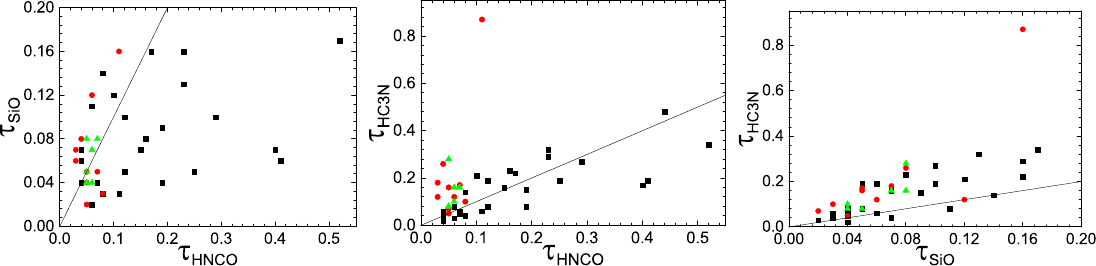}
    \caption{Same as Fig.~\ref{fig:Figure_25}, but for optical depth. The slopes of the black solid lines are 1.}
    \label{fig:Figure_26}
\end{figure*}

\begin{figure*}
  \centering
	\includegraphics[width = 1.0\linewidth]{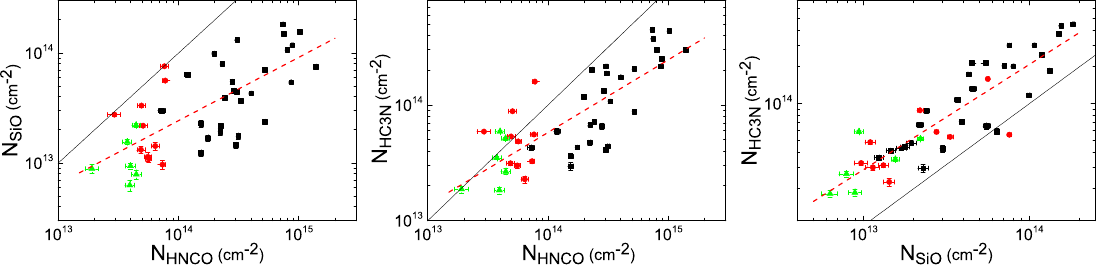}
    \caption{Same as Fig.~\ref{fig:Figure_25}, but for column density. The slopes of the black solid lines are 1.}
    \label{fig:Figure_27}
\end{figure*}

\begin{figure*}
  \centering
	\includegraphics[width = 1.0\linewidth]{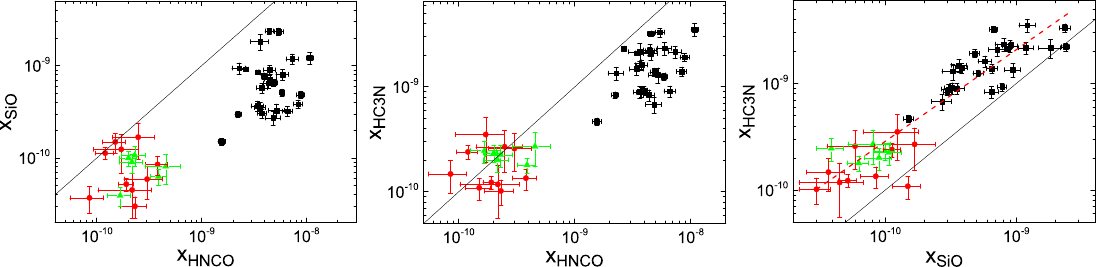}
    \caption{Same as Fig.~\ref{fig:Figure_25}, but for fractional abundance. The slopes of the black solid lines are 1.}
    \label{fig:Figure_abundance}
\end{figure*}

\begin{figure*}
  \centering
	\includegraphics[width = 1.0\linewidth]{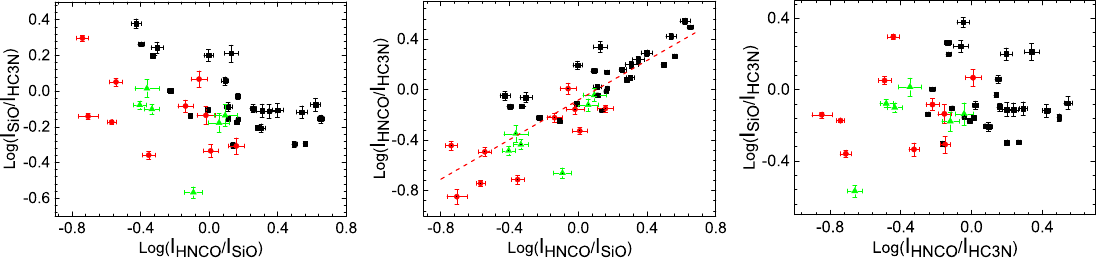}
    \caption{Left: plot of the integrated intensity ratios of SiO to HC$_{3}$N vs. the integrated intensity ratios of HNCO to SiO. Middle: plot of the integrated intensity ratios of HNCO to HC$_{3}$N vs. the integrated intensity ratios of HNCO to SiO. Right: plot of the integrated intensity ratios of SiO to HC$_{3}$N vs. the integrated intensity ratios of HNCO to HC$_{3}$N. For the color of the plotted intensity ratios, see the caption to Fig.~\ref{fig:Figure_24}. The red dashed line in the central panel represents the least-squares fitting result.}
    \label{fig:Figure_28}
\end{figure*}

\begin{figure*}
  \centering
	\includegraphics[width = 1.0\linewidth]{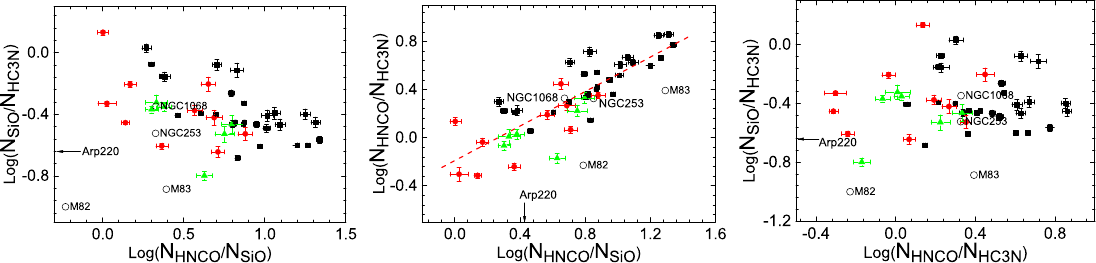}
    \caption{Same as Fig.~\ref{fig:Figure_28}, but for column density. The open circles represent the starburst galaxies of M\,83, NGC\,253, and M\,82, and the AGN and starburst galaxy NGC\,1068. The black arrows point to the ultraluminous infrared galaxy Arp\,220. The red dashed line in the central panel indicates the least-squares fit.}
    \label{fig:Figure_29}
\end{figure*}

From the left panel of Fig.~\ref{fig:Figure_28}, we deduce no correlation between the integrated intensities of ${I_{\rm HNCO}/I_{\rm SiO}}$ and ${I_{\rm SiO}/I_{\rm HC_{3}N}}$. It is interesting to find from Figs.~\ref{fig:Figure_28} and \ref{fig:Figure_29} that ${I_{\rm HNCO}/I_{\rm SiO}}$, ${I_{\rm HNCO}/I_{\rm HC_{3}N}}$, ${N_{\rm HNCO}/N_{\rm SiO}}$, and ${N_{\rm HNCO}/N_{\rm HC_{3}N}}$ ratios for Bubble sources present two separate groups, one of which (G322.159+00.635, G327.293$-$00.579, G345.004$-$00.224, G351.443+00.659, and G351.775$-$00.537) has smaller ratios, while the others (G008.671$-$00.357, G010.473+00.028, G326.653+00.618, G350.101+00.083, and G351.582$-$00.352) have higher ratios, compared to those of the NMSFR sources. Inflow and outflow activities have been simultaneously detected in three (G345.004$-$00.224, G351.443+00.659, G351.775$-$00.537) out of five sources in the group with smaller ratio (see Sect.~\ref{sec:outflow_inflow}). This may indicate that the combined action of star formation activities and H{\sc\,ii} region (bubble) feedback will enhance the abundances of SiO and HC$_{3}$N. The middle panel of Fig.~\ref{fig:Figure_28} plots ${I_{\rm HNCO}/I_{\rm HC3N}}$ as a function of ${I_{\rm HNCO}/I_{\rm SiO}}$. A strong positive correlation is found, and the fitted linear relationship is of the form ${log(I_{\rm HNCO}/I_{\rm HC_{3}N})} = (-0.08 \pm 0.03)+(0.78 \pm 0.08) \times {log(I_{\rm HNCO}/I_{\rm SiO})}$, with a Pearson's r = 0.84. This is due to the good linear correlation between ${I_{\rm SiO}}$ and ${I_{\rm HC_{3}N}}$ shown in Fig.~\ref{fig:Figure_25}. Moreover, the integrated intensity ratios of our three molecules, shown in the middle panel of Fig.~\ref{fig:Figure_28}, increase from the Bubble sources with small ratios to the NMSFR sources, then to the Bubble sources with higher ratios and finally to the CMZ sources. The integrated intensity ratio of ${I_{\rm HNCO}/I_{\rm HC_{3}N}}$ shows no obvious correlation with ${I_{\rm SiO}/I_{\rm HC_{3}N}}$ in the right panel of Fig.~\ref{fig:Figure_28}. As shown in the left panel of Fig.~\ref{fig:Figure_29}, there is a weak hint that the abundance ratio of ${N_{\rm SiO}/N_{\rm HC_{3}N}}$ decreases with an increasing ${N_{\rm HNCO}/N_{\rm SiO}}$ abundance ratio. Instead, a high correlation coefficient, 0.84, is obtained for the abundance ratio of ${N_{\rm HNCO}/N_{\rm HC_{3}N}}$ versus ${N_{\rm HNCO}/N_{\rm SiO}}$ for the whole sample in the middle panel of Fig.~\ref{fig:Figure_29}. The red dashed line indicates the least-squares fit expressed as ${log(N_{\rm HNCO}/N_{\rm HC_{3}N})} = (-0.19 \pm 0.06)+(0.72 \pm 0.07) \times {log(N_{\rm HNCO}/N_{\rm SiO})}$ with r = 0.84.  The abundance ratio of ${N_{\rm SiO}/N_{\rm HC_{3}N}}$ shows no obvious trend as a function of the ${N_{\rm HNCO}/N_{\rm HC_{3}N}}$ abundance ratio.

The average abundance ratios of ${N_{\rm HNCO}/N_{\rm SiO}}$ and ${N_{\rm HNCO}/N_{\rm HC_{3}N}}$ we find in the CMZ and the Bubble sources are 8.79 and 3.33, and 3.30 and 1.35, respectively. These are quite similar to those derived by \citet{2015A&A...579A.101A} towards starburst galaxies. But their average ${N_{\rm SiO}/N_{\rm HC_{3}N}}$ abundance ratio is almost three times lower than the corresponding values in the CMZ. Interestingly, abundance ratios in the AGN NGC\,1068 appear to be similar to the NMSFR sources. Beside the central region of our Galaxy we suggest that star formation regions having extended SiO and HNCO emissions in the Galactic disk can also be testbeds for studies of the chemistry of the interstellar medium in the nuclear regions of galaxies. It would be worthwhile to carry out high-sensitivity observations and to perform large sample statistics in the future.

We find that the mean ${I_{\rm HNCO}/I_{\rm SiO}}$ ratios and their standard deviations obtained from the Bubble, NMSFR and CMZ sources are 0.64$\pm$0.43, 0.75$\pm$0.38, and 1.76$\pm$1.15, respectively. As already mentioned (Sect.~\ref{sec:previ}), \citet{2017A&A...597A..11K} found that the mean ratios of ${I_{\rm HNCO}/I_{\rm SiO}}$ are 0.35$\pm$0.13 and 2.5$\pm$1.3 in the heavily shocked East Knot and mildly shocked West Knot of NGC1068, respectively. In the IRDC clumps studied by \citet{2014A&A...562A...3M}, the ${I_{\rm HNCO}/I_{\rm SiO}}$ ratio was found to be 2.37$\pm$0.74, comparable to our average value within the error limits towards the CMZ category. Moreover, the median abundance ratio ${N_{\rm HNCO}/N_{\rm SiO}}$, 6.47, found by \citet{2014A&A...562A...3M}, is very similar to 7.38 derived in this work towards the CMZ group of sources. This is explained by the fact that most of the clumps with simultaneously detected HNCO and SiO lines \citep[by][]{2014A&A...562A...3M} are also located in the Galactic center region.

\begin{figure*}
  \centering
	\includegraphics[width = 1.0\linewidth]{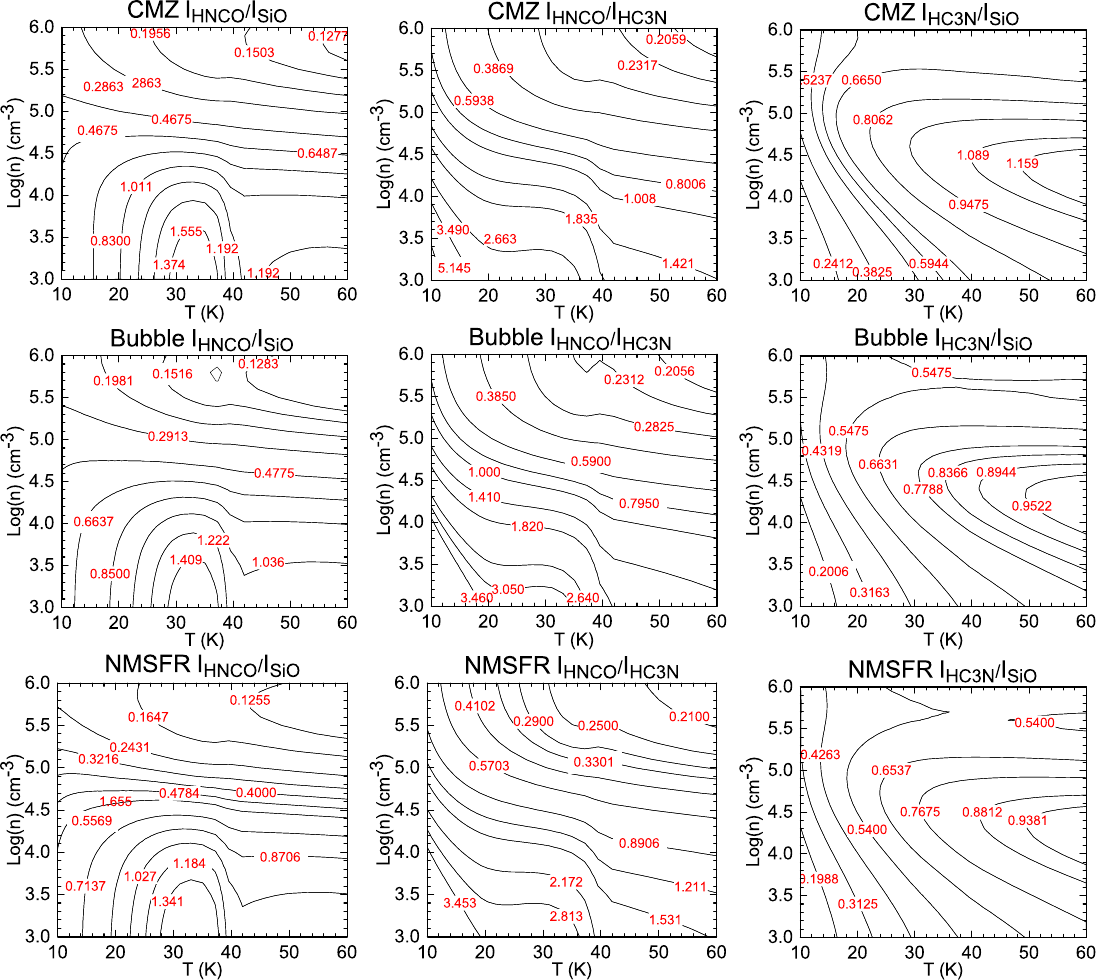}
    \caption{Integrated intensity ratios of ${I_{\rm HNCO}/I_{\rm SiO}}$, ${I_{\rm HNCO}/I_{\rm HC_{3}N}}$ and ${I_{\rm HC_{3}N}/I_{\rm SiO}}$ from RADEX models for sources of CMZ, Bubble, and NMSFR category with varying temperature from 10 K to 60 K and hydrogen number density from 10$^{3}$ cm$^{-3}$ to 10$^{6}$ cm$^{-3}$. The corresponding averaged input HNCO line widths and column densities are 17.49, 5.20 and 3.62 km s$^{-1}$, and 4.37$\times$10$^{14}$, 5.81$\times$10$^{13}$ and 3.75$\times$10$^{13}$ cm$^{-2}$. While the input line widths are the same for SiO and HC$_{3}$N, their column densities used for the plots are 9.58$\times$10$^{12}$ and 3.49$\times$10$^{13}$, 2.88$\times$10$^{13}$ and 5.81$\times$10$^{13}$, and 6.97$\times$10$^{13}$  and 1.60$\times$10$^{14}$ cm$^{-2}$, respectively. Horizontal isophotes indicate density, while vertical isophotes show kinetic temperature tracers.}
    \label{fig:Figure_30}
\end{figure*}

In order to characterize the integrated intensity ratios of ${I_{\rm HNCO}/I_{\rm SiO}}$, ${I_{\rm HNCO}/I_{\rm HC_{3}N}}$, ${I_{\rm HC_{3}N}/I_{\rm SiO}}$ and their relationship with hydrogen column density and gas kinetic temperature, we have applied a RADEX \citep{2007A&A...468..627V} analysis. We separately analyzed sources in the NMSFR, Bubble and CMZ category, using averaged HNCO line widths of 3.92, 5.61 and 17.82 \,km\,s$^{-1}$ , respectively. The corresponding input values for the HNCO column densities are 3.75$\times$10$^{13}$, 5.81$\times$10$^{13}$ and 4.37$\times$10$^{14}$ cm$^{-2}$ as derived from the results of Table~\ref{tab:source_parameters_2}. The corresponding column densities for SiO and HC$_{3}$N are 9.58$\times$10$^{12}$ and 3.49$\times$10$^{13}$, 2.88$\times$10$^{13}$ and 5.81$\times$10$^{13}$, and 6.97$\times$10$^{13}$  and 1.60$\times$10$^{14}$ cm$^{-2}$. Our grid of models includes varying hydrogen number densities from 10$^{3}$ cm$^{-3}$ to 10$^{6}$ cm$^{-3}$ and temperatures from 10 K to 60 K. The results are displayed in Fig.~\ref{fig:Figure_30}. In each column of Fig.~\ref{fig:Figure_30}, we find that the integrated intensity ratios in different classes show similar trends. ${I_{\rm HNCO}/I_{\rm SiO}}$ and ${I_{\rm HNCO}/I_{\rm HC_{3}N}}$, shown in the left and middle column of Fig.~\ref{fig:Figure_30}, are most sensitive to the hydrogen volume density n, while ${I_{\rm HC_{3}N}/I_{\rm SiO}}$,  shown in the right column of Fig.~\ref{fig:Figure_30} is more sensitive to the kinetic temperature $T_{\rm kin}$.

\begin{figure*}
  \centering
	\includegraphics[width = 1.0\linewidth]{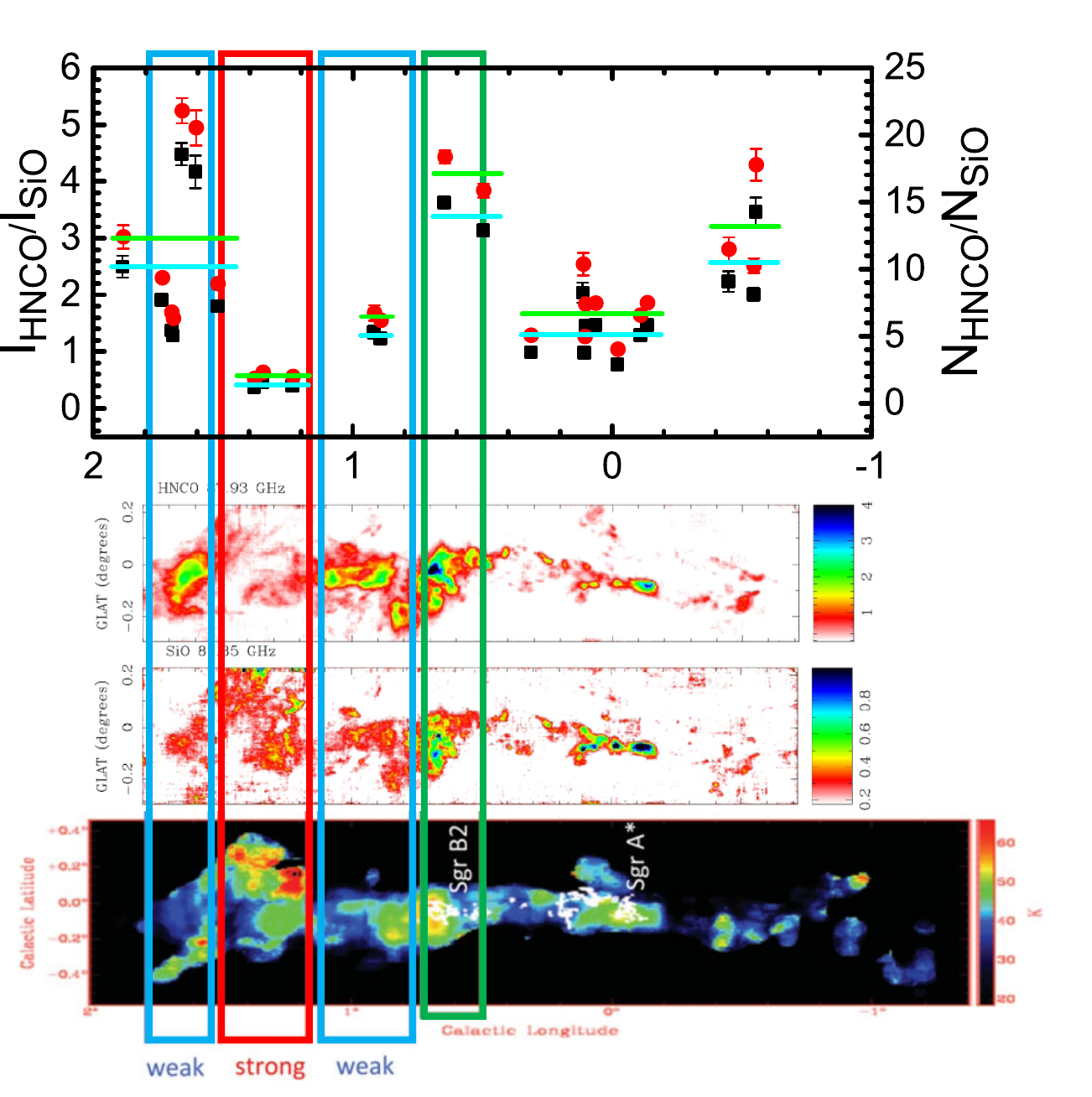}
    \caption{Integrated intensity ratios ${I_{\rm HNCO}/I_{\rm SiO}}$ (black squares) and abundance ratios ${N_{\rm HNCO}/N_{\rm SiO}}$ (red points) for seven separate Galactic longitude segments (top panel; the abcissa is in units of degrees). The averaged ${I_{\rm HNCO}/I_{\rm SiO}}$ indicated by cyan segments from left to right are 2.51, 0.41, 1.29, 3.38, 1.31 and 2.57, respectively. The corresponding averaged ${N_{\rm HNCO}/N_{\rm SiO}}$ ratios indicated by green segments are 12.29, 2.06, 6.46, 17.10, 6.66 and 13.16. The central two panels show the peak brightness images of the HNCO 4$_{\rm 04}$-3$_{\rm 03}$ and SiO 2-1 lines in units of K in the CMZ, taken from Fig. 1 of \citet{2014IAUS..303..104O}. The bottom panel shows the gas temperature map derived from NH$_{3}$ \citep[also][]{2014IAUS..303..104O}. The regions with weak and strong shocks are marked in blue and red rectangles, respectively. The green rectangle centers on the star forming region Sgr B2 and marks the likely point where the $x_{1}$ orbits accrete on the 100 pc ring, the location of the $x_{2}$ orbits.}
    \label{fig:Figure_31}
\end{figure*}

The Central Molecular Zone contains, for its limited volume, a large amount of molecular gas, existing under extreme physical conditions with respect to pressure and kinetic temperature \citep{2016A&A...586A..50G}. From the Mopra telescope HNCO, SiO and NH$_{3}$ mapping data towards the Central Molecular Zone of our Galaxy, \citet{2014IAUS..303..104O} found that there exists a distinct pattern of alternating strong SiO or HNCO lines. They suggested that the difference is due to the strength of the shocks. In order to quantitatively study this phenomenon, we plot the integrated intensity ratio ${I_{\rm HNCO}/I_{\rm SiO}}$ (black squares) and abundance ratios ${N_{\rm HNCO}/N_{\rm SiO}}$ (red points) for seven separate Galactic longitude segments. The averaged ${I_{\rm HNCO}/I_{\rm SiO}}$ indicated by cyan segments in Fig.~\ref{fig:Figure_31} from left to right are 2.51, 0.41, 1.29, 3.38, 1.31 and 2.57, respectively. The corresponding averaged ${N_{\rm HNCO}/N_{\rm SiO}}$ ratios indicated by green segments are 12.29, 2.06, 6.46, 17.10, 6.66 and 13.16. This variation is compatible with alternating stronger and weaker shocks affected by the motion of clouds and the dynamics of the $x_{1}$ and $x_{2}$ orbits in the Central Molecular Zone \citep[e.g.][]{2019MNRAS.486.3307D}, and is also compatible with the data from \citet{2014IAUS..303..104O}.

\subsection{Origin of the SiO emission: Outflow activity versus H\,{\footnotesize II}/SNR-induced shocks} \label{sec:outflow_inflow}
SiO is an excellent tracer of recent shock activity (e.g. active outflows). When Si is liberated from dust grains, it then reacts with other species observable from the ground to form SiO within $\sim$ 10$^{4}$ yr \citep{1997IAUS..182..199P}. \textbf{\citet{2016A&A...589A..29W} indeed proposed that most detected SiO emission is likely to be a result of outflow activity. However, models and observations also suggest that SiO emission could be caused by the UV illumination of ices in Photon Dominated Regions \citep[PDR;][]{2001A&A...372..291S,2004ApJ...601..952S} or by the expansion of HII regions or Supernova Remnants \citep[SNR;][]{2019ApJ...881L..42C,2020MNRAS.499.1666C}, and not by an outflow.} In \citet{2001A&A...372..291S}, narrow SiO emission (average line width of 2.0 km s$^{-1}$) with abundances of the order of 10$^{-11}$ -- 10$^{-10}$ is produced directly by the photo-desorption of a small fraction of silicon from the mantles of dust grains by the intense UV radiation field originating from the massive stars in the Trapezium. However, here we propose a different scenario: molecular gas affected by the expansion of an H{\sc\,ii} region or an SNR could induce low-velocity shocks in the surrounding molecular environment that inject SiO into the gas phase. This is more in line with the proposed scenario by \citet{2019ApJ...881L..42C} for spectrally narrow SiO emission (average line width of 1.6 km s$^{-1}$) detected toward IRDC G034.77$-$00.55, where the observed narrow SiO emission is produced by low-velocity shocks induced by the expansion of a supernova remnant. As star formation properties in the CMZ are complicated, here we focus on dynamical processes (i.e. inflow and outflow activities) in our Bubble and NMSFR sources.

\textbf{To estimate the fraction of sources where SiO is produced by outflow activity, we use HCO$^+$ which is a suitable species} to trace both inflow and outflow activities by analyzing its line profile \citep{2004MNRAS.351.1054R,2005A&A...442..949F,2015MNRAS.450.1926H,2016MNRAS.461.2288H}. To identify HCO$^{+}$ outflow features, we followed the method that was presented by \citet{2005ApJ...628L..57W}. HCO$^{+}$ (1-0) spectra near the source position should have high-velocity wings and position-velocity (P-V) diagrams should show intensities of the wing emission decreasing smoothly towards the edge of the mapped region. Also, we have checked the optically thin H$^{13}$CO$^{+}$ 1--0 or HNCO 4$_{04}$--3$_{03}$ lines to avoid multiple velocity components causing wing emission. As a result, we identify 4 outflow candidates (G008.671$-$00.357, G345.004$-$00.224, G351.443$+$00.659, G351.775$-$00.537) from 10 sources in the Bubble category and 4 outflow candidates (G329.030$-$00.202, G331.708$+$00.583, G331.709$+$00.602, G335.586$-$00.289) from the 6 sources in the NMSFR category. The detection rates of outflow candidates in the Bubble and NMSFR categories are 40\% and 67\%, respectively. The difference between 40\% and 67\% is not significant in a statistical sense and may be an effect of sensitivity limitations of the Mopra observations. The wing emission is weak and it may not have been detected in all sources. Therefore, it is necessary to investigate the outflow detection rate on the basis of a larger sample and with higher sensitivity to check if surrounding bubbles or cloud heating by shock waves can yield not only outflow activity but also an enhancement of SiO emission.

\begin{figure*}
  \centering
	\includegraphics[width = 0.76\linewidth]{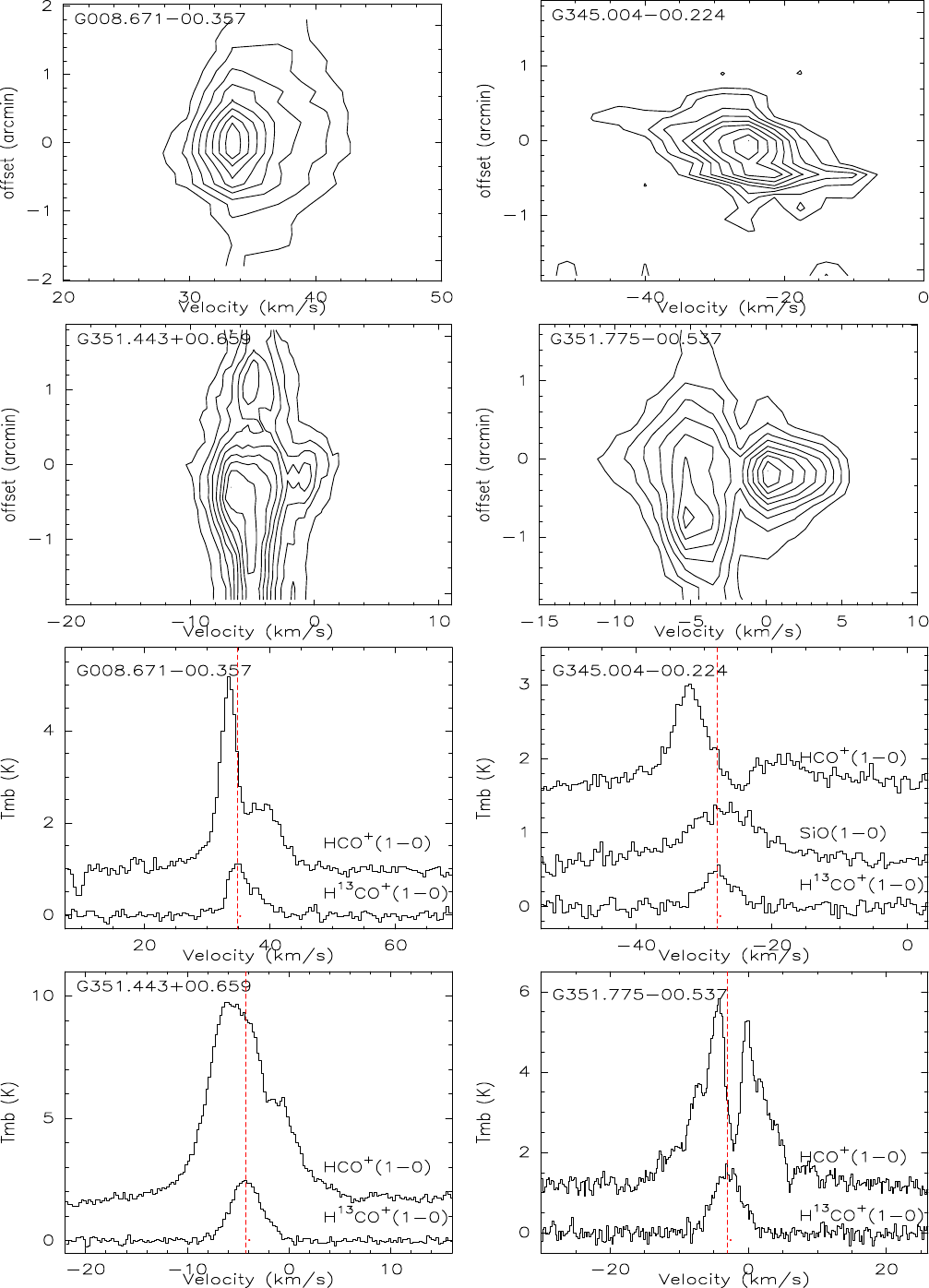}
    \caption{The upper four panels show the HCO$^{+}$ (1--0) P-V diagrams of the three outflow candidates G008.671$-$00.357, G351.443+00.659 and G351.775$-$00.537, and the SiO (2--1) P-V diagrams of the outflow candidate G345.004$-$00.224 in the Bubble category. All the P-V diagrams were cut along the east-west direction except for the P-V diagram of G351.443+00.659 which was cut along the south-north direction. Here the four cardinal directions, north, south, east and west, are meant with respect to the Galactic coordinates ($l^{\rm II}$, $b^{\rm II}$), and not with respect to Right Ascension and Declination. The lower four panels show the extracted spectra of HCO$^{+}$(1--0) and H$^{13}$CO$^{+}$(1--0) from the yellow cross position (see Figs.~\ref{fig:Figure_A15} and \ref{fig:Figure_A17} -- \ref{fig:Figure_A19}) of each outflow candidate. Red dashed lines indicate the central velocity of the sources.}
    \label{fig:Figure_32}
\end{figure*}

\begin{figure*}
  \centering
	\includegraphics[width = 0.8\linewidth]{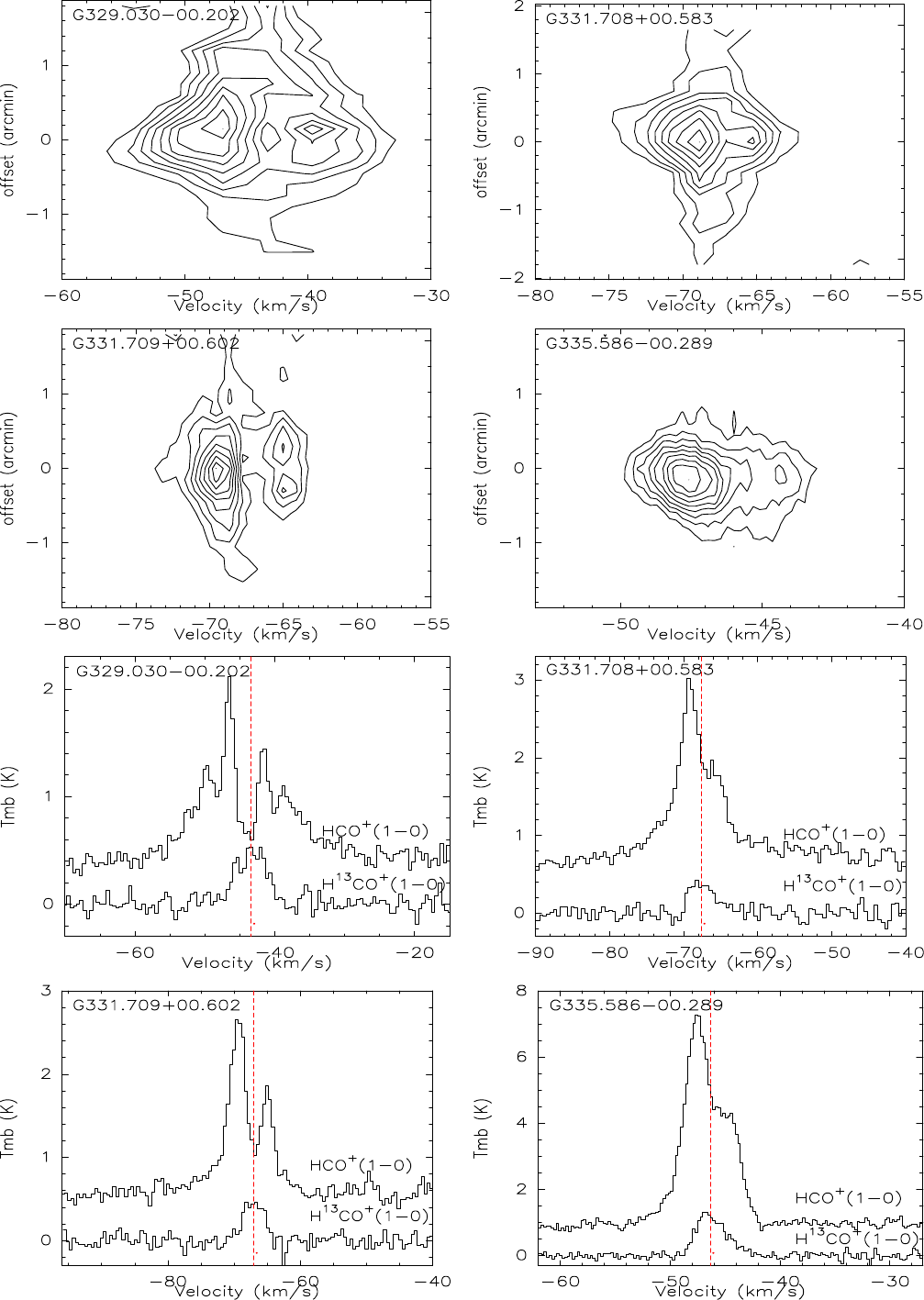}
    \caption{The upper four panels show the HCO$^{+}$ (1--0) P-V diagrams of four outflow candidates in the NMSFR category. All the P-V diagrams were cut along the east-west direction.  Here the four cardinal directions, north, south, east and west, are meant with respect to the Galactic coordinates ($l^{\rm II}$, $b^{\rm II}$), and not with respect to Right Ascension and Declination. The lower four panels show the extracted spectra of HCO$^{+}$(1--0) and H$^{13}$CO$^{+}$(1--0) from the yellow cross position (see Figs.~\ref{fig:Figure_A20} and \ref{fig:Figure_A21}) of each outflow candidate. Red dashed lines indicate the central velocity of the sources.}
    \label{fig:Figure_33}
\end{figure*}

The mean SiO abundances in the Bubble and NMSFR categories are 9.37$\times$10$^{-11}$ and 8.70$\times$10$^{-11}$, respectively. In combination with the abundances derived by \citet{2001A&A...372..291S} from SiO 2--1 in the photon dominated region of the Orion Bar, in the range 3$-$7$\times$10$^{-11}$, this suggests that the pressure of photoionized gas from the surrounding bubbles or cloud heating by shock waves yields an enhancement of SiO emission for sources in the Bubble category besides the outflow activity. The HCO$^{+}$ (1-0) position velocity (P-V) diagrams of these eight outflow candidates showing distinct wing emission are shown in Figs.~\ref{fig:Figure_32} and~\ref{fig:Figure_33}. Among these, G345.004-00.224 was identified by \citet{2014MNRAS.440.1213Y}, while G008.671$-$00.357, G329.030$-$00.202, G331.708$+$00.583, G331.709$+$00.602, G335.586$-$00.289, G351.443$+$00.659 and G351.775$-$00.537 are newly identified in this work. As the HCO$^{+}$ spectrum in G345.004$-$00.224 exhibits an obvious self-absorption dip, the wide SiO 2--1 line emission has been added to show outflow activities in Fig. \ref{fig:Figure_32}. All the P-V diagrams were cut along the $l$ direction in the Galactic coordinate system except for the P-V diagram of G351.443+00.659 which was cut along Galactic latitude where the HCO$^{+}$ 1--0 wing emission is more pronounced.

\citet{2012MNRAS.422.1098R} proposed that inflow and outflow motions should be closely related and interact with each other throughout the star-formation process. We followed the inflow identification criterion used in \citet{2015MNRAS.450.1926H}. Inflow motions have been studied by investigating the profile of the optically thick HCO$^{+}$ 1--0 line, which would then show a double peak with a brighter blue peak or a skewed single blue peak. Meanwhile, the optically thin single peak of the H$^{13}$CO$^{+}$ 1--0 line should be located at the dip of the optically thick HCO$^{+}$ 1--0 line to rule out a double peak caused by two velocity components along the line of sight. After checking the spatial variation of optically thick line profile asymmetries across the mapped region, we identified eight reliable inflow candidates. The extracted spectra of HCO$^{+}$ 1--0 and H$^{13}$CO$^{+}$ 1--0 from the yellow cross of each source are shown in the lower panels of \ref{fig:Figure_32} and \ref{fig:Figure_33}. Five (G008.671$-$00.357, G331.708$+$00.583, G331.709$+$00.602, G335.586$-$00.289 and G351.443$+$00.659) of them have been identified as inflow candidates by \citet{2015MNRAS.450.1926H,2016MNRAS.461.2288H} and the other three (G329.030$-$00.202, G345.004$-$00.224 and G351.775$-$00.537) are newly identified in this work. Interestingly, all these inflow candidates have been detected to simultaneously show outflow activities. Therefore, our results are also consistent with the coexistence of in- and outflowing molecular gas.

\section{Conclusions} \label{sec:conclusions}
We have performed a study of 43 southern star-forming regions in HNCO 4$_{04}$--3$_{03}$, SiO 2--1 and HC$_{3}$N 10--9, based on the MALT90 survey, the 870 $\mu$m ATLASGAL survey and Herschel 160 $\mu$m, 250 $\mu$m, 350 $\mu$m, 500 $\mu$m data. Our sample was divided into three categories: 27 sources in the Central Molecular Zone (CMZ) of the Galaxy, 10 sources associated with expanding bubbles of ionized gas, and 6 "normal" star forming (NMSFR) clouds. The spatial distributions of the three measured molecular lines and their properties are analysed. Integrated intensity ratios, abundance ratios, outflow activity and inflow activity are discussed. Our main results can be summarized as follows.

(i) The integrated intensity images show that the distribution of all three molecular lines are compact and consistent with condensed dust structures except for SiO in G000.110+00.148, G000.497+00.021, G001.610-00.172, G001.655-00.062, G001.694-00.385 and G003.240+00.635. The derived angular diameters of the three species indicate that HC$_{3}$N traces denser gas than HNCO and SiO.

(ii) The dust temperature images show that all the 27 CMZ sources are cold with $\thicksim$14 K towards the central 870 $\mu$m continuum emission regions, while the dust is warmer in the more diffuse surrounding regions. For CMZ sources, the variations of ${I_{\rm HNCO}/I_{\rm SiO}}$ and ${N_{\rm HNCO}/N_{\rm SiO}}$ as a function of Galactic longitude are compatible with alternating stronger and weaker shocks affected by the motion of clouds and the dynamics of the $x_{1}$ and $x_{2}$ orbits in the Central Molecular Zone.

(iii) From analyzing the centers of each cold dense clump in Section~\ref{sec:indiv}, the dust temperature $T_{\rm d}$, and the abundance ratios of ${N_{\rm HNCO}/N_{\rm SiO}}$ and ${N_{\rm HNCO}/N_{\rm HC_{3}N}}$ show a decreasing trend with increasing column densities, while ${N_{\rm HC_{3}N}/N_{\rm SiO}}$ reveals the opposite. This characteristic property appears to be most common for sources in the Central Molecular Zone. In Bubble and NMSFR category sources, dust temperature and $N_{\rm HC3N}/N_{\rm SiO}$ roughly follow a similar trend towards the center, where stars are forming.

(iv) The line widths of HNCO and HC$_{3}$N correlate well with each other and ratios are close to unity for the 43 star-forming regions. SiO lines tend to be wider than lines from HNCO and HC$_{3}$N, apparently being most sensitive to the presence of outflowing gas.  HC$_{3}$N provides, in the average, the most compact spatial distributions. SiO and HC$_{3}$N may have similar excitation mechanisms as is suggested by the good correlation between line widths, integrated intensities, column densities, and fractional abundances of these two species, though the critical density of SiO is about four times higher than that of HC$_{3}$N at 20 K.

(v) Eight star-forming sources, G008.671$-$00.357, G345.004$-$00.224, G351.443$+$00.659 and  G351.775$-$00.537 in the Bubble category, and G329.030$-$00.202, G331.708$+$00.583, G331.709$+$00.602 and G335.586$-$00.289 in the NMSFR category (see Sect.~\ref{sec:sampl}) are found to have simultaneously outflow and inflow motions.
And seven new outflow candidates (G008.671$-$00.357, G329.030$-$00.202, G331.708$+$00.583, G331.709$+$00.602, G335.586$-$00.289, G351.443$+$00.659 and G351.775$-$00.537) and three new inflow candidates (G329.030$-$00.202, G345.004$-$00.224 and G351.775$-$00.537) have been identified in this work.

\bigskip
\acknowledgments
We thank the referee for careful comments of this paper. This work was mainly funded by the National Natural Science foundation of China (NSFC) under grant 11703073. It was also partially funded by the NSFC under grants 11433008, 11973076, 11703074, and 11603063, The CAS "Light of West China" Program 2016-QNXZ-B-22, 2018-XBQNXZ-B-024. The "TianShan Youth Plan" under grant 2018Q084. The Heaven Lake Hundred-Talent Program of Xinjiang Uygur Autonomous Region of China. C.~Henkel has been funded by Chinese Academy of Sciences President's International Fellowship Initiative with grant No. 2021VMA0009. Moreover, this work is sponsored (in part) by the Chinese Academy of Sciences (CAS), through a grant to the CAS South America Center for Astronomy (CASSACA) in Santiago, Chile. AS acknowledges funding through Fondecyt Regular (project code 1180350) and Chilean Centro de Excelencia en Astrof\'isica y Tecnolog\'ias Afines (CATA) BASAL grant AFB-170002.

This research has made use of the data products from the MALT90 survey, the SIMBAD data base, operated at CDS, Strasbourg, France, the data from \emph{Herschel}, a European Space Agency space observatory with science instruments provided by European led consortia, and the ATLASGAL survey, which is a collaboration between the Max-Planck-Gesellschaft, the European Southern Observatory (ESO) and the Universidad de Chile.

\software{GILDAS/CLASS \citep{2005sf2a.conf..721P,2013ascl.soft05010G}, Matplotlib \citep{2007CSE.....9...90H}, astropy \citep{2013A&A...558A..33A}}

\begin{table*}
  \begin{center}
  \caption{Derived parameters for sources. Quantities in parentheses give the uncertainties. The columns are as follows: (1) MALT90 names; (2) and (3) Galactic longitude and Galactic latitude of the yellow crosses on maps of Figs.~\ref{fig:Figure_A1} to \ref{fig:Figure_A22}; (4) molecular species; (5) beam deconvolved angular diameter; (6) dust temperature $T_{\rm d}$; (7) centroid velocities; (8) main beam brightness temperatures; (9) full width to half power line width; (10) optical thickness of the line emission; (11) source category (see Sect.~\ref{sec:sampl}). The central Galactic coordinate of each map is used for the source name (e.g. MALT90 names in column 1 of Table~\ref{tab:source_parameters_1}).}

  \tiny
  \label{tab:source_parameters_1}
  \begin{tabular}{lrrrrrcccrr}
  \hline
  \hline
\multicolumn{1}{c}{Source$^{\rm{a}}$} & \multicolumn{1}{c}{l}   & \multicolumn{1}{c}{b}   & \multicolumn{1}{c}{MOL} & \multicolumn{1}{c}{$\theta$}       &   \multicolumn{1}{c}{T$_{\rm d}$} & \multicolumn{1}{c}{v$_{\rm LSR}$}   & \multicolumn{1}{c}{T$_{\rm MB}$} & \multicolumn{1}{c}{$\Delta$v}        & \multicolumn{1}{c}{$\tau$} & \multicolumn{1}{c}{comment}\\
                                      & \multicolumn{1}{c}{deg} & \multicolumn{1}{c}{deg} &                         & \multicolumn{1}{c}{$\prime\prime$} & \multicolumn{1}{c}{$K$}           &  \multicolumn{1}{c}{$km\ s^{-1}$\,} & \multicolumn{1}{c}{K}            & \multicolumn{1}{c}{$km\ s^{-1}$\,}   &                            &                            \\
\multicolumn{1}{c}{(1)}               & \multicolumn{1}{c}{(2)} & \multicolumn{1}{c}{(3)} & \multicolumn{1}{c}{(4)} & \multicolumn{1}{c}{(5)}            & \multicolumn{1}{c}{(6)}           & \multicolumn{1}{c}{(7)}             & \multicolumn{1}{c}{(8)}          & \multicolumn{1}{c}{(9)}              & \multicolumn{1}{c}{(10)}   & \multicolumn{1}{c}{(11)}   \\
  \hline
  G000.067$-$00.077                &    0.065     &   $-$0.078     &  HNCO               & 116 &   15.1     &   48.47 $\pm$ 0.31, 51.62 $\pm$ 2.19   &  1.89, 0.61  &    20.85 $\pm$ 1.37, 44.63 $\pm$ 5.19   &    0.17   &   CMZ    \\
                                   &    0.065     &   $-$0.078     &  SiO                & 118 &            &   50.56 $\pm$ 0.43, 75.73 $\pm$ 1.88   &  1.73, 0.32  &    23.17 $\pm$ 1.02, 17.20 $\pm$ 4.12   &    0.16   &          \\
                                   &    0.065     &   $-$0.078     &  HC$_{3}$N          & 105 &            &           49.50 $\pm$ 0.20       &      2.40   &    24.39 $\pm$ 1.81              &    0.22   &          \\
  G000.104$-$00.080                &    0.107     &   $-$0.085     &  HNCO               & 98  &   15.2     &   52.66 $\pm$ 0.09, 53.53 $\pm$ 0.05      &  2.50, 0.56  &    15.91 $\pm$ 0.07, 34.12 $\pm$ 1.83     &    0.23   &   CMZ    \\
                                   &    0.107     &   $-$0.085     &  SiO                & 107 &            &   55.18 $\pm$ 0.21, 45.45 $\pm$ 2.22   &  1.74, 0.72  &    18.52 $\pm$ 0.93, 36.81 $\pm$ 1.88    &    0.16   &          \\
                                   &    0.107     &   $-$0.085     &  HC$_{3}$N          & 54  &            &   52.85 $\pm$ 0.11, 36.55 $\pm$ 0.82    &  3.01, 0.26  &    18.23 $\pm$ 0.32, 4.83 $\pm$ 2.27     &    0.29   &          \\
  G000.106$-$00.001                &    0.104     &   $-$0.004     &  HNCO               & 114 &   15.4     &   51.07 $\pm$ 0.22, 53.68 $\pm$ 1.24   &  1.34, 0.62  &    13.61 $\pm$ 1.14, 34.01 $\pm$ 3.19   &    0.12   &   CMZ    \\
                                   &    0.104     &   $-$0.004     &  SiO                & 119 &            &           51.00 $\pm$ 0.20       &     1.16    &    24.11 $\pm$ 2.30              &    0.10   &          \\
                                   &    0.104     &   $-$0.004     &  HC$_{3}$N          & 90  &            &   50.83 $\pm$ 0.17, 67.98 $\pm$ 1.09   &  2.12, 0.24  &    17.27 $\pm$ 0.37, 20.20 $\pm$ 1.28   &    0.19   &          \\
  G000.110$+$00.148                &    0.113     &      0.152     &  HNCO               & 113 &   15.2     &           99.85 $\pm$ 0.34      &     0.76    &    15.07 $\pm$ 1.14              &    0.07   &   CMZ    \\
                                   &    0.113     &      0.152     &  SiO                & 140 &            &           99.59 $\pm$ 0.76      &     0.45    &    12.72 $\pm$ 1.50              &    0.04   &          \\
                                   &    0.113     &      0.152     &  HC$_{3}$N          & 105 &            &           98.91 $\pm$ 0.31      &     0.69    &    10.95 $\pm$ 0.78              &    0.06   &          \\
  G000.314$-$00.100                &    0.314     &   $-$0.096     &  HNCO               & 90  &   15.8     &   73.88 $\pm$ 1.69, 82.80 $\pm$ 0.59   &  0.44, 0.40  &    43.90 $\pm$ 2.68, 7.20 $\pm$ 1.71    &    0.04   &   CMZ    \\
                                   &    0.314     &   $-$0.096     &  SiO                & 97  &            &   80.79 $\pm$ 1.24, 81.56 $\pm$ 1.87  &  0.48, 0.11  &    40.89 $\pm$ 2.35, 6.92 $\pm$ 3.26    &    0.04   &          \\
                                   &    0.314     &   $-$0.096     &  HC$_{3}$N          & 73  &            &   78.03 $\pm$ 2.09, 81.21 $\pm$ 0.63   &  0.30, 0.27  &    38.50 $\pm$ 3.00, 3.98 $\pm$ 1.78    &    0.02   &          \\
  G000.497$+$00.021                &    0.497     &      0.018     &  HNCO               & 177 &   15.0     &   30.56 $\pm$ 0.46, 12.58 $\pm$ 0.46    &  2.65, 0.23  &    22.54 $\pm$ 0.46, 27.82 $\pm$ 0.46     &    0.25   &   CMZ    \\
                                   &    0.497     &      0.018     &  SiO                & 186 &            &   35.56 $\pm$ 0.64, 16.12 $\pm$ 0.82    &  0.62, 0.37  &    20.46 $\pm$ 2.67, 21.17 $\pm$ 2.08   &    0.05   &          \\
                                   &    0.497     &      0.018     &  HC$_{3}$N          & 151 &            &   32.48 $\pm$ 0.26, 11.51 $\pm$ 2.22   &  2.06, 0.22  &    18.46 $\pm$ 0.60, 16.05 $\pm$ 4.89    &    0.19   &          \\
  G000.645$+$00.027                &    0.647     &      0.032     &  HNCO               & 190 &   14.9     &          54.31 $\pm$ 0.09        &    3.94     &    26.89 $\pm$ 0.66               &    0.41   &   CMZ    \\
                                   &    0.647     &      0.032     &  SiO                & 196 &            &          56.02 $\pm$ 0.64       &    0.68     &    45.60 $\pm$ 1.43              &    0.06   &          \\
                                   &    0.647     &      0.032     &  HC$_{3}$N          & 146 &            &          52.53 $\pm$ 0.20       &    1.99     &    27.60 $\pm$ 1.87              &    0.19   &          \\
  G000.892$+$00.143                &    0.894     &      0.142     &  HNCO               & 70  &   14.5     &          76.80 $\pm$ 0.39       &    0.76     &    26.25 $\pm$ 0.81              &    0.07   &   CMZ    \\
                                   &    0.894     &      0.142     &  SiO                & 86  &            &          77.45 $\pm$ 0.63       &    0.50     &    31.06 $\pm$ 1.41              &    0.04   &          \\
                                   &    0.894     &      0.142     &  HC$_{3}$N          & 71  &            &          76.84 $\pm$ 0.49       &    0.55     &    25.56 $\pm$ 1.09              &    0.05   &          \\
  G000.908$+$00.116                &    0.919     &      0.116     &  HNCO               & 82  &   14.4     &    64.64 $\pm$ 0.81, 75.78 $\pm$ 6.23  &  0.61, 0.21  &    12.44 $\pm$ 2.23, 23.40 $\pm$ 6.54   &    0.06   &   CMZ    \\
                                   &    0.919     &      0.116     &  SiO                & 79  &            &    66.18 $\pm$ 3.22, 75.83 $\pm$ 2.15 &  0.17, 0.14  &    27.32 $\pm$ 1.06, 32.29 $\pm$ 2.49   &    0.02   &          \\
                                   &    0.919     &      0.116     &  HC$_{3}$N          & 79  &            &    63.68 $\pm$ 0.69, 78.60 $\pm$ 3.42  &  0.28, 0.13  &    9.67 $\pm$ 1.48, 27.51 $\pm$ 7.74    &    0.03   &          \\
  G001.226$+$00.059                &    1.233     &      0.061     &  HNCO               & 149 &   14.0     &         103.87 $\pm$ 0.48       &     0.61    &    18.42 $\pm$ 1.79              &    0.06   &   CMZ    \\
                                   &    1.233     &      0.061     &  SiO                & 130 &            &         113.73 $\pm$ 0.33       &     1.15    &    36.32 $\pm$ 1.70              &    0.11   &          \\
                                   &    1.233     &      0.061     &  HC$_{3}$N          & 75  &            &         109.38 $\pm$ 0.34       &     0.83    &    23.22 $\pm$ 1.39              &    0.08   &          \\
  G001.344$+$00.258                &    1.348     &      0.262     &  HNCO               & 112 &   13.7     &         105.58 $\pm$ 0.49       &     0.82    &    28.85 $\pm$ 1.76              &    0.08   &   CMZ    \\
                                   &    1.348     &      0.262     &  SiO                & 110 &            &         102.74 $\pm$ 0.47       &     1.41    &    38.43 $\pm$ 1.25              &    0.14   &          \\
                                   &    1.348     &      0.262     &  HC$_{3}$N          & 88  &            &         100.65 $\pm$ 0.26       &     1.36    &    25.42 $\pm$ 0.98              &    0.14   &          \\
  G001.381$+$00.201                &    1.381     &      0.187     &  HNCO               & 165 &   13.8     &         102.80 $\pm$ 0.87       &     0.40    &    21.89 $\pm$ 2.42              &    0.04   &   CMZ    \\
                                   &    1.381     &      0.187     &  SiO                & 156 &            &         100.16 $\pm$ 0.65       &     0.69    &    36.29 $\pm$ 2.47              &    0.07   &          \\
                                   &    1.381     &      0.187     &  HC$_{3}$N          & 113 &            &         100.14 $\pm$ 1.05      &     0.37    &    25.99 $\pm$ 2.48              &    0.04   &          \\
  G001.510$+$00.155                &    1.522     &      0.144     &  HNCO               & 134 &   13.3     &         78.00 $\pm$ 0.19        &     1.75    &    15.15 $\pm$ 0.45              &    0.19   &   CMZ    \\
                                   &    1.522     &      0.144     &  SiO                & 159 &            &         79.33 $\pm$ 0.46        &     0.84    &    18.11 $\pm$ 1.15              &    0.09   &          \\
                                   &    1.522     &      0.144     &  HC$_{3}$N          & 100 &            &         77.52 $\pm$ 0.25        &     1.43    &    13.15 $\pm$ 0.75               &    0.15   &          \\
  G001.610$-$00.172                &    1.607     &   $-$0.174     &  HNCO               & 159 &   13.1     &     46.85 $\pm$ 0.18, 37.09 $\pm$ 0.11  &  1.69, 0.77  &    9.08 $\pm$ 0.38, 10.68 $\pm$ 0.65      &    0.19   &   CMZ    \\
                                   &    1.607     &   $-$0.174     &  SiO                & 107 &            &     47.05 $\pm$ 0.81, 38.65 $\pm$ 0.64  &  0.42, 0.40  &    7.91 $\pm$ 1.75), 6.02 $\pm$ 1.98     &    0.04   &          \\
                                   &    1.607     &   $-$0.174     &  HC$_{3}$N          & 89  &            &     48.46 $\pm$ 0.25, 41.51 $\pm$ 0.66  &  0.75, 0.42  &    4.63 $\pm$ 0.57, 7.75 $\pm$ 1.43      &    0.08   &          \\
  \hline
  \end{tabular}
  \end{center}
\end{table*}

\setcounter{table}{0}
\begin{table*}
  \begin{center}
  \caption{{\em - continued.}}
  \tiny
  \begin{tabular}{lrrrrrcccrr}
  \hline
  \hline
\multicolumn{1}{c}{Source$^{\rm{a}}$} & \multicolumn{1}{c}{l}   & \multicolumn{1}{c}{b}   & \multicolumn{1}{c}{MOL} & \multicolumn{1}{c}{$\theta$}       &   \multicolumn{1}{c}{T$_{\rm d}$} & \multicolumn{1}{c}{v$_{\rm LSR}$}   & \multicolumn{1}{c}{T$_{\rm MB}$} & \multicolumn{1}{c}{$\Delta$v}        & \multicolumn{1}{c}{$\tau$} & \multicolumn{1}{c}{comment}\\
                                      & \multicolumn{1}{c}{deg} & \multicolumn{1}{c}{deg} &                         & \multicolumn{1}{c}{$\prime\prime$} & \multicolumn{1}{c}{$K$}           &  \multicolumn{1}{c}{$km\ s^{-1}$\,} & \multicolumn{1}{c}{K}            & \multicolumn{1}{c}{$km\ s^{-1}$\,}   &                            &                            \\
\multicolumn{1}{c}{(1)}               & \multicolumn{1}{c}{(2)} & \multicolumn{1}{c}{(3)} & \multicolumn{1}{c}{(4)} & \multicolumn{1}{c}{(5)}            & \multicolumn{1}{c}{(6)}           & \multicolumn{1}{c}{(7)}             & \multicolumn{1}{c}{(8)}          & \multicolumn{1}{c}{(9)}              & \multicolumn{1}{c}{(10)}   & \multicolumn{1}{c}{(11)}   \\
  \hline
  G001.655$-$00.062                &    1.661     &   $-$0.053     &  HNCO               & 200 &   12.4     &     52.21 $\pm$ 0.07, 44.66 $\pm$ 0.03    &  3.08, 1.58  &    10.55 $\pm$ 0.51, 8.58 $\pm$ 0.25      &    0.40   &   CMZ    \\
  G1.87$-$SMM 1$^{\ast}$           &              &                &                     &     &            &                          &             &                            &           &          \\
                                   &    1.661     &   $-$0.053     &  SiO                & 196 &            &     55.81 $\pm$ 0.34, 44.84 $\pm$ 0.49  &  0.67, 0.55  &    7.62 $\pm$ 0.91, 8.77 $\pm$ 1.01      &    0.07   &          \\
                                   &    1.661     &   $-$0.053     &  HC$_{3}$N          & 150 &            &     51.74 $\pm$ 0.50, 44.77 $\pm$ 2.23 &  1.48, 0.43  &    7.00 $\pm$ 0.24,8.54 $\pm$ 3.54      &    0.17   &          \\
  G001.694$-$00.385                &    1.694     &   $-$0.388     &  HNCO               & 144 &   13.1     &       $-$35.36 $\pm$ 0.28       &     1.37    &    18.88 $\pm$ 0.91               &    0.15   &   CMZ    \\
                                   &    1.694     &   $-$0.388     &  SiO                & 37  &            &       $-$24.52 $\pm$ 0.83       &     0.63    &    29.95 $\pm$ 1.72              &    0.07   &          \\
                                   &    1.694     &   $-$0.388     &  HC$_{3}$N          & 83  &            &       $-$34.31 $\pm$ 0.27       &     1.51    &    13.87 $\pm$ 0.65               &    0.16   &          \\
  G001.699$-$00.366                &    1.700     &   $-$0.367     &  HNCO               & 101 &   13.0     &       $-$32.62 $\pm$ 0.21       &     1.47    &    16.13 $\pm$ 0.52               &    0.16   &   CMZ    \\
                                   &    1.700     &   $-$0.367     &  SiO                & 133 &            &$-$24.31 $\pm$ 0.34, $-$40.97 $\pm$ 3.31&  0.74, 0.27  &    13.69 $\pm$ 1.22, 30.37 $\pm$ 7.69   &    0.08   &          \\
                                   &    1.700     &   $-$0.367     &  HC$_{3}$N          & 86  &            &       $-$31.67 $\pm$ 0.14       &     2.05    &    16.76 $\pm$ 0.36               &    0.23   &          \\
  G001.734$-$00.410                &    1.737     &   $-$0.412     &  HNCO               & 110 &   12.5     &       $-$38.74 $\pm$ 0.17       &     2.36    &    14.98 $\pm$ 0.77               &    0.29   &   CMZ    \\
                                   &    1.737     &   $-$0.412     &  SiO                & 112 &            &       $-$36.85 $\pm$ 0.43       &     0.92    &    19.49 $\pm$ 0.85               &    0.10   &          \\
                                   &    1.737     &   $-$0.412     &  HC$_{3}$N          & 90  &            &       $-$37.88 $\pm$ 0.08        &     2.24    &    11.87 $\pm$ 0.27               &    0.27   &          \\
  G001.883$-$00.062                &    1.887     &   $-$0.063     &  HNCO               & 88  &   13.5     &    35.79 $\pm$ 0.30, 44.45 $\pm$ 3.21  &  1.18, 0.28  &    7.94 $\pm$ 0.58, 12.82 $\pm$ 4.20     &    0.12   &   CMZ    \\
  G1.87$-$SMM 23$^{\ast}$          &              &                &                     &     &            &                          &             &                            &           &          \\
                                   &    1.887     &   $-$0.063     &  SiO                & 75  &            &    37.32 $\pm$ 0.58, 42.55 $\pm$ 0.80   &  0.55, 0.28  &    5.85 $\pm$ 1.20, 4.36 $\pm$ 1.96     &    0.05   &          \\
                                   &    1.887     &   $-$0.063     &  HC$_{3}$N          & 60  &            &    36.17 $\pm$ 0.30, 41.15 $\pm$ 0.51   &  0.84, 0.38  &    6.67 $\pm$ 0.89, 3.03 $\pm$ 1.03      &    0.08   &          \\
  G003.240$+$00.635                &    3.241     &      0.635     &  HNCO               & 44  &   12.6     &        41.39 $\pm$ 0.81         &     0.36    &    18.38 $\pm$ 1.90              &    0.04   &   CMZ    \\
                                   &    3.241     &      0.635     &  SiO                & 91  &            &        42.50 $\pm$ 0.52         &     0.59    &    20.05 $\pm$ 1.47              &    0.06   &          \\
                                   &    3.241     &      0.635     &  HC$_{3}$N          & 106 &            &    42.23 $\pm$ 0.51   &  0.45  &    16.18 $\pm$ 1.27     &    0.05   &          \\
  G003.338$+$00.419                &    3.339     &      0.425     &  HNCO               & 132 &   12.4     &        28.80 $\pm$ 0.44         &     0.90    &    21.24 $\pm$ 1.52              &    0.10   &   CMZ    \\
                                   &    3.339     &      0.425     &  SiO                & 152 &            &    35.84 $\pm$ 0.75, 16.03 $\pm$ 0.82   &  1.02, 0.58  &    23.80 $\pm$ 1.82, 14.74 $\pm$ 1.56   &    0.12   &          \\
                                   &    3.339     &      0.425     &  HC$_{3}$N          & 120 &            &        29.22 $\pm$ 0.18         &     1.75    &    17.44 $\pm$ 0.60               &    0.21   &          \\
  G359.445$-$00.054                &  359.446     &   $-$0.058     &  HNCO               & 117 &   15.7     &      $-$102.71 $\pm$ 0.22       &     1.29    &    15.96 $\pm$ 0.73               &    0.11   &   CMZ    \\
                                   &  359.446     &   $-$0.058     &  SiO                & 115 &            &      $-$103.16 $\pm$ 0.71       &     0.39    &    17.72 $\pm$ 1.95              &    0.03   &          \\
                                   &  359.446     &   $-$0.058     &  HC$_{3}$N          & 113 &            &      $-$101.97 $\pm$ 0.36       &     0.71    &    13.48 $\pm$ 0.82               &    0.06   &          \\
  G359.453$-$00.112                &  359.455     &   $-$0.112     &  HNCO               & 107 &   15.6     &$-$52.00 $\pm$ 0.19, $-$61.60 $\pm$ 0.42 &  1.43, 0.50  &    8.36 $\pm$ 0.53, 5.32 $\pm$ 0.81       &    0.12   &   CMZ    \\
                                   &  359.455     &   $-$0.112     &  SiO                & 106 &            &      $-$51.69 $\pm$ 0.40        &     0.66    &    13.44 $\pm$ 1.02              &    0.05   &          \\
                                   &  359.455     &   $-$0.112     &  HC$_{3}$N          & 103 &            &$-$51.07 $\pm$ 0.33, $-$62.12 $\pm$ 0.90 &  1.01, 0.26  &    12.17 $\pm$ 0.97, 6.43 $\pm$ 1.28     &    0.08   &          \\
  G359.565$-$00.161                &  359.551     &   $-$0.166     &  HNCO               & 84  &   15.8     &        $-$57.78 $\pm$ 0.34      &     0.92    &    16.85 $\pm$ 1.15              &    0.08   &   CMZ    \\
                                   &  359.551     &   $-$0.166     &  SiO                & 87  &            &        $-$55.33 $\pm$ 0.93      &     0.36    &    19.86 $\pm$ 1.26              &    0.03   &          \\
                                   &  359.551     &   $-$0.166     &  HC$_{3}$N          & 67  &            &        $-$54.69 $\pm$ 0.72      &     0.51    &    17.57 $\pm$ 1.35              &    0.04   &          \\
  G359.868$-$00.085                &  359.866     &   $-$0.082     &  HNCO               & 177 &   15.3     &        9.21 $\pm$ 0.05, 0.53 $\pm$ 0.17  &  4.89, 1.24  &    9.55 $\pm$ 0.13, 6.27 $\pm$ 0.27       &    0.52   &   CMZ    \\
                                   &  359.866     &   $-$0.082     &  SiO                & 149 &            &           5.82 $\pm$ 0.21       &     1.86    &    22.23 $\pm$ 0.49               &    0.17   &          \\
                                   &  359.866     &   $-$0.082     &  HC$_{3}$N          & 148 &            &     8.69 $\pm$ 0.26, 0.58 $\pm$ 0.60    &  3.49, 1.29  &    12.57 $\pm$ 0.49, 10.28 $\pm$ 1.82    &    0.34   &          \\
  G359.895$-$00.069                &  359.892     &   $-$0.076     &  HNCO               & 134 &   15.2     &     17.40 $\pm$ 0.06, 5.11 $\pm$ 0.66    &  4.32, 0.38  &    16.64 $\pm$ 0.14, 15.36 $\pm$ 0.20     &    0.44   &   CMZ    \\
                                   &  359.892     &   $-$0.076     &  SiO                & 137 &            &          15.20 $\pm$ 0.10       &     2.78    &    21.34 $\pm$ 0.23               &    0.26   &          \\
                                   &  359.892     &   $-$0.076     &  HC$_{3}$N          & 137 &            &          16.24 $\pm$ 0.05        &     4.59    &    18.49 $\pm$ 0.13               &    0.48   &          \\
  G359.977$-$00.072                &  359.980     &   $-$0.070     &  HNCO               & 135 &   16.6     &     49.88 $\pm$ 0.14, 36.37 $\pm$ 0.56  &  2.75, 0.69  &    14.93 $\pm$ 0.34, 12.65 $\pm$ 0.91     &    0.23   &   CMZ    \\
                                   &  359.980     &   $-$0.070     &  SiO                & 114 &            &     51.80 $\pm$ 0.02, 37.00 $\pm$ 0.53   &  1.66, 1.20  &    25.39 $\pm$ 0.70, 21.08 $\pm$ 0.37     &    0.13   &          \\
                                   &  359.980     &   $-$0.070     &  HC$_{3}$N          & 108 &            &     48.50 $\pm$ 0.47, 38.42 $\pm$ 1.33 &  3.63, 0.66  &    20.92 $\pm$ 0.52, 25.50 $\pm$ 4.12    &    0.32   &          \\
  G008.671$-$00.357                &    8.677     &   $-$0.360     &  HNCO               & 75  &   14.8     &           36.09 $\pm$ 0.19      &     0.79    &    5.69 $\pm$ 0.44               &    0.07   &   Bubble \\
                                   &    8.677     &   $-$0.360     &  SiO                & 67  &            &           35.99 $\pm$ 0.32      &     0.55    &    7.39 $\pm$ 0.85                &    0.05   &          \\
                                   &    8.677     &   $-$0.360     &  HC$_{3}$N          & 69  &            &           35.46 $\pm$ 0.08       &     1.86    &    4.88 $\pm$ 0.19                &    0.17   &          \\
  G010.473$+$00.028                &   10.480     &      0.030     &  HNCO               & 65  &   15.5     &           65.79 $\pm$ 0.23      &     0.63    &    6.37 $\pm$ 0.61                &    0.05   &   Bubble \\
                                   &   10.480     &      0.030     &  SiO                & 49  &            &           65.82 $\pm$ 1.36     &     0.21    &    23.41 $\pm$ 3.13              &    0.02   &          \\
                                   &   10.480     &      0.030     &  HC$_{3}$N          & 18  &            &           66.31 $\pm$ 0.19      &     0.79    &    6.96 $\pm$ 0.55                &    0.07   &          \\
  G322.159$+$00.635$^{\dag}$       &  322.159     &      0.637     &  HNCO               & 96  &   19.4     &        $-$56.94 $\pm$ 0.27      &     0.45    &    4.02 $\pm$ 0.61                &    0.03   &   Bubble \\
                                   &  322.159     &      0.637     &  SiO                & 14  &            &        $-$56.34 $\pm$ 0.15      &     1.12    &    7.75 $\pm$ 0.44                &    0.07   &          \\
                                   &  322.159     &      0.637     &  HC$_{3}$N          & 43  &            &        $-$56.89 $\pm$ 0.05       &     2.67    &    4.99 $\pm$ 0.11                &    0.18   &          \\

  \hline
  \end{tabular}
  \end{center}
  \normalsize
\end{table*}

\setcounter{table}{0}
\begin{table*}
  \begin{center}
  \caption{{\em - continued.}}
  \tiny
  \begin{tabular}{lrrrrrcccrr}
  \hline
  \hline
\multicolumn{1}{c}{Source$^{\rm{a}}$} & \multicolumn{1}{c}{l}   & \multicolumn{1}{c}{b}   & \multicolumn{1}{c}{MOL} & \multicolumn{1}{c}{$\theta$}       &   \multicolumn{1}{c}{T$_{\rm d}$} & \multicolumn{1}{c}{v$_{\rm LSR}$}   & \multicolumn{1}{c}{T$_{\rm MB}$} & \multicolumn{1}{c}{$\Delta$v}        & \multicolumn{1}{c}{$\tau$} & \multicolumn{1}{c}{comment}\\
                                      & \multicolumn{1}{c}{deg} & \multicolumn{1}{c}{deg} &                         & \multicolumn{1}{c}{$\prime\prime$} & \multicolumn{1}{c}{$K$}           &  \multicolumn{1}{c}{$km\ s^{-1}$\,} & \multicolumn{1}{c}{K}            & \multicolumn{1}{c}{$km\ s^{-1}$\,}   &                            &                            \\
\multicolumn{1}{c}{(1)}               & \multicolumn{1}{c}{(2)} & \multicolumn{1}{c}{(3)} & \multicolumn{1}{c}{(4)} & \multicolumn{1}{c}{(5)}            & \multicolumn{1}{c}{(6)}           & \multicolumn{1}{c}{(7)}             & \multicolumn{1}{c}{(8)}          & \multicolumn{1}{c}{(9)}              & \multicolumn{1}{c}{(10)}   & \multicolumn{1}{c}{(11)}   \\
  \hline
  G326.653$+$00.618                &  326.640     &      0.615     &  HNCO               & 88  &   15.3     &        $-$39.78 $\pm$ 0.27      &     0.60    &    5.88 $\pm$ 0.69                &    0.05   &   Bubble \\
                                   &  326.640     &      0.615     &  SiO                & 89  &            &        $-$39.45 $\pm$ 0.32      &     0.58    &    6.95 $\pm$ 1.13               &    0.05   &          \\
                                   &  326.640     &      0.615     &  HC$_{3}$N          & 98  &            &        $-$39.02 $\pm$ 0.06       &     1.79    &    3.30 $\pm$ 0.16                &    0.16   &          \\
  G327.293$-$00.579$^{\dag}$       &  327.296     &   $-$0.577     &  HNCO               & 25  &   16.9     &        $-$44.73 $\pm$ 0.24      &     0.58    &    5.55 $\pm$ 0.67                &    0.04   &   Bubble \\
                                   &  327.296     &   $-$0.577     &  SiO                & 34  &            &        $-$44.29 $\pm$ 0.21      &     1.01    &    7.64 $\pm$ 0.60                &    0.08   &          \\
                                   &  327.296     &   $-$0.577     &  HC$_{3}$N          & 52  &            &        $-$44.88 $\pm$ 0.05       &     3.10    &    5.76 $\pm$ 0.11                &    0.26   &          \\
  G345.004$-$00.224$^{\dag}$       &  345.007     &   $-$0.220     &  HNCO               & 32  &   16.5     &        $-$26.43 $\pm$ 0.36      &     0.44    &    8.02 $\pm$ 0.85                &    0.03   &   Bubble \\
                                   &  345.007     &   $-$0.220     &  SiO                & 50  &            & $-$24.91 $\pm$ 0.34, $-$34.45 $\pm$ 0.36&  0.77, 0.42  &    9.23 $\pm$ 0.96                &    0.06   &          \\
                                   &  345.007     &   $-$0.220     &  HC$_{3}$N          & 30  &            &        $-$27.07 $\pm$ 0.11      &     1.46    &    7.11 $\pm$ 0.27                &    0.12   &          \\
  G350.101$+$00.083$^{\dag}$       &  350.110     &      0.096     &  HNCO               & 55  &   16.0     &        $-$68.53 $\pm$ 0.35      &     0.57    &    8.07 $\pm$ 0.70                &    0.05   &   Bubble \\
                                   &  350.110     &      0.096     &  SiO                & 58  &            &        $-$69.38 $\pm$ 0.60      &     0.46    &    12.46 $\pm$ 1.65              &    0.04   &          \\
                                   &  350.110     &      0.096     &  HC$_{3}$N          & 44  &            &        $-$69.84 $\pm$ 0.25      &     0.66    &    6.97 $\pm$ 0.76                &    0.05   &          \\
  G351.443$+$00.659                &  351.451     &      0.654     &  HNCO               & 67  &   15.7     &        $-$4.41 $\pm$ 0.11       &     1.29    &    4.32 $\pm$ 0.29                &    0.11   &   Bubble \\
                                   &  351.451     &      0.654     &  SiO                & 49  &            &$-$3.84 $\pm$ 0.09, $-$3.86 $\pm$ 0.46    &  1.90, 0.76  &    5.58 $\pm$ 0.40, 14.73 $\pm$ 1.90     &    0.16   &          \\
                                   &  351.451     &      0.654     &  HC$_{3}$N          & 64  &            &        $-$4.23 $\pm$ 0.13           &     7.29    &    4.02 $\pm$ 0.05                 &    0.87   &          \\
  G351.582$-$00.352                &  351.579     &   $-$0.355     &  HNCO               & 69  &   17.9     &        $-$95.72 $\pm$ 0.11      &     1.09    &    4.60 $\pm$ 0.23                &    0.08   &   Bubble \\
                                   &  351.579     &   $-$0.355     &  SiO                & 58  &            &        $-$96.03 $\pm$ 0.44      &     0.47    &    6.68 $\pm$ 1.48               &    0.03   &          \\
                                   &  351.579     &   $-$0.355     &  HC$_{3}$N          & 27  &            &        $-$95.51 $\pm$ 0.10      &     1.40    &    4.93 $\pm$ 0.24                &    0.10   &          \\
  G351.775$-$00.537                &  351.773     &   $-$0.539     &  HNCO               & 35  &   20.7     & $-$2.64 $\pm$ 0.16, $-$6.84 $\pm$ 0.28  &  1.01, 0.47  &    3.58 $\pm$ 0.42, 1.77 $\pm$ 0.61       &    0.06   &   Bubble \\
                                   &  351.773     &   $-$0.539     &  SiO                & 39  &            & $-$2.83 $\pm$ 0.56, 5.40 $\pm$ 1.22    &  1.97, 0.71  &    9.45 $\pm$ 94, 7.75 $\pm$ 1.48      &    0.12   &          \\
                                   &  351.773     &   $-$0.539     &  HC$_{3}$N          & 32  &            &        $-$2.41 $\pm$ 0.10       &     1.94    &    6.22 $\pm$ 0.28                &    0.12   &          \\
  G329.030$-$00.202$^{\dag}$       &  329.031     &   $-$0.201     &  HNCO               & 55  &   15.4     &        $-$44.47 $\pm$ 0.17      &     0.83    &    3.86 $\pm$ 0.37                &    0.07   &   NMSFR  \\
                                   &  329.031     &   $-$0.201     &  SiO                & 49  &            & $-$43.17 $\pm$ 0.33, $-$50.46 $\pm$ 0.62&  0.90, 0.41  &    6.99 $\pm$ 0.58, 6.16 $\pm$ 1.89      &    0.08   &          \\
                                   &  329.031     &   $-$0.201     &  HC$_{3}$N          & 51  &            &        $-$44.08 $\pm$ 0.08       &     1.79    &    4.95 $\pm$ 0.22                &    0.16   &          \\
  G331.708$+$00.583                &  331.709     &      0.582     &  HNCO               & $-$ &   14.0     &        $-$67.61 $\pm$ 0.22      &     0.50    &    2.48 $\pm$ 0.59                &    0.05   &   NMSFR  \\
                                   &  331.709     &      0.582     &  SiO                & $-$ &            &        $-$66.99 $\pm$ 0.47      &     0.39    &    7.89 $\pm$ 1.39               &    0.04   &          \\
                                   &  331.709     &      0.582     &  HC$_{3}$N          & $-$ &            &        $-$66.84 $\pm$ 0.15      &     0.79    &    4.23 $\pm$ 0.37                &    0.08   &          \\
  G331.709$+$00.602                &  331.709     &      0.602     &  HNCO               & $-$ &   14.4     &        $-$66.93 $\pm$ 0.25      &     0.61    &    4.90 $\pm$ 0.71                &    0.06   &   NMSFR  \\
                                   &  331.709     &      0.602     &  SiO                & $-$ &            &        $-$67.22 $\pm$ 0.32      &     0.45    &    4.60 $\pm$ 0.72                &    0.04   &          \\
                                   &  331.709     &      0.602     &  HC$_{3}$N          & $-$ &            &        $-$67.62 $\pm$ 0.11      &     1.03    &    3.34 $\pm$ 0.27                &    0.10   &          \\
  G335.586$-$00.289$^{\dag}$       &  335.584     &   $-$0.288     &  HNCO               & 35  &   16.7     &        $-$46.43 $\pm$ 0.15      &     0.78    &    3.26 $\pm$ 0.46                &    0.06   &   NMSFR  \\
                                   &  335.584     &   $-$0.288     &  SiO                & 39  &            & $-$46.08 $\pm$ 0.16, $-$52.48 $\pm$ 0.59&  0.91, 0.32  &    4.55 $\pm$ 0.45, 4.92 $\pm$ 1.17      &    0.07   &          \\
                                   &  335.584     &   $-$0.288     &  HC$_{3}$N          & 30  &            &        $-$46.45 $\pm$ 0.06       &     1.98    &    3.41 $\pm$ 0.15                &    0.16   &          \\
  G348.754$-$00.941$^{\dag}$       &  348.761     &   $-$0.948     &  HNCO               & 86  &   13.7     &        $-$14.06 $\pm$ 0.32      &     0.56    &    5.44 $\pm$ 0.76                &    0.05   &   NMSFR  \\
                                   &  348.761     &   $-$0.948     &  SiO                & 48  &            &        $-$12.66 $\pm$ 0.32      &     0.56    &    5.45 $\pm$ 1.03               &    0.05   &          \\
                                   &  348.761     &   $-$0.948     &  HC$_{3}$N          & 32  &            &        $-$13.24 $\pm$ 0.19      &     0.85    &    4.73 $\pm$ 0.50                &    0.08   &          \\
  G351.157$+$00.701                &  351.155     &      0.709     &  HNCO               & 75  &   17.5     &        $-$6.48 $\pm$ 0.18       &     0.75    &    3.57 $\pm$ 0.54                &    0.05   &   NMSFR  \\
                                   &  351.155     &      0.709     &  SiO                & 34  &            &        $-$6.48 $\pm$ 0.12       &     1.05    &    3.17 $\pm$ 0.34                &    0.08   &          \\
                                   &  351.155     &      0.709     &  HC$_{3}$N          & 41  &            &        $-$6.66 $\pm$ 0.04        &     3.54    &    3.42 $\pm$ 0.09                 &    0.28   &          \\
  \hline
  \end{tabular}
  \smallskip
  \leftline{\scriptsize{\emph{Note}. $^{\rm{a}}$ $\ast$ indicates the source name is adopted from \citet{2014A&A...562A...3M}. $\dag$ indicates a source is identified as an IRDC by \citet{2009A&A...505..405P}.}}
  \end{center}
  \normalsize
\end{table*}

\clearpage

\begin{table*}
  \begin{center}
  \caption{Derived parameters for sources. Quantities in parentheses give the uncertainties. The columns are as follows: (1) MALT90 names; (2) and (3) Galactic longitude and Galactic latitude of the yellow crosses on maps of Figs.~\ref{fig:Figure_A1} to \ref{fig:Figure_A22}; (4) molecular species;  (5) velocity range; (6) integrated intensity is derived by integrating over the velocity range indicated in column 5;  (7) beam-averaged column densities of the molecules discussed in this paper; (8)beam-averaged H$_{\rm 2}$ column density; (9) fractional abundances relative to H$_{\rm 2}$; (10) source category (see Sect.~\ref{sec:sampl}).}

  \tiny
  \label{tab:source_parameters_2}
  \begin{tabular}{lrrrcrrrrr}
  \hline
  \hline
\multicolumn{1}{c}{Source$^{\rm{a}}$} & \multicolumn{1}{c}{l}   & \multicolumn{1}{c}{b}   & \multicolumn{1}{c}{MOL} &     \multicolumn{1}{c}{v$_{\rm range}$} &  \multicolumn{1}{c}{$\int {T_{\rm MB}} dv$\,}  &  \multicolumn{1}{c}{N}                          & \multicolumn{1}{c}{N$_{\rm H_{2}}$}            & \multicolumn{1}{c}{$x$}                 & \multicolumn{1}{c}{comment}\\
                                      & \multicolumn{1}{c}{deg} & \multicolumn{1}{c}{deg} &                         &    \multicolumn{1}{c}{$km\ s^{-1}$\,}   &  \multicolumn{1}{c}{$K \cdot km\ s^{-1}$\,}    &  \multicolumn{1}{c}{$\times 10^{13} cm^{-2}$\,} & \multicolumn{1}{c}{$\times 10^{22} cm^{-2}$\,} & \multicolumn{1}{c}{$\times 10^{-10}$\,} &                            \\
\multicolumn{1}{c}{(1)}               & \multicolumn{1}{c}{(2)} & \multicolumn{1}{c}{(3)} & \multicolumn{1}{c}{(4)} &   \multicolumn{1}{c}{(5)}               &  \multicolumn{1}{c}{(6)}                       &  \multicolumn{1}{c}{(7)}                       & \multicolumn{1}{c}{(8)}                       & \multicolumn{1}{c}{(9)}                & \multicolumn{1}{c}{(10)}   \\
  \hline
  G000.067$-$00.077                &    0.065     &   $-$0.078     &  HNCO           &  16.86, 96.65    &    70.65 $\pm$ 1.54  &     88.15 $\pm$ 1.92     &   18.13 $\pm$ 1.56   &   48.62 $\pm$ 4.34  &   CMZ    \\
                                   &    0.065     &   $-$0.078     &  SiO            &  26.22, 91.03    &    48.09 $\pm$ 1.32  &     11.84 $\pm$ 0.32     &                 &    6.53 $\pm$ 0.59  &          \\
                                   &    0.065     &   $-$0.078     &  HC$_{3}$N      &  32.59, 69.68    &    51.26 $\pm$ 0.98  &     25.16 $\pm$ 0.48     &                 &   13.88 $\pm$ 1.23  &          \\
  G000.104$-$00.080                &    0.107     &   $-$0.085     &  HNCO           &  32.75, 73.68    &    59.60 $\pm$ 0.65  &     74.70 $\pm$ 0.81     &   16.72 $\pm$ 2.11   &   44.68 $\pm$ 5.66  &   CMZ    \\
                                   &    0.107     &   $-$0.085     &  SiO            &  17.44, 76.46    &    60.75 $\pm$ 0.79  &     15.01 $\pm$ 0.19     &                 &    8.98 $\pm$ 1.13  &          \\
                                   &    0.107     &   $-$0.085     &  HC$_{3}$N      &  26.07, 72.01    &    76.95 $\pm$ 0.76  &     37.60 $\pm$ 0.37     &                 &   22.49 $\pm$ 2.84  &          \\
  G000.106$-$00.001                &    0.104     &   $-$0.004     &  HNCO           &  27.84, 84.77    &    41.18 $\pm$ 0.67  &     52.09 $\pm$ 0.84     &    8.76 $\pm$ 1.22   &   59.44 $\pm$ 8.34  &   CMZ    \\
                                   &    0.104     &   $-$0.004     &  SiO            &  30.31, 71.54    &    28.39 $\pm$ 0.53  &      7.06 $\pm$ 0.13     &                 &    8.05 $\pm$ 1.13  &          \\
                                   &    0.104     &   $-$0.004     &  HC$_{3}$N      &  34.31, 72.77    &    42.33 $\pm$ 0.49  &     20.50 $\pm$ 0.23     &                 &   23.39 $\pm$ 3.27  &          \\
  G000.110$+$00.148                &    0.113     &      0.152     &  HNCO           &  80.21, 121.25   &    13.96 $\pm$ 0.56  &     17.50 $\pm$ 0.70     &    4.81 $\pm$ 0.54   &   36.37 $\pm$ 4.39  &   CMZ    \\
                                   &    0.113     &      0.152     &  SiO            &  87.88, 113.59   &     6.85 $\pm$ 0.52  &      1.69 $\pm$ 0.12     &                 &    3.52 $\pm$ 0.48  &          \\
                                   &    0.113     &      0.152     &  HC$_{3}$N      &  90.06, 110.86   &     8.82 $\pm$ 0.39  &      4.31 $\pm$ 0.19     &                 &    8.96 $\pm$ 1.09  &          \\
  G000.314$-$00.100                &    0.314     &   $-$0.096     &  HNCO           &  47.32, 104.77   &    21.71 $\pm$ 0.82  &     27.97 $\pm$ 1.05     &    7.09 $\pm$ 0.40   &   39.46 $\pm$ 2.69  &   CMZ    \\
                                   &    0.314     &   $-$0.096     &  SiO            &  37.29, 112.58   &    21.95 $\pm$ 1.07  &      5.52 $\pm$ 0.26     &                 &    7.79 $\pm$ 0.58  &          \\
                                   &    0.314     &   $-$0.096     &  HC$_{3}$N      &  51.79, 116.48   &    13.82 $\pm$ 0.86  &      6.58 $\pm$ 0.40     &                 &    9.29 $\pm$ 0.78  &          \\
  G000.497$+$00.021                &    0.497     &      0.018     &  HNCO           &  3.51, 53.42     &    69.41 $\pm$ 0.72  &     86.21 $\pm$ 0.89     &   16.59 $\pm$ 2.84   &   51.96 $\pm$ 8.93  &   CMZ    \\
                                   &    0.497     &      0.018     &  SiO            &  0.70, 51.31     &    22.14 $\pm$ 0.73  &      5.44 $\pm$ 0.17     &                 &    3.28 $\pm$ 0.57  &          \\
                                   &    0.497     &      0.018     &  HC$_{3}$N      &  2.81, 53.42     &    43.80 $\pm$ 0.66  &     21.60 $\pm$ 0.32     &                 &   13.02 $\pm$ 2.24  &          \\
  G000.645$+$00.027                &    0.647     &      0.032     &  HNCO           &  21.76, 81.95    &   112.17 $\pm$ 0.84  &    138.69 $\pm$ 1.03     &   15.67 $\pm$ 1.34   &   88.49 $\pm$ 7.62  &   CMZ    \\
                                   &    0.647     &      0.032     &  SiO            &  22.45, 93.99    &    30.90 $\pm$ 0.81  &      7.56 $\pm$ 0.19     &                 &    4.83 $\pm$ 0.43  &          \\
                                   &    0.647     &      0.032     &  HC$_{3}$N      &  26.92, 77.82    &    60.69 $\pm$ 0.76  &     30.08 $\pm$ 0.37     &                 &   19.19 $\pm$ 1.66  &          \\
  G000.892$+$00.143                &    0.894     &      0.142     &  HNCO           &  57.39, 98.18    &    19.97 $\pm$ 0.52  &     24.25 $\pm$ 0.63     &    3.30 $\pm$ 0.41   &   73.46 $\pm$ 9.52  &   CMZ    \\
                                   &    0.894     &      0.142     &  SiO            &  54.08, 108.93   &    16.16 $\pm$ 0.61  &      3.91 $\pm$ 0.14     &                 &   11.84 $\pm$ 1.56  &          \\
                                   &    0.894     &      0.142     &  HC$_{3}$N      &  54.36, 92.40    &    14.11 $\pm$ 0.49  &      7.14 $\pm$ 0.24     &                 &   21.62 $\pm$ 2.84  &          \\
  G000.908$+$00.116                &    0.919     &      0.116     &  HNCO           &  51.24, 88.10    &    12.73 $\pm$ 0.66  &     15.39 $\pm$ 0.79     &    3.53 $\pm$ 0.32   &   43.58 $\pm$ 4.60  &   CMZ    \\
                                   &    0.919     &      0.116     &  SiO            &  41.22, 93.71    &     9.49 $\pm$ 0.69  &      2.29 $\pm$ 0.16     &                 &    6.48 $\pm$ 0.76  &          \\
                                   &    0.919     &      0.116     &  HC$_{3}$N      &  56.05, 88.90    &     5.82 $\pm$ 0.46  &      2.96 $\pm$ 0.23     &                 &    8.38 $\pm$ 1.01  &          \\
  G001.226$+$00.059                &    1.233     &      0.061     &  HNCO           &  77.97, 136.29   &    16.64 $\pm$ 0.67  &     19.76 $\pm$ 0.79     &    5.43 $\pm$ 1.05   &   36.39 $\pm$ 7.20  &   CMZ    \\
                                   &    1.233     &      0.061     &  SiO            &  79.95, 147.82   &    41.61 $\pm$ 0.73  &      9.92 $\pm$ 0.17     &                 &   18.27 $\pm$ 3.55  &          \\
                                   &    1.233     &      0.061     &  HC$_{3}$N      &  88.52, 131.35   &    22.60 $\pm$ 0.48  &     11.76 $\pm$ 0.24     &                 &   21.66 $\pm$ 4.22  &          \\
  G001.344$+$00.258                &    1.348     &      0.262     &  HNCO           &  75.10, 131.08   &    26.14 $\pm$ 0.71  &     30.63 $\pm$ 0.83     &    5.65 $\pm$ 0.52   &   54.24 $\pm$ 5.25  &   CMZ    \\
                                   &    1.348     &      0.262     &  SiO            &  66.91, 136.99   &    55.98 $\pm$ 0.86  &     13.23 $\pm$ 0.20     &                 &   23.43 $\pm$ 2.20  &          \\
                                   &    1.348     &      0.262     &  HC$_{3}$N      &  76.01, 124.25   &    35.37 $\pm$ 0.60  &     18.75 $\pm$ 0.31     &                 &   33.21 $\pm$ 3.13  &          \\
  G001.381$+$00.201                &    1.381     &      0.187     &  HNCO           &  75.30, 123.23   &    10.08 $\pm$ 0.61  &     11.86 $\pm$ 0.71     &    2.68 $\pm$ 0.17   &   44.31 $\pm$ 3.94  &   CMZ    \\
                                   &    1.381     &      0.187     &  SiO            &  60.05, 133.58   &    26.93 $\pm$ 0.76  &      6.38 $\pm$ 0.18     &                 &   23.84 $\pm$ 1.69  &          \\
                                   &    1.381     &      0.187     &  HC$_{3}$N      &  69.85, 121.05   &    11.24 $\pm$ 0.62  &      5.92 $\pm$ 0.32     &                 &   22.11 $\pm$ 1.89  &          \\
  G001.510$+$00.155                &    1.522     &      0.144     &  HNCO           &  64.27, 95.39    &    28.51 $\pm$ 0.61  &     32.83 $\pm$ 0.70     &    3.04 $\pm$ 0.40   &  107.93 $\pm$ 7.35  &   CMZ    \\
                                   &    1.522     &      0.144     &  SiO            &  66.14, 98.90    &    15.81 $\pm$ 0.63  &      3.69 $\pm$ 0.14     &                 &   12.15 $\pm$ 1.70  &          \\
                                   &    1.522     &      0.144     &  HC$_{3}$N      &  66.37, 92.11    &    19.75 $\pm$ 0.61  &     10.76 $\pm$ 0.33     &                 &   35.36 $\pm$ 4.88  &          \\
  G001.610$-$00.172                &    1.607     &   $-$0.174     &  HNCO           &  29.14, 55.45    &    26.24 $\pm$ 0.49  &     29.96 $\pm$ 0.55     &    4.53 $\pm$ 0.50   &   66.07 $\pm$ 7.42  &   CMZ    \\
                                   &    1.607     &   $-$0.174     &  SiO            &  30.13, 55.94    &     6.29 $\pm$ 0.42  &      1.46 $\pm$ 0.09     &                 &    3.22 $\pm$ 0.41  &          \\
                                   &    1.607     &   $-$0.174     &  HC$_{3}$N      &  34.80, 55.94    &     7.47 $\pm$ 0.37  &      4.13 $\pm$ 0.20     &                 &    9.10 $\pm$ 1.10  &          \\

  \hline
  \end{tabular}
  \end{center}
  \normalsize
\end{table*}

\setcounter{table}{1}
\begin{table*}
  \begin{center}
  \caption{{\em - continued.}}
  \tiny
  \begin{tabular}{lrrrcrrrrr}
  \hline
  \hline
\multicolumn{1}{c}{Source$^{\rm{a}}$} & \multicolumn{1}{c}{l}   & \multicolumn{1}{c}{b}   & \multicolumn{1}{c}{MOL} &     \multicolumn{1}{c}{v$_{\rm range}$} &  \multicolumn{1}{c}{$\int {T_{\rm MB}} dv$\,}  &  \multicolumn{1}{c}{N}                          & \multicolumn{1}{c}{N$_{\rm H_{2}}$}            & \multicolumn{1}{c}{$x$}                 & \multicolumn{1}{c}{comment}\\
                                      & \multicolumn{1}{c}{deg} & \multicolumn{1}{c}{deg} &                         &    \multicolumn{1}{c}{$km\ s^{-1}$\,}   &  \multicolumn{1}{c}{$K \cdot km\ s^{-1}$\,}    &  \multicolumn{1}{c}{$\times 10^{13} cm^{-2}$\,} & \multicolumn{1}{c}{$\times 10^{22} cm^{-2}$\,} & \multicolumn{1}{c}{$\times 10^{-10}$\,} &                            \\
\multicolumn{1}{c}{(1)}               & \multicolumn{1}{c}{(2)} & \multicolumn{1}{c}{(3)} & \multicolumn{1}{c}{(4)} &   \multicolumn{1}{c}{(5)}               &  \multicolumn{1}{c}{(6)}                       &  \multicolumn{1}{c}{(7)}                       & \multicolumn{1}{c}{(8)}                       & \multicolumn{1}{c}{(9)}                & \multicolumn{1}{c}{(10)}   \\
  \hline

    G001.655$-$00.062                &    1.661     &   $-$0.053     &  HNCO           &  36.36, 62.59    &    46.96 $\pm$ 0.47  &     52.08 $\pm$ 0.52     &    6.23 $\pm$ 0.61   &   83.63 $\pm$ 8.29  &   CMZ    \\
  G1.87$-$SMM 1$^{\ast}$           &              &                &                 &                  &                &                    &                 &               &          \\
                                   &    1.661     &   $-$0.053     &  SiO            &  38.79, 64.65    &    10.48 $\pm$ 0.44  &      2.39 $\pm$ 0.10     &                 &    3.84 $\pm$ 0.41  &          \\
                                   &    1.661     &   $-$0.053     &  HC$_{3}$N      &  39.36, 60.72    &    14.98 $\pm$ 0.39  &      8.76 $\pm$ 0.22     &                 &   14.06 $\pm$ 1.43  &          \\
  G001.694$-$00.385                &    1.694     &   $-$0.388     &  HNCO           &  -50.97, -22.28  &    25.33 $\pm$ 0.69  &     28.92 $\pm$ 0.78     &    6.38 $\pm$ 0.83   &   45.31 $\pm$ 6.03  &   CMZ    \\
                                   &    1.694     &   $-$0.388     &  SiO            &  -61.15, -15.18  &    19.66 $\pm$ 0.93  &      4.57 $\pm$ 0.21     &                 &    7.16 $\pm$ 0.99  &          \\
                                   &    1.694     &   $-$0.388     &  HC$_{3}$N      &  -47.89, -20.12  &    24.02 $\pm$ 0.70  &     13.27 $\pm$ 0.38     &                 &   20.79 $\pm$ 2.77  &          \\
  G001.699$-$00.366                &    1.700     &   $-$0.367     &  HNCO           &  -53.14, -19.61  &    26.66 $\pm$ 0.55  &     30.31 $\pm$ 0.62     &    6.70 $\pm$ 0.40   &   45.26 $\pm$ 2.91  &   CMZ    \\
                                   &    1.700     &   $-$0.367     &  SiO            &  -59.97, -10.30  &    19.35 $\pm$ 0.63  &      4.48 $\pm$ 0.14     &                 &    6.70 $\pm$ 0.46  &          \\
                                   &    1.700     &   $-$0.367     &  HC$_{3}$N      &  -49.41, -10.85  &    38.80 $\pm$ 0.54  &     21.60 $\pm$ 0.30     &                 &   32.25 $\pm$ 2.02  &          \\
  G001.734$-$00.410                &    1.737     &   $-$0.412     &  HNCO           &  -57.64, -24.24  &    36.06 $\pm$ 0.43  &     40.15 $\pm$ 0.47     &   11.75 $\pm$ 1.53   &   34.18 $\pm$ 4.49  &   CMZ    \\
                                   &    1.737     &   $-$0.412     &  SiO            &  -54.63, -20.53  &    18.80 $\pm$ 0.44  &      4.30 $\pm$ 0.10     &                 &    3.66 $\pm$ 0.48  &          \\
                                   &    1.737     &   $-$0.412     &  HC$_{3}$N      &  -49.99, -25.17  &    30.04 $\pm$ 0.36  &     17.41 $\pm$ 0.20     &                 &   14.82 $\pm$ 1.94  &          \\
  G001.883$-$00.062                &    1.887     &   $-$0.063     &  HNCO           &  26.73, 49.74    &    13.22 $\pm$ 0.36  &     15.36 $\pm$ 0.41     &    4.07 $\pm$ 0.38   &   37.69 $\pm$ 3.69  &   CMZ    \\
  G1.87$-$SMM 23$^{\ast}$          &              &                &                 &                  &                &                    &                 &               &          \\
                                   &    1.887     &   $-$0.063     &  SiO            &  30.10, 44.78    &     5.29 $\pm$ 0.38  &      1.24 $\pm$ 0.08     &                 &    3.05 $\pm$ 0.36  &          \\
                                   &    1.887     &   $-$0.063     &  HC$_{3}$N      &  28.52, 43.79    &     6.74 $\pm$ 0.34  &      3.62 $\pm$ 0.18     &                 &    8.89 $\pm$ 0.94  &          \\
  G003.240$+$00.635                &    3.241     &      0.635     &  HNCO           &  29.37, 56.47    &     6.47 $\pm$ 0.45  &      7.23 $\pm$ 0.50     &    3.17 $\pm$ 0.41   &   22.80 $\pm$ 3.40  &   CMZ    \\
                                   &    3.241     &      0.635     &  SiO            &  23.88, 59.90    &    13.07 $\pm$ 0.52  &      3.00 $\pm$ 0.11     &                 &    9.44 $\pm$ 1.30  &          \\
                                   &    3.241     &      0.635     &  HC$_{3}$N      &  28.68, 56.81    &     7.44 $\pm$ 0.45  &      4.28 $\pm$ 0.25     &                 &   13.48 $\pm$ 1.95  &          \\
  G003.338$+$00.419                &    3.339     &      0.425     &  HNCO           &  9.00, 55.42     &    20.86 $\pm$ 0.72  &     23.13 $\pm$ 0.79     &    8.73 $\pm$ 0.30   &   26.51 $\pm$ 1.29  &   CMZ    \\
                                   &    3.339     &      0.425     &  SiO            &  1.87, 60.94     &    35.05 $\pm$ 0.74  &      7.99 $\pm$ 0.16     &                 &    9.16 $\pm$ 0.37  &          \\
                                   &    3.339     &      0.425     &  HC$_{3}$N      &  6.09, 56.72     &    34.77 $\pm$ 0.74  &     20.32 $\pm$ 0.43     &                 &   23.29 $\pm$ 0.94  &          \\
  G359.445$-$00.054                &  359.446     &   $-$0.058     &  HNCO           &  -121.35, -84.20 &    24.52 $\pm$ 0.58  &     31.45 $\pm$ 0.74     &    6.51 $\pm$ 1.02   &   48.32 $\pm$ 7.70  &   CMZ    \\
                                   &  359.446     &   $-$0.058     &  SiO            &  -121.35, -92.22 &     7.07 $\pm$ 0.47  &      1.77 $\pm$ 0.11     &                 &    2.72 $\pm$ 0.46  &          \\
                                   &  359.446     &   $-$0.058     &  HC$_{3}$N      &  -112.13, -92.52 &     9.24 $\pm$ 0.42  &      4.42 $\pm$ 0.20     &                 &    6.79 $\pm$ 1.11  &          \\
  G359.453$-$00.112                &  359.455     &   $-$0.112     &  HNCO           &  -65.95, -42.46  &    17.49 $\pm$ 0.49  &     22.33 $\pm$ 0.62     &   14.37 $\pm$ 1.09   &   15.53 $\pm$ 1.26  &   CMZ    \\
                                   &  359.455     &   $-$0.112     &  SiO            &  -62.96, -43.10  &     8.70 $\pm$ 0.41  &      2.18 $\pm$ 0.10     &                 &    1.51 $\pm$ 0.13  &          \\
                                   &  359.455     &   $-$0.112     &  HC$_{3}$N      &  -66.38, -42.46  &    14.00 $\pm$ 0.46  &      6.72 $\pm$ 0.22     &                 &    4.68 $\pm$ 0.38  &          \\
  G359.565$-$00.161                &  359.551     &   $-$0.166     &  HNCO           &  -75.68, -43.94  &    17.28 $\pm$ 0.53  &     22.26 $\pm$ 0.68     &    3.79 $\pm$ 0.18   &   58.67 $\pm$ 3.44  &   CMZ    \\
                                   &  359.551     &   $-$0.166     &  SiO            &  -70.39, -35.12  &     7.73 $\pm$ 0.57  &      1.94 $\pm$ 0.14     &                 &    5.12 $\pm$ 0.45  &          \\
                                   &  359.551     &   $-$0.166     &  HC$_{3}$N      &  -72.51, -43.23  &     9.99 $\pm$ 0.50  &      4.76 $\pm$ 0.23     &                 &   12.54 $\pm$ 0.88  &          \\
  G359.868$-$00.085                &  359.866     &   $-$0.082     &  HNCO           &  -9.51, 23.01    &    63.64 $\pm$ 0.54  &     80.13 $\pm$ 0.67     &   35.98 $\pm$ 2.52   &   22.27 $\pm$ 1.57  &   CMZ    \\
                                   &  359.866     &   $-$0.082     &  SiO            &  -14.66, 30.87   &    43.22 $\pm$ 0.69  &     10.71 $\pm$ 0.17     &                 &    2.98 $\pm$ 0.21  &          \\
                                   &  359.866     &   $-$0.082     &  HC$_{3}$N      &  -11.95, 22.20   &    61.99 $\pm$ 0.62  &     30.16 $\pm$ 0.30     &                 &    8.38 $\pm$ 0.59  &          \\
  G359.895$-$00.069                &  359.892     &   $-$0.076     &  HNCO           &  -4.15, 35.34    &    81.24 $\pm$ 0.55  &    101.83 $\pm$ 0.68     &   26.93 $\pm$ 3.12   &   37.81 $\pm$ 4.40  &   CMZ    \\
                                   &  359.892     &   $-$0.076     &  SiO            &  -12.52, 37.35   &    62.86 $\pm$ 0.58  &     15.53 $\pm$ 0.14     &                 &    5.77 $\pm$ 0.67  &          \\
                                   &  359.892     &   $-$0.076     &  HC$_{3}$N      &  -7.83, 35.67    &    89.48 $\pm$ 0.52  &     43.73 $\pm$ 0.25     &                 &   16.24 $\pm$ 1.88  &          \\
  G359.977$-$00.072                &  359.980     &   $-$0.070     &  HNCO           &  26.79, 63.94    &    55.09 $\pm$ 0.79  &     73.64 $\pm$ 1.05     &   21.39 $\pm$ 0.87   &   34.43 $\pm$ 1.48  &   CMZ    \\
                                   &  359.980     &   $-$0.070     &  SiO            &  19.59, 76.07    &    70.96 $\pm$ 0.93  &     18.29 $\pm$ 0.23     &                 &    8.55 $\pm$ 0.36  &          \\
                                   &  359.980     &   $-$0.070     &  HC$_{3}$N      &  21.87, 71.52    &    97.20 $\pm$ 0.89  &     44.99 $\pm$ 0.41     &                 &   21.03 $\pm$ 0.87  &          \\
  G008.671$-$00.357                &    8.677     &   $-$0.360     &  HNCO           &  31.47, 40.86    &     4.59 $\pm$ 0.27  &      5.65 $\pm$ 0.33     &   18.69 $\pm$ 7.45   &    3.02 $\pm$ 1.21  &   Bubble \\
                                   &    8.677     &   $-$0.360     &  SiO            &  28.63, 41.97    &     4.50 $\pm$ 0.35  &      1.10 $\pm$ 0.08     &                 &    0.59 $\pm$ 0.23  &          \\
                                   &    8.677     &   $-$0.360     &  HC$_{3}$N      &  31.47, 42.47    &     9.71 $\pm$ 0.31  &      4.84 $\pm$ 0.15     &                 &    2.59 $\pm$ 1.03  &          \\
  G010.473$+$00.028                &   10.480     &      0.030     &  HNCO           &  58.39, 71.82    &     4.37 $\pm$ 0.28  &      5.55 $\pm$ 0.35     &   25.54 $\pm$ 6.74   &    2.17 $\pm$ 1.15  &   Bubble \\
                                   &   10.480     &      0.030     &  SiO            &  55.54, 84.44    &     4.57 $\pm$ 0.47  &      1.14 $\pm$ 0.11     &                 &    0.45 $\pm$ 0.23  &          \\
                                   &   10.480     &      0.030     &  HC$_{3}$N      &  59.21, 76.30    &     6.23 $\pm$ 0.35  &      3.00 $\pm$ 0.16     &                 &    1.18 $\pm$ 0.62  &          \\
  G322.159$+$00.635$^{\dag}$       &  322.159     &      0.637     &  HNCO           &  -59.93, -51.67  &     1.93 $\pm$ 0.25  &      2.93 $\pm$ 0.37     &   24.45 $\pm$ 3.76   &    1.20 $\pm$ 0.24  &   Bubble \\
                                   &  322.159     &      0.637     &  SiO            &  -63.59, -45.31  &     9.83 $\pm$ 0.32  &      2.76 $\pm$ 0.08     &                 &    1.13 $\pm$ 0.17  &          \\
                                   &  322.159     &      0.637     &  HC$_{3}$N      &  -61.83, -52.08  &    13.57 $\pm$ 0.25  &      5.88 $\pm$ 0.10     &                 &    2.40 $\pm$ 0.37  &          \\

  \hline
  \end{tabular}
  \end{center}
  \normalsize
\end{table*}

\setcounter{table}{1}
\begin{table*}
  \begin{center}
  \caption{{\em - continued.}}
  \tiny
  \begin{tabular}{lrrrcrrrrr}
  \hline
  \hline
\multicolumn{1}{c}{Source$^{\rm{a}}$} & \multicolumn{1}{c}{l}   & \multicolumn{1}{c}{b}   & \multicolumn{1}{c}{MOL} &     \multicolumn{1}{c}{v$_{\rm range}$} &  \multicolumn{1}{c}{$\int {T_{\rm MB}} dv$\,}  &  \multicolumn{1}{c}{N}                          & \multicolumn{1}{c}{N$_{\rm H_{2}}$}            & \multicolumn{1}{c}{$x$}                 & \multicolumn{1}{c}{comment}\\
                                      & \multicolumn{1}{c}{deg} & \multicolumn{1}{c}{deg} &                         &    \multicolumn{1}{c}{$km\ s^{-1}$\,}   &  \multicolumn{1}{c}{$K \cdot km\ s^{-1}$\,}    &  \multicolumn{1}{c}{$\times 10^{13} cm^{-2}$\,} & \multicolumn{1}{c}{$\times 10^{22} cm^{-2}$\,} & \multicolumn{1}{c}{$\times 10^{-10}$\,} &                            \\
\multicolumn{1}{c}{(1)}               & \multicolumn{1}{c}{(2)} & \multicolumn{1}{c}{(3)} & \multicolumn{1}{c}{(4)} &   \multicolumn{1}{c}{(5)}               &  \multicolumn{1}{c}{(6)}                       &  \multicolumn{1}{c}{(7)}                       & \multicolumn{1}{c}{(8)}                       & \multicolumn{1}{c}{(9)}                & \multicolumn{1}{c}{(10)}   \\
  \hline
     G326.653$+$00.618                &  326.640     &      0.615     &  HNCO           &  -46.18, -34.80  &     3.87 $\pm$ 0.30  &      4.87 $\pm$ 0.37     &   25.46 $\pm$ 3.52   &    1.91 $\pm$ 0.30  &   Bubble \\
                                   &  326.640     &      0.615     &  SiO            &  -48.69, -28.43  &     5.32 $\pm$ 0.37  &      1.32 $\pm$ 0.09     &                 &    0.52 $\pm$ 0.08  &          \\
                                   &  326.640     &      0.615     &  HC$_{3}$N      &  -43.34, -36.30  &     6.44 $\pm$ 0.23  &      3.13 $\pm$ 0.11     &                 &    1.23 $\pm$ 0.17  &          \\
  G327.293$-$00.579$^{\dag}$       &  327.296     &   $-$0.577     &  HNCO           &  -51.43, -39.94  &     3.74 $\pm$ 0.28  &      5.07 $\pm$ 0.37     &   59.52 $\pm$ 10.28   &    0.85 $\pm$ 0.30  &   Bubble \\
                                   &  327.296     &   $-$0.577     &  SiO            &  -53.44, -35.71  &     8.43 $\pm$ 0.38  &      2.19 $\pm$ 0.09     &                 &    0.37 $\pm$ 0.12  &          \\
                                   &  327.296     &   $-$0.577     &  HC$_{3}$N      &  -50.62, -36.72  &    19.24 $\pm$ 0.30  &      8.82 $\pm$ 0.13     &                 &    1.48 $\pm$ 0.51  &          \\
  G345.004$-$00.224$^{\dag}$       &  345.007     &   $-$0.220     &  HNCO           &  -34.85, -20.90  &     3.70 $\pm$ 0.28  &      4.92 $\pm$ 0.37     &   19.83 $\pm$ 8.50   &    2.48 $\pm$ 1.08  &   Bubble \\
                                   &  345.007     &   $-$0.220     &  SiO            &  -43.62, -11.96  &    12.96 $\pm$ 0.54  &      3.33 $\pm$ 0.13     &                 &    1.68 $\pm$ 0.72  &          \\
                                   &  345.007     &   $-$0.220     &  HC$_{3}$N      &  -35.63, -19.17  &    11.49 $\pm$ 0.32  &      5.34 $\pm$ 0.14     &                 &    2.69 $\pm$ 1.15  &          \\
  G350.101$+$00.083$^{\dag}$       &  350.110     &      0.096     &  HNCO           &  -74.50, -62.16  &     4.93 $\pm$ 0.36  &      6.41 $\pm$ 0.46     &   16.87 $\pm$ 3.76   &    3.80 $\pm$ 0.89  &   Bubble \\
                                   &  350.110     &      0.096     &  SiO            &  -77.85, -58.98  &     5.65 $\pm$ 0.45  &      1.43 $\pm$ 0.11     &                 &    0.85 $\pm$ 0.20  &          \\
                                   &  350.110     &      0.096     &  HC$_{3}$N      &  -75.74, -62.16  &     4.83 $\pm$ 0.34  &      2.28 $\pm$ 0.16     &                 &    1.35 $\pm$ 0.31  &          \\
  G351.443$+$00.659                &  351.451     &      0.654     &  HNCO           &  -8.29, 2.03     &     6.05 $\pm$ 0.31  &      7.76 $\pm$ 0.39     &   45.35 $\pm$ 10.38   &    1.71 $\pm$ 0.78  &   Bubble \\
                                   &  351.451     &      0.654     &  SiO            &  -16.13, 5.85    &    22.42 $\pm$ 0.48  &      5.62 $\pm$ 0.12     &                 &    1.24 $\pm$ 0.56  &          \\
                                   &  351.451     &      0.654     &  HC$_{3}$N      &  -11.73, 2.41    &    33.35 $\pm$ 0.35  &     15.95 $\pm$ 0.16     &                 &    3.52 $\pm$ 1.61  &          \\
  G351.582$-$00.352                &  351.579     &   $-$0.355     &  HNCO           &  -100.73, -90.92 &     5.18 $\pm$ 0.25  &      7.35 $\pm$ 0.35     &   31.87 $\pm$ 8.80   &    2.31 $\pm$ 0.64  &   Bubble \\
                                   &  351.579     &   $-$0.355     &  SiO            &  -104.31, -92.48 &     3.60 $\pm$ 0.35  &      0.97 $\pm$ 0.09     &                 &    0.30 $\pm$ 0.08  &          \\
                                   &  351.579     &   $-$0.355     &  HC$_{3}$N      &  -101.04, -90.92 &     7.30 $\pm$ 0.30  &      3.26 $\pm$ 0.13     &                 &    1.02 $\pm$ 0.28  &          \\
  G351.775$-$00.537                &  351.773     &   $-$0.539     &  HNCO           &  -8.04, 1.36     &     4.73 $\pm$ 0.35  &      7.62 $\pm$ 0.56     &   50.86 $\pm$ 6.09   &    1.50 $\pm$ 0.37  &   Bubble \\
                                   &  351.773     &   $-$0.539     &  SiO            &  -13.53, 11.97   &    25.98 $\pm$ 0.59  &      7.58 $\pm$ 0.17     &                 &    1.49 $\pm$ 0.35  &          \\
                                   &  351.773     &   $-$0.539     &  HC$_{3}$N      &  -8.53, 3.80     &    13.09 $\pm$ 0.40  &      5.57 $\pm$ 0.17     &                 &    1.09 $\pm$ 0.26  &          \\
  G329.030$-$00.202$^{\dag}$       &  329.031     &   $-$0.201     &  HNCO           &  -47.95, -39.93  &     3.50 $\pm$ 0.27  &      4.43 $\pm$ 0.34     &   22.03 $\pm$ 3.86   &    2.01 $\pm$ 0.38  &   NMSFR  \\
                                   &  329.031     &   $-$0.201     &  SiO            &  -56.70, -36.64  &     8.90 $\pm$ 0.37  &      2.21 $\pm$ 0.09     &                 &    1.00 $\pm$ 0.18  &          \\
                                   &  329.031     &   $-$0.201     &  HC$_{3}$N      &  -55.25, -38.65  &    10.63 $\pm$ 0.35  &      5.15 $\pm$ 0.16     &                 &    2.34 $\pm$ 0.41  &          \\
  G331.708$+$00.583                &  331.709     &      0.582     &  HNCO           &  -69.26, -62.52  &     1.61 $\pm$ 0.21  &      1.91 $\pm$ 0.24     &    8.32 $\pm$ 1.88   &    2.30 $\pm$ 0.60  &   NMSFR  \\
                                   &  331.709     &      0.582     &  SiO            &  -74.56, -59.08  &     3.72 $\pm$ 0.35  &      0.89 $\pm$ 0.08     &                 &    1.07 $\pm$ 0.26  &          \\
                                   &  331.709     &      0.582     &  HC$_{3}$N      &  -70.41, -61.80  &     3.59 $\pm$ 0.25  &      1.87 $\pm$ 0.13     &                 &    2.25 $\pm$ 0.53  &          \\
  G331.709$+$00.602                &  331.709     &      0.602     &  HNCO           &  -70.63, -60.74  &     3.26 $\pm$ 0.30  &      3.94 $\pm$ 0.36     &   10.13 $\pm$ 1.53   &    3.89 $\pm$ 0.68  &   NMSFR  \\
                                   &  331.709     &      0.602     &  SiO            &  -71.11, -58.98  &     2.63 $\pm$ 0.33  &      0.63 $\pm$ 0.07     &                 &    0.63 $\pm$ 0.12  &          \\
                                   &  331.709     &      0.602     &  HC$_{3}$N      &  -70.79, -63.77  &     3.59 $\pm$ 0.25  &      1.83 $\pm$ 0.12     &                 &    1.80 $\pm$ 0.30  &          \\
  G335.586$-$00.289$^{\dag}$       &  335.584     &   $-$0.288     &  HNCO           &  -49.02, -41.51  &     2.79 $\pm$ 0.24  &      3.75 $\pm$ 0.32     &   17.33 $\pm$ 3.99   &    2.16 $\pm$ 0.53  &   NMSFR  \\
                                   &  335.584     &   $-$0.288     &  SiO            &  -56.53, -41.92  &     6.02 $\pm$ 0.31  &      1.56 $\pm$ 0.08     &                 &    0.90 $\pm$ 0.21  &          \\
                                   &  335.584     &   $-$0.288     &  HC$_{3}$N      &  -52.31, -42.13  &     7.58 $\pm$ 0.28  &      3.50 $\pm$ 0.12     &                 &    2.02 $\pm$ 0.47  &          \\
  G348.754$-$00.941$^{\dag}$       &  348.761     &   $-$0.948     &  HNCO           &  -22.22, -10.53  &     3.81 $\pm$ 0.35  &      4.46 $\pm$ 0.41     &    9.80 $\pm$ 3.42   &    4.55 $\pm$ 1.64  &   NMSFR  \\
                                   &  348.761     &   $-$0.948     &  SiO            &  -16.89, -6.91   &     3.34 $\pm$ 0.35  &      0.79 $\pm$ 0.08     &                 &    0.81 $\pm$ 0.29  &          \\
                                   &  348.761     &   $-$0.948     &  HC$_{3}$N      &  -16.54, -4.68   &     5.02 $\pm$ 0.32  &      2.66 $\pm$ 0.16     &                 &    2.71 $\pm$ 0.96  &          \\
  G351.157$+$00.701                &  351.155     &      0.709     &  HNCO           &  -9.69, -2.76    &     2.86 $\pm$ 0.27  &      3.99 $\pm$ 0.37     &   23.82 $\pm$ 5.66   &    1.67 $\pm$ 0.42  &   NMSFR  \\
                                   &  351.155     &      0.709     &  SiO            &  -9.18, -2.38    &     3.54 $\pm$ 0.25  &      0.94 $\pm$ 0.06     &                 &    0.39 $\pm$ 0.09  &          \\
                                   &  351.155     &      0.709     &  HC$_{3}$N      &  -10.82, -1.87   &    13.08 $\pm$ 0.28  &      5.90 $\pm$ 0.12     &                 &    2.47 $\pm$ 0.59  &          \\

  \hline
  \end{tabular}
  \smallskip
  \leftline{\scriptsize{\emph{Note}. $^{\rm{a}}$ $\ast$ indicates the source name is adopted from \citet{2014A&A...562A...3M}. $\dag$ indicates a source is identified as an IRDC by \citet{2009A&A...505..405P}.}}
  \end{center}
  \normalsize
\end{table*}

\begin{table*}
  \begin{center}
  \tiny
  \caption{The integrated intensity and abundance ratios of HNCO, SiO and HC$_{3}$N.}

  \label{tab:source_ratios}
  \begin{tabular}{lrrrrrrr}
  \hline
  \hline
  Source  &  ${I_{\rm HNCO}/I_{\rm SiO}}$  & ${I_{\rm HNCO}/I_{\rm HC_{3}N}}$  &  ${I_{\rm SiO}/I_{\rm HC_{3}N}}$   &  ${N_{\rm HNCO}/N_{\rm SiO}}$ & ${N_{\rm HNCO}/N_{\rm HC_{3}N}}$ &  ${N_{\rm SiO}/N_{\rm HC_{3}N}}$  &  comment  \\
  \hline
    G000.067$-$00.077   &   1.47 $\pm$ 0.05   &   1.38 $\pm$ 0.04   &   0.94 $\pm$ 0.03   &    7.45 $\pm$ 0.26   &   3.50 $\pm$ 0.10   &   0.47 $\pm$ 0.02   &  CMZ    \\
  G000.104$-$00.080   &   0.98 $\pm$ 0.02   &   0.77 $\pm$ 0.01   &   0.79 $\pm$ 0.01   &    4.98 $\pm$ 0.08   &   1.99 $\pm$ 0.03   &   0.40 $\pm$ 0.01   &  CMZ    \\
  G000.106$-$00.001   &   1.45 $\pm$ 0.04   &   0.97 $\pm$ 0.02   &   0.67 $\pm$ 0.01   &    7.38 $\pm$ 0.18   &   2.54 $\pm$ 0.05   &   0.34 $\pm$ 0.01   &  CMZ    \\
  G000.110$+$00.148   &   2.04 $\pm$ 0.17   &   1.58 $\pm$ 0.09   &   0.78 $\pm$ 0.07   &   10.36 $\pm$ 0.84   &   4.06 $\pm$ 0.24   &   0.39 $\pm$ 0.03   &  CMZ    \\
  G000.314$-$00.100   &   0.99 $\pm$ 0.06   &   1.57 $\pm$ 0.11   &   1.59 $\pm$ 0.13   &    5.07 $\pm$ 0.31   &   4.25 $\pm$ 0.30   &   0.84 $\pm$ 0.06   &  CMZ    \\
  G000.497$+$00.021   &   3.14 $\pm$ 0.11   &   1.58 $\pm$ 0.03   &   0.51 $\pm$ 0.02   &   15.85 $\pm$ 0.52   &   3.99 $\pm$ 0.07   &   0.25 $\pm$ 0.01   &  CMZ    \\
  G000.645$+$00.027   &   3.63 $\pm$ 0.10   &   1.85 $\pm$ 0.03   &   0.51 $\pm$ 0.01   &   18.35 $\pm$ 0.48   &   4.61 $\pm$ 0.07   &   0.25 $\pm$ 0.01   &  CMZ    \\
  G000.892$+$00.143   &   1.24 $\pm$ 0.06   &   1.42 $\pm$ 0.06   &   1.15 $\pm$ 0.06   &    6.20 $\pm$ 0.27   &   3.40 $\pm$ 0.14   &   0.55 $\pm$ 0.03   &  CMZ    \\
  G000.908$+$00.116   &   1.34 $\pm$ 0.12   &   2.19 $\pm$ 0.21   &   1.63 $\pm$ 0.18   &    6.72 $\pm$ 0.58   &   5.20 $\pm$ 0.48   &   0.77 $\pm$ 0.08   &  CMZ    \\
  G001.226$+$00.059   &   0.40 $\pm$ 0.02   &   0.74 $\pm$ 0.03   &   1.84 $\pm$ 0.05   &    1.99 $\pm$ 0.09   &   1.68 $\pm$ 0.08   &   0.84 $\pm$ 0.02   &  CMZ    \\
  G001.344$+$00.258   &   0.47 $\pm$ 0.01   &   0.74 $\pm$ 0.02   &   1.58 $\pm$ 0.04   &    2.32 $\pm$ 0.07   &   1.63 $\pm$ 0.05   &   0.71 $\pm$ 0.02   &  CMZ    \\
  G001.381$+$00.201   &   0.37 $\pm$ 0.02   &   0.90 $\pm$ 0.07   &   2.40 $\pm$ 0.15   &    1.86 $\pm$ 0.12   &   2.00 $\pm$ 0.16   &   1.08 $\pm$ 0.07   &  CMZ    \\
  G001.510$+$00.155   &   1.80 $\pm$ 0.08   &   1.44 $\pm$ 0.05   &   0.80 $\pm$ 0.04   &    8.90 $\pm$ 0.39   &   3.05 $\pm$ 0.11   &   0.34 $\pm$ 0.02   &  CMZ    \\
  G001.610$-$00.172   &   4.17 $\pm$ 0.29   &   3.51 $\pm$ 0.19   &   0.84 $\pm$ 0.07   &   20.52 $\pm$ 1.32   &   7.25 $\pm$ 0.38   &   0.35 $\pm$ 0.03   &  CMZ    \\
  G001.655$-$00.062   &   4.48 $\pm$ 0.19   &   3.13 $\pm$ 0.09   &   0.70 $\pm$ 0.03   &   21.79 $\pm$ 0.94   &   5.95 $\pm$ 0.16   &   0.27 $\pm$ 0.01   &  CMZ    \\
  G001.694$-$00.385   &   1.29 $\pm$ 0.07   &   1.05 $\pm$ 0.04   &   0.82 $\pm$ 0.05   &    6.33 $\pm$ 0.34   &   2.18 $\pm$ 0.09   &   0.34 $\pm$ 0.02   &  CMZ    \\
  G001.699$-$00.366   &   1.38 $\pm$ 0.05   &   0.69 $\pm$ 0.02   &   0.50 $\pm$ 0.02   &    6.77 $\pm$ 0.25   &   1.40 $\pm$ 0.03   &   0.21 $\pm$ 0.01   &  CMZ    \\
  G001.734$-$00.410   &   1.92 $\pm$ 0.05   &   1.20 $\pm$ 0.02   &   0.63 $\pm$ 0.02   &    9.34 $\pm$ 0.24   &   2.31 $\pm$ 0.04   &   0.25 $\pm$ 0.01   &  CMZ    \\
  G001.883$-$00.062   &   2.50 $\pm$ 0.19   &   1.96 $\pm$ 0.11   &   0.78 $\pm$ 0.07   &   12.39 $\pm$ 0.86   &   4.24 $\pm$ 0.24   &   0.34 $\pm$ 0.03   &  CMZ    \\
  G003.240$+$00.635   &   0.50 $\pm$ 0.04   &   0.87 $\pm$ 0.08   &   1.76 $\pm$ 0.13   &    2.41 $\pm$ 0.19   &   1.69 $\pm$ 0.15   &   0.70 $\pm$ 0.05   &  CMZ    \\
  G003.338$+$00.419   &   0.60 $\pm$ 0.02   &   0.60 $\pm$ 0.02   &   1.01 $\pm$ 0.03   &    2.89 $\pm$ 0.11   &   1.14 $\pm$ 0.05   &   0.39 $\pm$ 0.01   &  CMZ    \\
  G359.445$-$00.054   &   3.47 $\pm$ 0.24   &   2.65 $\pm$ 0.14   &   0.77 $\pm$ 0.06   &   17.77 $\pm$ 1.18   &   7.12 $\pm$ 0.36   &   0.40 $\pm$ 0.03   &  CMZ    \\
  G359.453$-$00.112   &   2.01 $\pm$ 0.11   &   1.25 $\pm$ 0.05   &   0.62 $\pm$ 0.04   &   10.24 $\pm$ 0.55   &   3.32 $\pm$ 0.14   &   0.32 $\pm$ 0.02   &  CMZ    \\
  G359.565$-$00.161   &   2.24 $\pm$ 0.18   &   1.73 $\pm$ 0.10   &   0.77 $\pm$ 0.07   &   11.47 $\pm$ 0.90   &   4.68 $\pm$ 0.27   &   0.41 $\pm$ 0.04   &  CMZ    \\
  G359.868$-$00.085   &   1.47 $\pm$ 0.03   &   1.03 $\pm$ 0.01   &   0.70 $\pm$ 0.01   &    7.48 $\pm$ 0.13   &   2.66 $\pm$ 0.03   &   0.36 $\pm$ 0.01   &  CMZ    \\
  G359.895$-$00.069   &   1.29 $\pm$ 0.01   &   0.91 $\pm$ 0.01   &   0.70 $\pm$ 0.01   &    6.56 $\pm$ 0.07   &   2.33 $\pm$ 0.02   &   0.36 $\pm$ 0.01   &  CMZ    \\
  G359.977$-$00.072   &   0.78 $\pm$ 0.02   &   0.57 $\pm$ 0.01   &   0.73 $\pm$ 0.01   &    4.03 $\pm$ 0.08   &   1.64 $\pm$ 0.03   &   0.41 $\pm$ 0.01   &  CMZ    \\
  G008.671$-$00.357   &   1.02 $\pm$ 0.10   &   0.47 $\pm$ 0.03   &   0.46 $\pm$ 0.04   &    5.14 $\pm$ 0.48   &   1.17 $\pm$ 0.08   &   0.23 $\pm$ 0.02   &  Bubble \\
  G010.473$+$00.028   &   0.96 $\pm$ 0.12   &   0.70 $\pm$ 0.06   &   0.73 $\pm$ 0.09   &    4.87 $\pm$ 0.56   &   1.85 $\pm$ 0.15   &   0.38 $\pm$ 0.04   &  Bubble \\
  G322.159$+$00.635   &   0.20 $\pm$ 0.03   &   0.14 $\pm$ 0.02   &   0.72 $\pm$ 0.03   &    1.06 $\pm$ 0.14   &   0.50 $\pm$ 0.06   &   0.47 $\pm$ 0.02   &  Bubble \\
  G326.653$+$00.618   &   0.73 $\pm$ 0.08   &   0.60 $\pm$ 0.05   &   0.83 $\pm$ 0.06   &    3.69 $\pm$ 0.38   &   1.56 $\pm$ 0.13   &   0.42 $\pm$ 0.03   &  Bubble \\
  G327.293$-$00.579   &   0.44 $\pm$ 0.04   &   0.19 $\pm$ 0.01   &   0.44 $\pm$ 0.02   &    2.32 $\pm$ 0.19   &   0.57 $\pm$ 0.04   &   0.25 $\pm$ 0.01   &  Bubble \\
  G345.004$-$00.224   &   0.29 $\pm$ 0.02   &   0.32 $\pm$ 0.03   &   1.13 $\pm$ 0.06   &    1.48 $\pm$ 0.13   &   0.92 $\pm$ 0.07   &   0.62 $\pm$ 0.03   &  Bubble \\
  G350.101$+$00.083   &   0.87 $\pm$ 0.09   &   1.02 $\pm$ 0.10   &   1.17 $\pm$ 0.12   &    4.48 $\pm$ 0.47   &   2.81 $\pm$ 0.28   &   0.63 $\pm$ 0.07   &  Bubble \\
  G351.443$+$00.659   &   0.27 $\pm$ 0.01   &   0.18 $\pm$ 0.01   &   0.67 $\pm$ 0.02   &    1.38 $\pm$ 0.08   &   0.49 $\pm$ 0.02   &   0.35 $\pm$ 0.01   &  Bubble \\
  G351.582$-$00.352   &   1.44 $\pm$ 0.16   &   0.71 $\pm$ 0.04   &   0.49 $\pm$ 0.05   &    7.58 $\pm$ 0.79   &   2.25 $\pm$ 0.14   &   0.30 $\pm$ 0.03   &  Bubble \\
  G351.775$-$00.537   &   0.18 $\pm$ 0.01   &   0.36 $\pm$ 0.03   &   1.98 $\pm$ 0.08   &    1.01 $\pm$ 0.08   &   1.37 $\pm$ 0.11   &   1.36 $\pm$ 0.05   &  Bubble \\
  G329.030$-$00.202   &   0.39 $\pm$ 0.03   &   0.33 $\pm$ 0.03   &   0.84 $\pm$ 0.04   &    2.00 $\pm$ 0.17   &   0.86 $\pm$ 0.07   &   0.43 $\pm$ 0.02   &  NMSFR  \\
  G331.708$+$00.583   &   0.43 $\pm$ 0.07   &   0.45 $\pm$ 0.07   &   1.04 $\pm$ 0.12   &    2.15 $\pm$ 0.33   &   1.02 $\pm$ 0.15   &   0.48 $\pm$ 0.05   &  NMSFR  \\
  G331.709$+$00.602   &   1.24 $\pm$ 0.19   &   0.91 $\pm$ 0.10   &   0.73 $\pm$ 0.11   &    6.25 $\pm$ 0.90   &   2.15 $\pm$ 0.24   &   0.34 $\pm$ 0.04   &  NMSFR  \\
  G335.586$-$00.289   &   0.46 $\pm$ 0.05   &   0.37 $\pm$ 0.03   &   0.79 $\pm$ 0.05   &    2.40 $\pm$ 0.24   &   1.07 $\pm$ 0.10   &   0.45 $\pm$ 0.03   &  NMSFR  \\
  G348.754$-$00.941   &   1.14 $\pm$ 0.16   &   0.76 $\pm$ 0.08   &   0.67 $\pm$ 0.08   &    5.65 $\pm$ 0.77   &   1.68 $\pm$ 0.18   &   0.30 $\pm$ 0.03   &  NMSFR  \\
  G351.157$+$00.701   &   0.81 $\pm$ 0.10   &   0.22 $\pm$ 0.02   &   0.27 $\pm$ 0.02   &    4.24 $\pm$ 0.48   &   0.68 $\pm$ 0.06   &   0.16 $\pm$ 0.01   &  NMSFR  \\
  \hline
  \end{tabular}
  \end{center}
  \normalsize
\end{table*}

\begin{table}
  \begin{center}
  \scriptsize
  \caption{Statistics of integrated intensity and abundance ratios.}

  \label{tab:source_statistics}
  \begin{tabular}{lrrrrr}
  \hline
  \hline

  \multicolumn{6}{c}{All sources\,}                                                    \\
  Quantity                             &  Mean  & Std$^{a}$  & Median  & Min. & Max.   \\
  \hline
  ${I_{\rm HNCO}/I_{\rm SiO}}$         &  1.36  & 1.07       & 1.14    & 0.18 & 4.48   \\
  ${I_{\rm HNCO}/I_{\rm HC_{3}N}}$     &  1.07  & 0.77       & 0.90    & 0.14 & 3.51   \\
  ${I_{\rm SiO}/I_{\rm HC_{3}N}}$      &  0.92  & 0.46       & 0.78    & 0.27 & 2.40   \\
  ${N_{\rm HNCO}/N_{\rm SiO}}$         &  6.82  & 5.33       & 5.65    & 1.01 & 21.79  \\
  ${N_{\rm HNCO}/N_{\rm HC_{3}N}}$     &  2.58  & 1.70       & 2.15    & 0.49 & 7.25   \\
  ${N_{\rm SiO}/N_{\rm HC_{3}N}}$      &  0.45  & 0.24       & 0.39    & 0.16 & 1.36   \\

  \hline
  \multicolumn{6}{c}{CMZ sources\,}                                                    \\
  Quantity                             &  Mean  & Std$^{a}$   & Median  & Min. & Max.   \\
  \hline
  ${I_{\rm HNCO}/I_{\rm SiO}}$         &  1.76  & 1.15       & 1.45    & 0.37 & 4.48   \\
  ${I_{\rm HNCO}/I_{\rm HC_{3}N}}$     &  1.42  & 0.75       & 1.25    & 0.57 & 3.51   \\
  ${I_{\rm SiO}/I_{\rm HC_{3}N}}$      &  0.98  & 0.49       & 0.78    & 0.50 & 2.40   \\
  ${N_{\rm HNCO}/N_{\rm SiO}}$         &  8.79  & 5.71       & 7.38    & 1.86 & 21.79  \\
  ${N_{\rm HNCO}/N_{\rm HC_{3}N}}$     &  3.33  & 1.69       & 3.05    & 1.14 & 7.25   \\
  ${N_{\rm SiO}/N_{\rm HC_{3}N}}$      &  0.46  & 0.22       & 0.39    & 0.21 & 1.08   \\

  \hline 					
  \multicolumn{6}{c}{Bubble sources\,}                                                 \\
  Quantity                             &  Mean  &  Std$^{a}$  & Median  & Min. & Max.   \\
  \hline
  ${I_{\rm HNCO}/I_{\rm SiO}}$         &  0.64  &  0.43      & 0.59    & 0.18 & 1.44   \\
  ${I_{\rm HNCO}/I_{\rm HC_{3}N}}$     &  0.47  &  0.29      & 0.42    & 0.14 & 1.02   \\
  ${I_{\rm SiO}/I_{\rm HC_{3}N}}$      &  0.86  &  0.47      & 0.73    & 0.44 & 1.98   \\
  ${N_{\rm HNCO}/N_{\rm SiO}}$         &  3.30  &  2.21      & 3.00    & 1.01 & 7.58   \\
  ${N_{\rm HNCO}/N_{\rm HC_{3}N}}$     &  1.35  &  0.78      & 1.27    & 0.49 & 2.81   \\
  ${N_{\rm SiO}/N_{\rm HC_{3}N}}$      &  0.50  &  0.33      & 0.40    & 0.23 & 1.36   \\

  \hline 					
  \multicolumn{6}{c}{NMSFR sources\,}                                                  \\
  Quantity                             &  Mean  & Std$^{a}$   & Median  & Min. & Max.   \\
  \hline
  ${I_{\rm HNCO}/I_{\rm SiO}}$         &  0.75  & 0.38       & 0.64    & 0.39 & 1.24   \\
  ${I_{\rm HNCO}/I_{\rm HC_{3}N}}$     &  0.51  & 0.27       & 0.41    & 0.22 & 0.91   \\
  ${I_{\rm SiO}/I_{\rm HC_{3}N}}$      &  0.72  & 0.25       & 0.76    & 0.27 & 1.04   \\
  ${N_{\rm HNCO}/N_{\rm SiO}}$         &  3.78  & 1.87       & 3.32    & 2.00 & 6.25   \\
  ${N_{\rm HNCO}/N_{\rm HC_{3}N}}$     &  1.24  & 0.56       & 1.05    & 0.68 & 2.15   \\
  ${N_{\rm SiO}/N_{\rm HC_{3}N}}$      &  0.36  & 0.12       & 0.39    & 0.16 & 0.48   \\
  \hline
  \end{tabular}
  \end{center}
  \footnotesize{$^a$ Standard deviation of the mean.}
\end{table}

\clearpage
\appendix{}
\restartappendixnumbering

\section{Individual targets analysis}
\label{sec:Individual_targets_analysis}
\begin{figure}
  \centering
	\includegraphics[width = 0.75\linewidth]{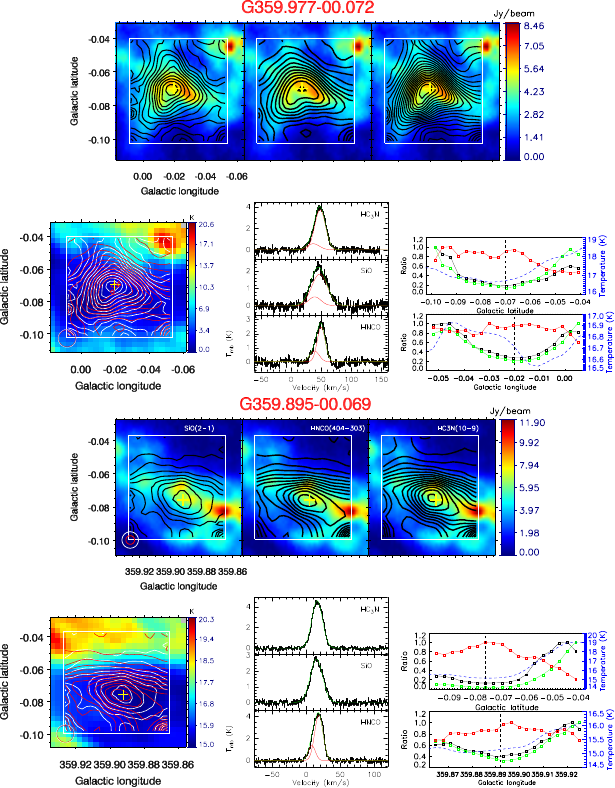}
    \caption{\label{figA1}Contour maps of SiO 2--1, HNCO 4$_{04}$--3$_{03}$ and HC$_{3}$N 10--9 integrated intensity superimposed on its 870 $\mu$m continuum emission map in the upper panels of G359.977$-$00.072 (contours start from 3$\sigma$ and have steps of 3$\sigma$, where 3$\sigma$ is 2.58, 2.10 and 2.22 K km s$^{-1}$, respectively) and G359.895$-$00.069 (starting with \textbf{steps of} 6$\sigma$, where 6$\sigma$ is 3.54, 3.30 and 2.94 K km s$^{-1}$, respectively). Lower panels of G359.977$-$00.072 and G359.895$-$00.069 left to right: contour maps of SiO 2-1 (white solid lines) and HNCO 4$_{04}$--3$_{03}$ (red solid lines) integrated intensity superimposed on its dust temperature map, the extracted beam averaged spectra of SiO 2--1, HNCO 4$_{04}$--3$_{03}$ and HC$_{3}$N 10--9 from the position marked by a yellow cross, and normalized variations in ${\rm N_{HNCO}/N_{SiO}}$ (black line), ${\rm N_{HNCO}/N_{HC_{3}N}}$ (green line), ${\rm N_{HC_{3}N}/N_{SiO}}$ (red line) and dust temperature (blue dashed line) along the Galactic longitude and Galactic latitude directions passing through the yellow cross (vertical dashed lines). The two Gaussian components and total fits to the spectra are shown with red and green lines, respectively}
    \label{fig:Figure_A1}
\end{figure}

\begin{figure}
  \centering
	\includegraphics[width = 0.85\linewidth]{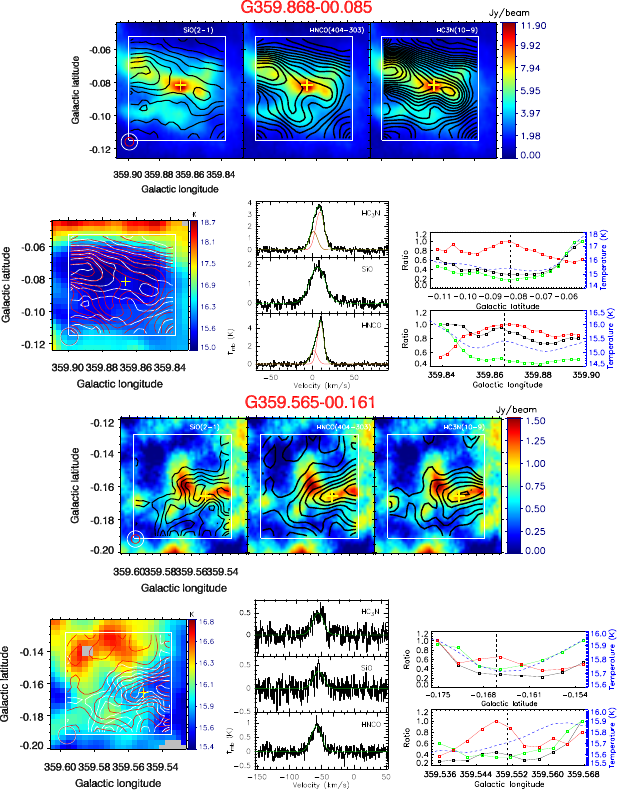}
    \caption{Same as Fig.~\ref{fig:Figure_A1}, but for G359.868$-$00.085 (contours start from 3$\sigma$ with steps of 6$\sigma$, where 6$\sigma$ is 3.66, 2.94 and 1.92 K km s$^{-1}$, respectively) and G359.565$-$00.161 (contours start from 3$\sigma$ with steps of 2$\sigma$, where 2$\sigma$ is 0.76, 0.96 and 0.80 K km s$^{-1}$, respectively).}
    \label{fig:Figure_A2}
\end{figure}

\begin{figure}
  \centering
	\includegraphics[width = 0.85\linewidth]{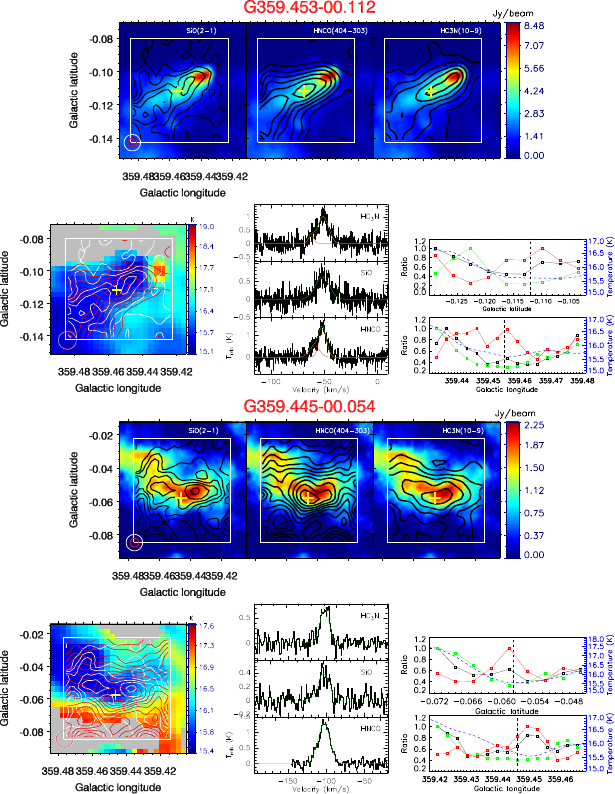}
    \caption{Same as Fig.~\ref{fig:Figure_A1}, but for G359.453$-$00.112 (contours start from 3$\sigma$ with steps of 3$\sigma$, where 3$\sigma$ is 1.23, 1.35 and 1.38 K km s$^{-1}$, respectively) and G359.445$-$00.054 (contours start from 3$\sigma$ with steps of 2$\sigma$, where 2$\sigma$ is 0.84, 1.00 and 0.80 K km s$^{-1}$, respectively).}
    \label{fig:Figure_A3}
\end{figure}

\begin{figure}
  \centering
	\includegraphics[width = 0.85\linewidth]{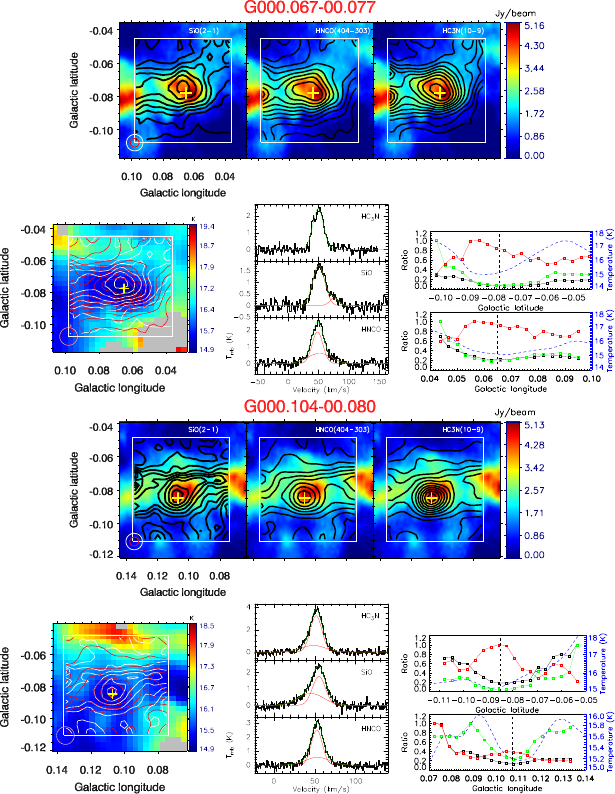}
    \caption{Same as Fig.~\ref{fig:Figure_A1}, but for G000.067$-$00.077 (contours start from 3$\sigma$ with steps of 3$\sigma$, where 3$\sigma$ is 3.00, 2.85 and 1.44 K km s$^{-1}$, respectively) and G000.104$-$00.080 (contours start from 3$\sigma$ with steps of 5$\sigma$, where 5$\sigma$ is 3.35, 3.50 and 3.15 K km s$^{-1}$, respectively).}
    \label{fig:Figure_A4}
\end{figure}

\begin{figure}
  \centering
	\includegraphics[width = 0.85\linewidth]{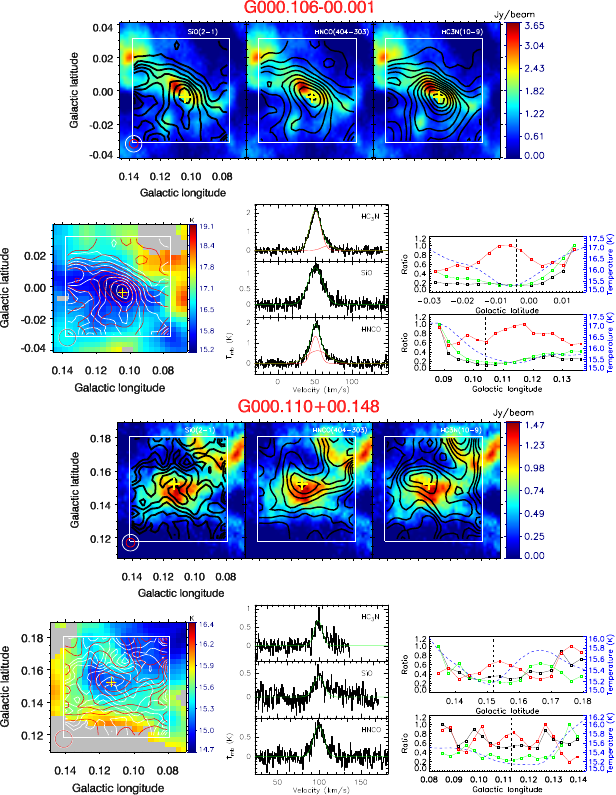}
    \caption{Same as Fig.~\ref{fig:Figure_A1}, but for G000.106$-$00.001 (contours start from 3$\sigma$ with steps of 4$\sigma$, where 4$\sigma$ is 2.08, 2.12 and 1.38 K km s$^{-1}$, respectively) and G000.110+00.148 (contours start from 3$\sigma$ with steps of 2$\sigma$, where 2$\sigma$ is 1.12, 1.04 and 0.74 K km s$^{-1}$, respectively).}
    \label{fig:Figure_A5}
\end{figure}

\begin{figure}
  \centering
	\includegraphics[width = 0.85\linewidth]{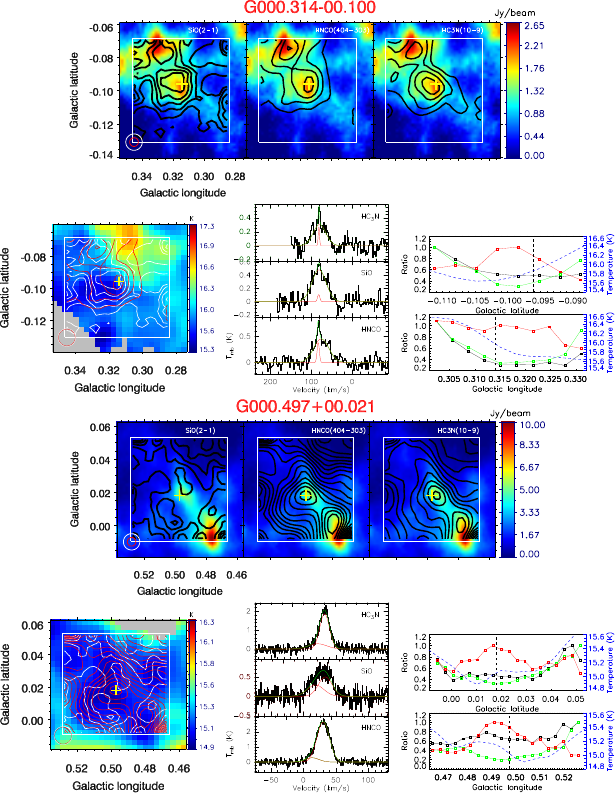}
    \caption{Same as Fig.~\ref{fig:Figure_A1}, but for G000.314$-$00.100 (contours start from 3$\sigma$ with steps of 2$\sigma$, where 2$\sigma$ is 1.84, 1.74 and 1.60 K km s$^{-1}$, respectively) and G000.497+00.021 (contours start from 3$\sigma$ with steps of 3$\sigma$, where 3$\sigma$ is 1.80, 1.89 and 1.18 K km s$^{-1}$, respectively).}
    \label{fig:Figure_A6}
\end{figure}

\begin{figure}
  \centering
	\includegraphics[width = 0.85\linewidth]{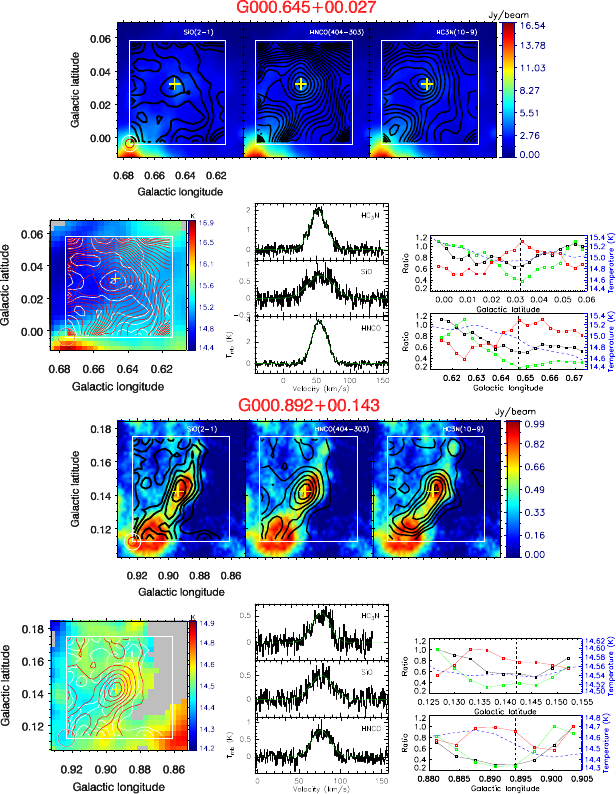}
    \caption{Same as Fig.~\ref{fig:Figure_A1}, but for G000.645+00.027 (contours start from 3$\sigma$ with steps of 3$\sigma$, where 3$\sigma$ is 2.37, 2.16 and 2.34 K km s$^{-1}$, respectively) and G000.892+00.143 (contours start from 3$\sigma$ with steps of 2$\sigma$, where 2$\sigma$ is 1.24, 1.24 and 0.96 K km s$^{-1}$, respectively).}
    \label{fig:Figure_A7}
\end{figure}

\begin{figure}
  \centering
	\includegraphics[width = 0.85\linewidth]{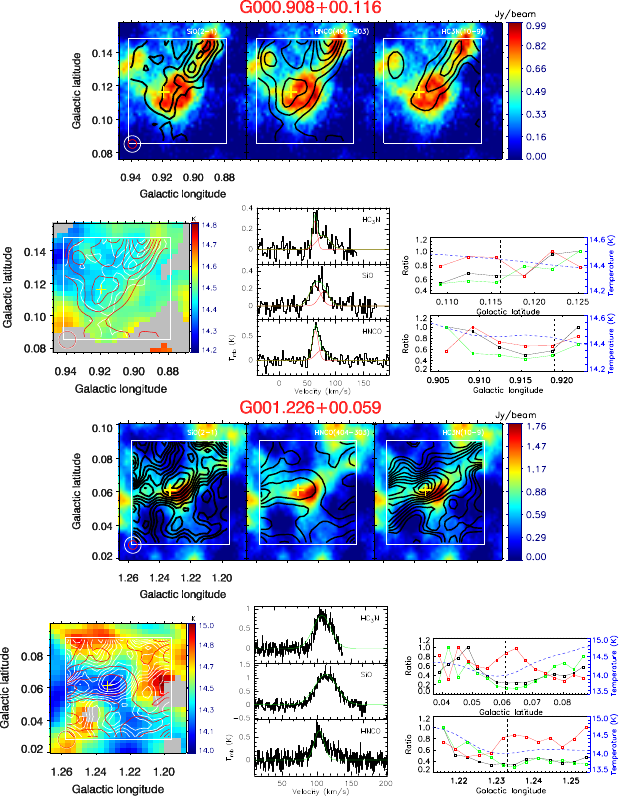}
    \caption{Same as Fig.~\ref{fig:Figure_A1}, but for G000.908+00.116 (contours start from 3$\sigma$ with steps of 2$\sigma$, where 2$\sigma$ is 2.43, 1.58 and 1.44 K km s$^{-1}$, respectively) and G001.226+00.059 (contours start from 3$\sigma$ with steps of 2$\sigma$, where 2$\sigma$ is 1.52, 1.46 and 0.96 K km s$^{-1}$, respectively).}
    \label{fig:Figure_A8}
\end{figure}

\begin{figure}
  \centering
	\includegraphics[width = 0.85\linewidth]{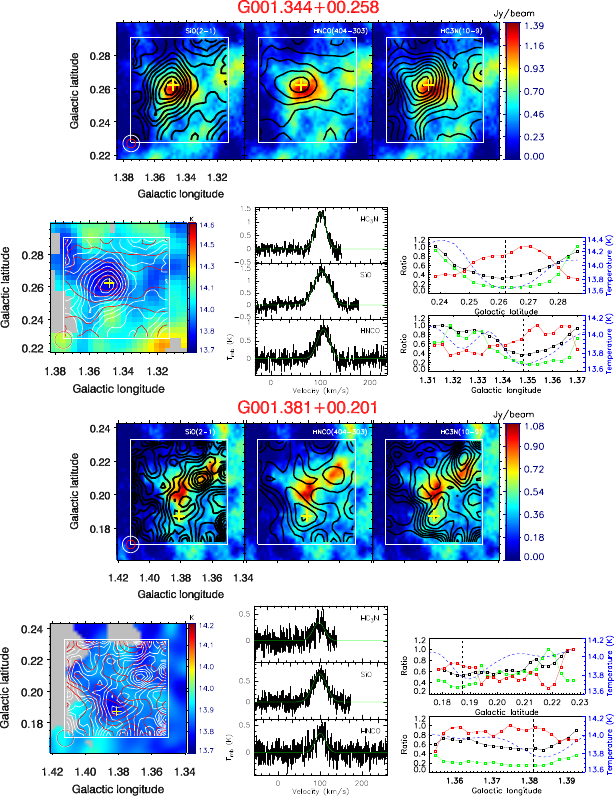}
    \caption{Same as Fig.~\ref{fig:Figure_A1}, but for G001.344+00.258 (contours start from 3$\sigma$ with steps of 3$\sigma$, where 3$\sigma$ is 2.19, 2.10 and 1.77 K km s$^{-1}$, respectively) and G001.381+00.201 (contours start from 3$\sigma$ with steps of 2$\sigma$, $\sigma$ and $\sigma$ for SiO 2--1, HNCO 4$_{04}$--3$_{03}$ and HC$_{3}$N 10--9, respectively. The corresponding values are 1.02, 0.58 and 0.45 K km s$^{-1}$).}
    \label{fig:Figure_A9}
\end{figure}

\begin{figure}
  \centering
	\includegraphics[width = 0.85\linewidth]{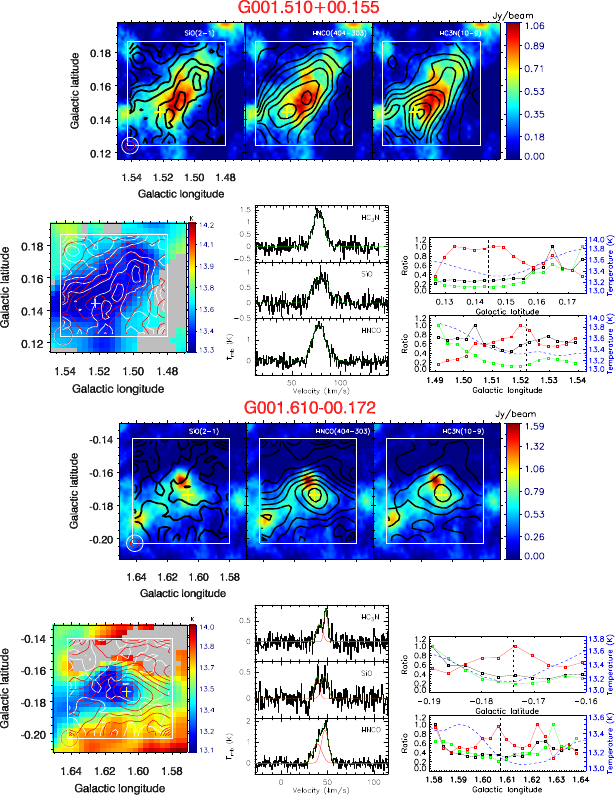}
    \caption{Same as Fig.~\ref{fig:Figure_A1}, but for G001.510+00.155 (contours start from 3$\sigma$ with steps of 3$\sigma$, where 3$\sigma$ is 1.68, 1.56 and 1.11 K km s$^{-1}$, respectively) and G001.610$-$00.172 (contours start from 3$\sigma$ with steps of 3$\sigma$, where 3$\sigma$ is 1.11, 1.11 and 0.78 K km s$^{-1}$, respectively).}
    \label{fig:Figure_A10}
\end{figure}

\begin{figure}
  \centering
	\includegraphics[width = 0.85\linewidth]{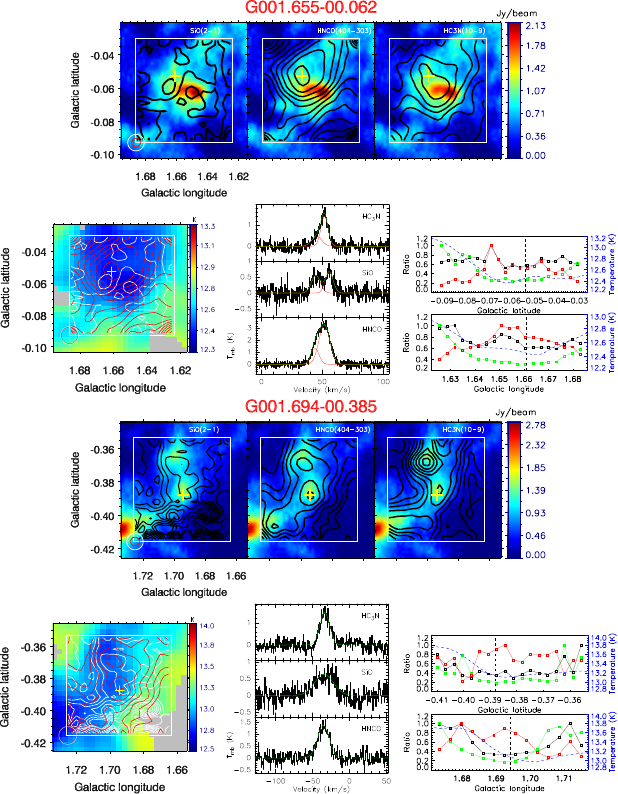}
    \caption{Same as Fig.~\ref{fig:Figure_A1}, but for G001.655$-$00.062 (contours start from 3$\sigma$ with steps of 3$\sigma$, where 3$\sigma$ is 1.26, 1.38 and 1.23 K km s$^{-1}$, respectively) and G001.694$-$00.385 (contours start from 3$\sigma$ with steps of 3$\sigma$, where 3$\sigma$ is 1.86, 2.01 and 1.71 K km s$^{-1}$, respectively).}
    \label{fig:Figure_A11}
\end{figure}

\begin{figure}
  \centering
	\includegraphics[width = 0.85\linewidth]{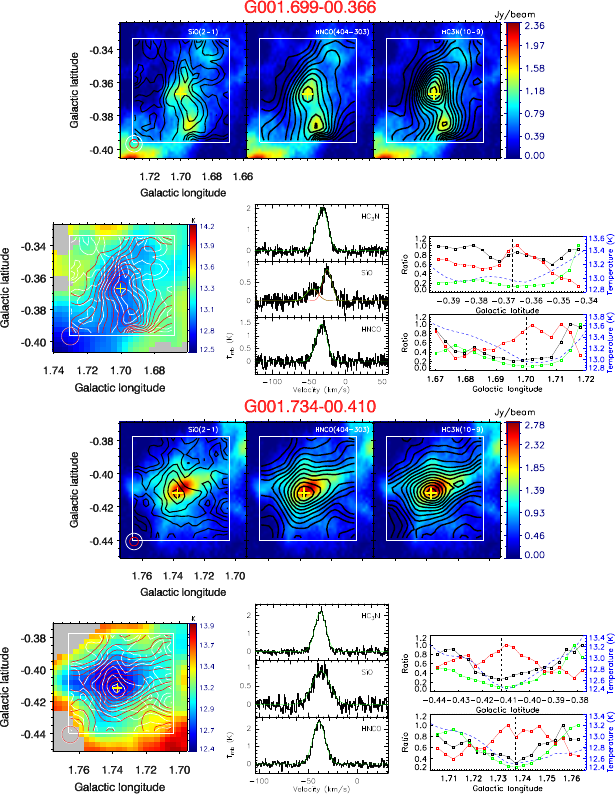}
    \caption{Same as Fig.~\ref{fig:Figure_A1}, but for G001.699$-$00.366 (contours start from 3$\sigma$ with steps of 3$\sigma$, where 3$\sigma$ is 1.74, 1.50 and 1.41 K km s$^{-1}$, respectively) and G001.734$-$00.410 (contours start from 3$\sigma$ with steps of 3$\sigma$, where 3$\sigma$ is 1.53, 1.35 and 1.08 K km s$^{-1}$, respectively).}
    \label{fig:Figure_A12}
\end{figure}

\begin{figure}
  \centering
	\includegraphics[width = 0.85\linewidth]{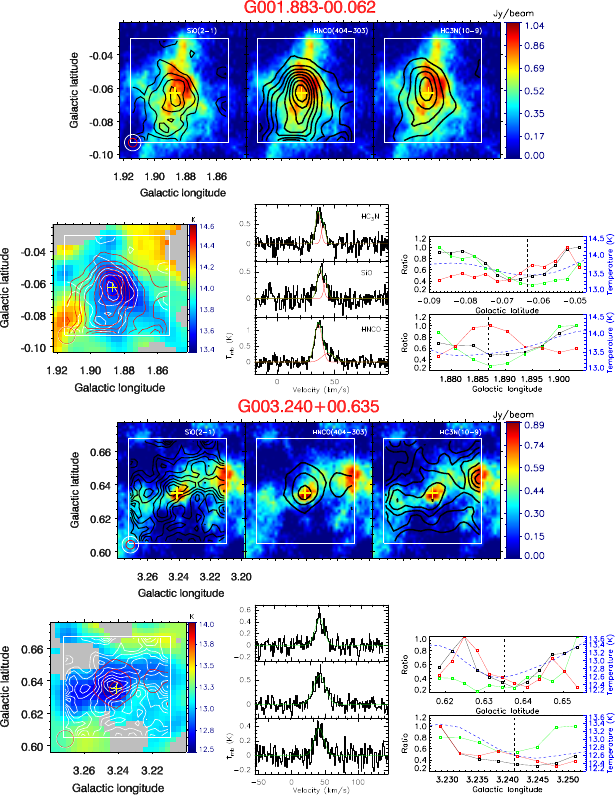}
    \caption{Same as Fig.~\ref{fig:Figure_A1}, but for G001.883$-$00.062 (contours start from 3$\sigma$ with steps of 2$\sigma$, where 2$\sigma$ is 0.72, 0.78 and 0.74 K km s$^{-1}$, respectively) and G003.240+00.635 (contours start from 3$\sigma$ with steps of 2$\sigma$, where 2$\sigma$ is 1.10, 0.98 and 0.82 K km s$^{-1}$, respectively).}
    \label{fig:Figure_A13}
\end{figure}

\begin{figure}
  \centering
	\includegraphics[width = 0.85\linewidth]{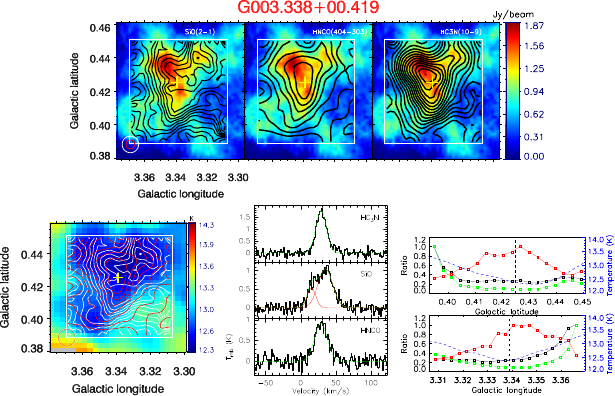}
    \caption{Same as Fig.~\ref{fig:Figure_A1}, but for G003.338+00.419 (contours start from 3$\sigma$ with steps of 2$\sigma$, where 2$\sigma$ is 1.40, 1.12 and 1.04 K km s$^{-1}$, respectively) and G359.445$-$00.054 (contours start from 3$\sigma$ with steps of 2$\sigma$, where 2$\sigma$ is 0.84, 1.00 and 0.80 K km s$^{-1}$, respectively).}
    \label{fig:Figure_A14}
\end{figure}

\begin{figure}
  \centering
	\includegraphics[width = 0.85\linewidth]{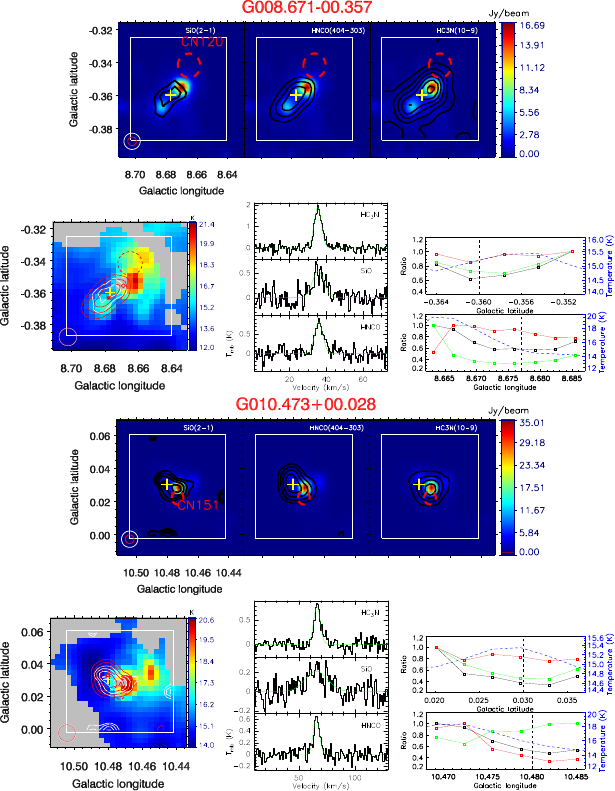}
    \caption{Same as Fig.~\ref{fig:Figure_A1}, but for  and G008.671$-$00.357 (contours start from 3$\sigma$ with steps of $\sigma$, $\sigma$ and 3$\sigma$ for SiO 2--1, HNCO 4$_{04}$--3$_{03}$ and HC$_{3}$N 10--9, respectively. The $\sigma$ values are 0.39, 0.39 and 0.25 K km s$^{-1}$) and G010.473+00.028 (contours start from 3$\sigma$ with steps of $\sigma$, $\sigma$ and 3$\sigma$ for SiO 2--1, HNCO 4$_{04}$--3$_{03}$ and HC$_{3}$N 10--9, respectively. The $\sigma$ values are 0.30, 0.31 and 0.34 K km s$^{-1}$).}
    \label{fig:Figure_A15}
\end{figure}

\begin{figure}
  \centering
	\includegraphics[width = 0.85\linewidth]{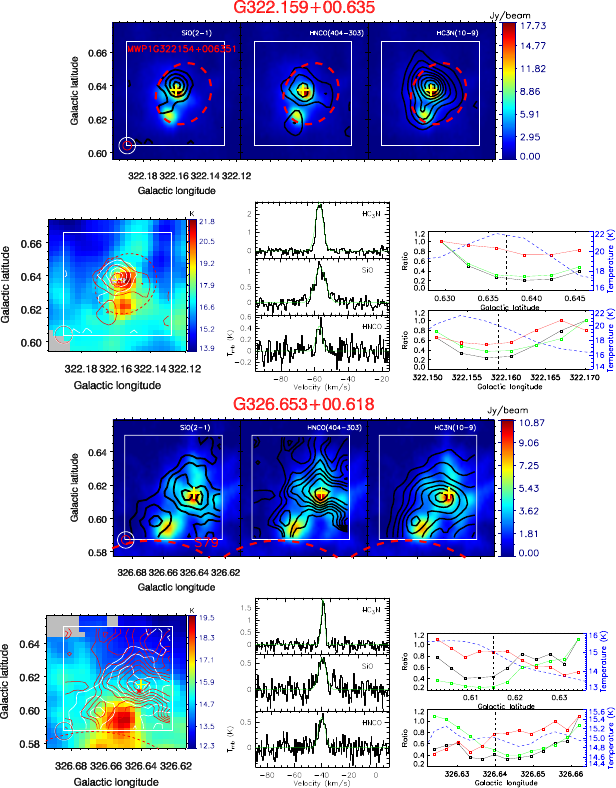}
    \caption{Same as Fig.~\ref{fig:Figure_A1}, but for  and G322.159+00.635 (contours start from 3$\sigma$ with steps of 3$\sigma$, $\sigma$ and 3$\sigma$ for SiO 2--1, HNCO 4$_{04}$--3$_{03}$ and HC$_{3}$N 10--9, respectively. The $\sigma$ values are 0.29, 0.19 and 0.24 K km s$^{-1}$) and G326.653+00.618 (contours start from 3$\sigma$ with steps of 3$\sigma$, $\sigma$ and 3$\sigma$ for SiO 2--1, HNCO 4$_{04}$--3$_{03}$ and HC$_{3}$N 10--9, respectively. The $\sigma$ values are 0.26, 0.21 and 0.22 K km s$^{-1}$).}
    \label{fig:Figure_A16}
\end{figure}

\begin{figure}
  \centering
	\includegraphics[width = 0.85\linewidth]{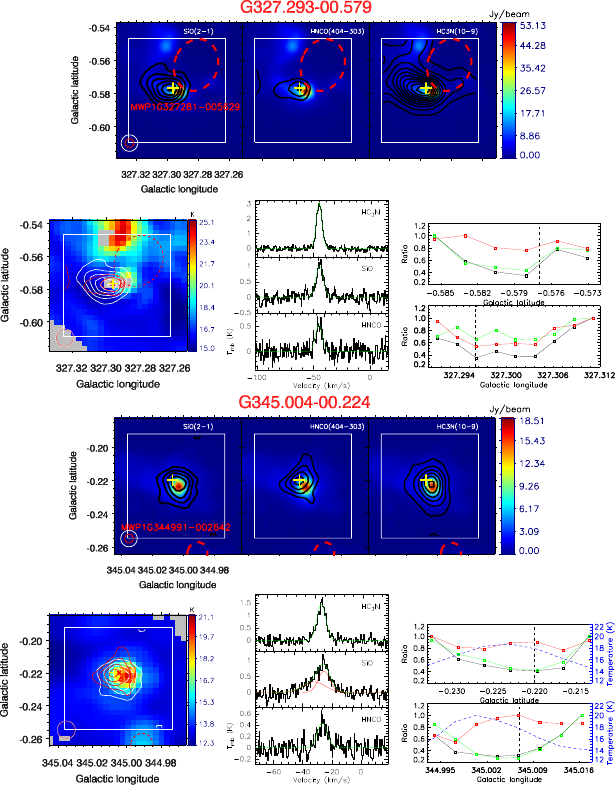}
    \caption{Same as Fig.~\ref{fig:Figure_A1}, but for  and G327.293$-$00.579 (contours start from 3$\sigma$ with steps of 3$\sigma$, $\sigma$ and 3$\sigma$ for SiO 2--1, HNCO 4$_{04}$--3$_{03}$ and HC$_{3}$N 10--9, respectively. The $\sigma$ values are 0.24, 0.22 and 0.29 K km s$^{-1}$) and G345.004$-$00.224 (contours start from 3$\sigma$ with steps of 2$\sigma$, $\sigma$ and 3$\sigma$ for SiO 2--1, HNCO 4$_{04}$--3$_{03}$ and HC$_{3}$N 10--9, respectively. The $\sigma$ values are 0.43, 0.26 and 0.26 K km s$^{-1}$).}
    \label{fig:Figure_A17}
\end{figure}

\begin{figure}
  \centering
	\includegraphics[width = 0.85\linewidth]{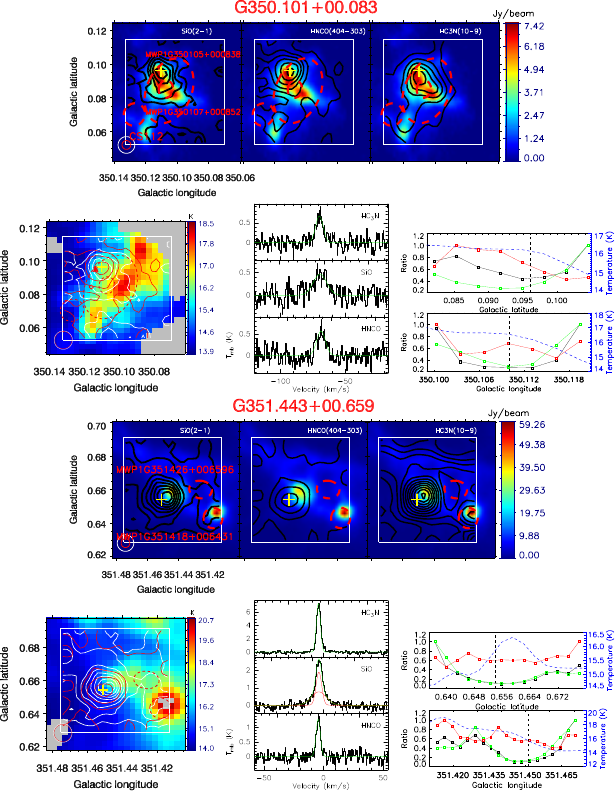}
    \caption{Same as Fig.~\ref{fig:Figure_A1}, but for  and G350.101+00.083 (contours start from 3$\sigma$ with steps of 2$\sigma$, 2$\sigma$ and 3$\sigma$ for SiO 2--1, HNCO 4$_{04}$--3$_{03}$ and HC$_{3}$N 10--9, respectively. The $\sigma$ values are 0.23, 0.20 and 0.21 K km s$^{-1}$) and G351.443+00.659 (contours start from 3$\sigma$ with steps of 6$\sigma$, 3$\sigma$ and 5$\sigma$ for SiO 2--1, HNCO 4$_{04}$--3$_{03}$ and HC$_{3}$N 10--9, respectively. The $\sigma$ values are 0.23, 0.23 and 0.24 K km s$^{-1}$).}
    \label{fig:Figure_A18}
\end{figure}

\begin{figure}
  \centering
	\includegraphics[width = 0.85\linewidth]{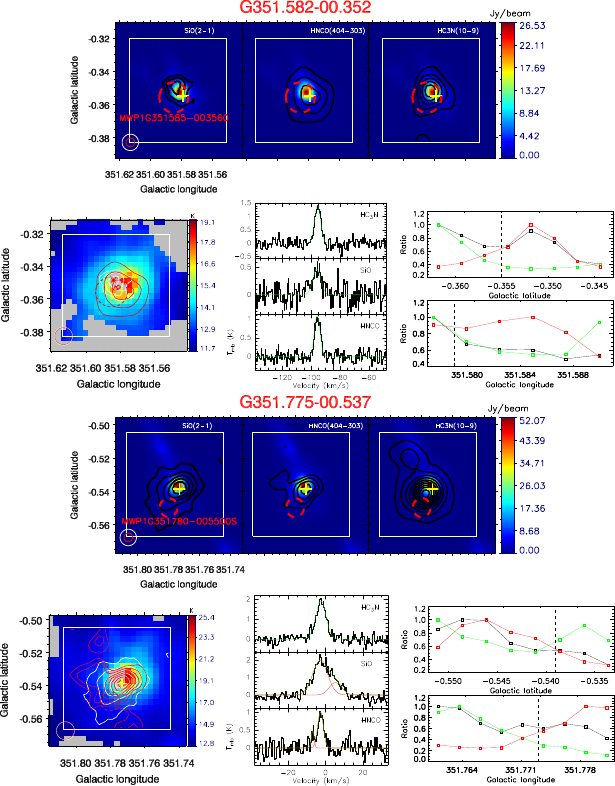}
    \caption{Same as Fig.~\ref{fig:Figure_A1}, but for  and G351.582$-$00.352 (contours start from 3$\sigma$ with steps of$\sigma$, 3$\sigma$ and 3$\sigma$ for SiO 2--1, HNCO 4$_{04}$--3$_{03}$ and HC$_{3}$N 10--9, respectively. The $\sigma$ values are 0.27, 0.24 and 0.27 K km s$^{-1}$) and G351.775$-$00.537 (contours start from 3$\sigma$ with steps of 3$\sigma$, $\sigma$ and 3$\sigma$ for SiO 2--1, HNCO 4$_{04}$--3$_{03}$ and HC$_{3}$N 10--9, respectively. The $\sigma$ values are 0.78, 0.35 and 0.38 K km s$^{-1}$).}
    \label{fig:Figure_A19}
\end{figure}

\begin{figure}
  \centering
	\includegraphics[width = 0.85\linewidth]{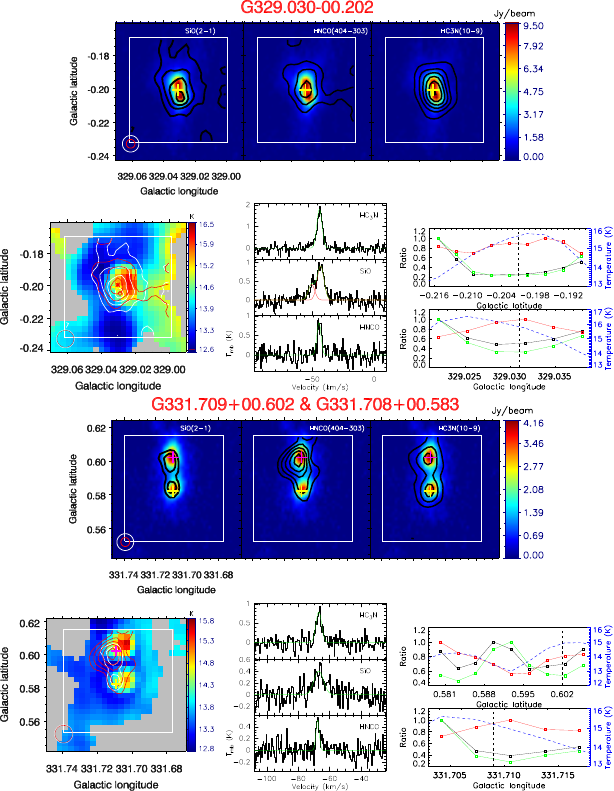}
    \caption{Same as Fig.~\ref{fig:Figure_A1}, but for  and G329.030$-$00.202 (contours start from 3$\sigma$ with steps of 3$\sigma$, 2$\sigma$ and 3$\sigma$ for SiO 2--1, HNCO 4$_{04}$--3$_{03}$ and HC$_{3}$N 10--9, respectively. The $\sigma$ values are 0.28, 0.21 and 0.28 K km s$^{-1}$) and G331.709+00.602 \& G331.708+00.583 (contours start from 3$\sigma$ with steps of $\sigma$, $\sigma$ and 3$\sigma$ for SiO 2--1, HNCO 4$_{04}$--3$_{03}$ and HC$_{3}$N 10--9, respectively. The $\sigma$ values are 0.32, 0.17 and 0.20 K km s$^{-1}$).}
    \label{fig:Figure_A20}
\end{figure}

\begin{figure}
  \centering
	\includegraphics[width = 0.85\linewidth]{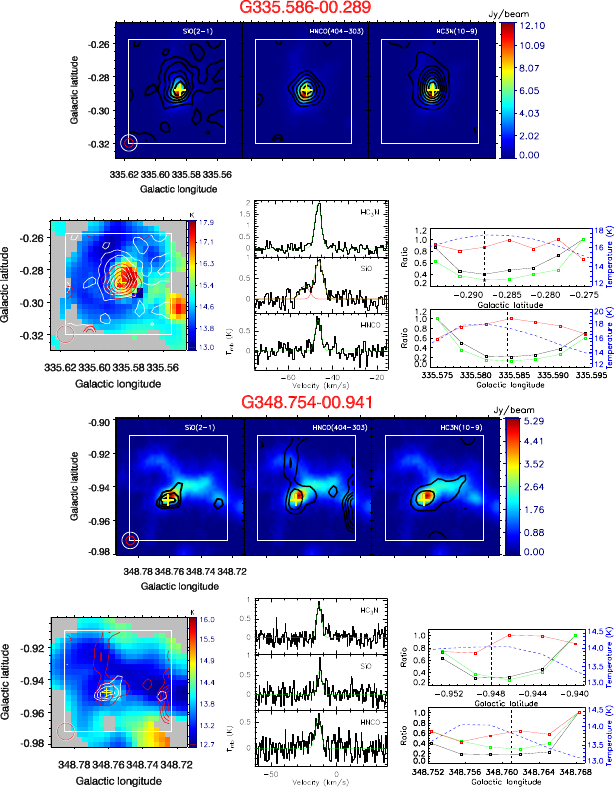}
    \caption{Same as Fig.~\ref{fig:Figure_A1}, but for  and G335.586$-$00.289 (contours start from 3$\sigma$ with steps of 3$\sigma$, $\sigma$ and 3$\sigma$ for SiO 2--1, HNCO 4$_{04}$--3$_{03}$ and HC$_{3}$N 10--9, respectively. The $\sigma$ values are 0.25, 0.17 and 0.16 K km s$^{-1}$) and G348.754$-$00.941 (contours start from 3$\sigma$ with steps of $\sigma$, $\sigma$ and 3$\sigma$ for SiO 2--1, HNCO 4$_{04}$--3$_{03}$ and HC$_{3}$N 10--9, respectively. The $\sigma$ values are 0.29, 0.26 and 0.22 K km s$^{-1}$).}
    \label{fig:Figure_A21}
\end{figure}

\begin{figure}
  \centering
	\includegraphics[width = 0.85\linewidth]{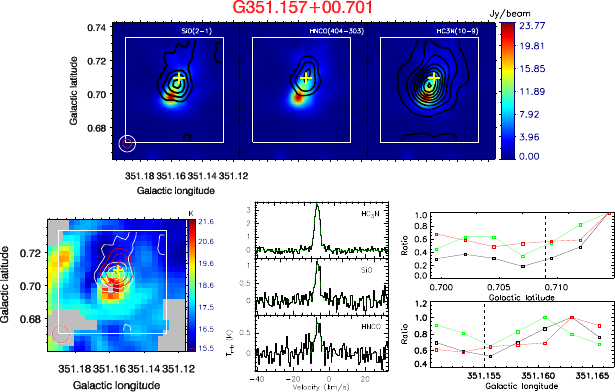}
    \caption{Same as Fig.~\ref{fig:Figure_A1}, but for G351.157+00.701 (contours start from 3$\sigma$ with steps of 3$\sigma$, $\sigma$ and 4$\sigma$ for SiO 2--1, HNCO 4$_{04}$--3$_{03}$ and HC$_{3}$N 10--9, respectively. The $\sigma$ values are 0.24, 0.26 and 0.26 K km s$^{-1}$).}
    \label{fig:Figure_A22}
\end{figure}

Below we give comments on individual sources. Figs.~\ref{fig:Figure_A1}--\ref{fig:Figure_A14}, Figs.~\ref{fig:Figure_A15}--\ref{fig:Figure_A19}, and Figs.~\ref{fig:Figure_A20}--\ref{fig:Figure_A22} show the CMZ, Bubble, and NMSFR samples, respectively. It should be understood that in the following the four cardinal directions, north, south, east and west, are meant with respect to the Galactic coordinates ($l^{\rm II}$, $b^{\rm II}$), and not with respect to Right Ascension and Declination.

\emph{G359.977$-$00.072} and \emph{G359.895$-$00.069}. G359.977$-$00.072 is well known under the name +50 km s$^{-1}$ cloud. For the spectral line maps of G359.977$-$00.072 and G359.895$-$00.069 shown in Fig.~\ref{fig:Figure_A1}, there are single cores towards the lower dust temperature region, but these have a slight offset from the maximum of the dust continuum emission. The SiO spectra in G359.977$-$00.072 are broader than those of HNCO and HC$_3$N. The abundance ratio of ${N_{\rm HC_{3}N}/N_{\rm SiO}}$ in these two sources shows a reverse trend to that of the dust temperature.

\emph{G359.868$-$00.085}. From the top panel of Fig.~\ref{fig:Figure_A2}, it can be seen that there is no obvious molecular emission core at the centrally located brightest clump, which is associated with water masers \citep{2002A&A...391..967S}. The SiO and HC$_{3}$N molecular emissions clearly show two cores on each side of the central clump. The abundance ratios of ${N_{\rm HNCO}/N_{\rm HC_{3}N}}$ and ${N_{\rm HC_{3}N}/N_{\rm SiO}}$ both show a similar trend as the dust temperature.

\emph{G359.565$-$00.161}. Molecular line emissions show similar morphologies in the lower panels of Fig.~\ref{fig:Figure_A2}. Weak molecular line emission is found towards the north-eastern high dust temperature region. All molecular line peaks have an offset from the maximum of the 870 $\mu$m dust continuum emission. The SiO 2--1 data are noisy. Near the yellow cross position, ${N_{\rm HC_{3}N}/N_{\rm SiO}}$ reaches maxima, while  ${N_{\rm HNCO}/N_{\rm SiO}}$ and ${N_{\rm HNCO}/N_{\rm HC_{3}N}}$ and the dust temperature are low.

\emph{G359.453$-$00.112}. As shown in the top panel of Fig.~\ref{fig:Figure_A3}, SiO, HNCO and HC$_{3}$N exhibit a single core elongated from north-west to south-east in the velocity-integrated intensity maps, consistent with the cometary shape of the low temperature dust continuum emission. In the north-western high temperature corner there is the H{\sc\,ii} region G359.4$-$0.1 (i.e. Sgr C) \citep{1997ApJ...488..224K}. An additional velocity component can be distinguished at the eastern side of the main component in all spectra. The ${N_{\rm HNCO}/N_{\rm HC_{3}N}}$ abundance ratio (green lines) is closely related to changes in the dust temperature.

\emph{G359.445$-$00.054}. G359.445$-$00.054 is part of the Sgr C Filament \citep{2011ApJ...741...95L}. We find one isolated core in HNCO and HC$_3$N and two compact cores in the SiO line map. Moreover, the cores are slightly offset from the maximum of dust continuum emission as it is shown in the lower panels of Fig.~\ref{fig:Figure_A3}. In the case of the SiO 2--1 emission, the structure is complicated, possibly because of low signal-to-noise ratios. There are two obviously saturated areas in the Hi-GAL data (visible in gray in the image shown in the lower left panel of G359.445$-$00.054), where the bubble MWP1G359450$-$000200S \citep{2012MNRAS.424.2442S} and the H{\sc\,ii} region Gal 359.43$-$00.09 \citep{2002ApJ...566..880G} are found to be located on either side of the low temperature filamentary structure of the dust. In addition, the velocity range appears to be widest for the HNCO 4$_{04}$--3$_{03}$ spectrum. With respect to temperature variations, the changes in abundance ratios are similar to source G001.734$-$00.410. Near the yellow cross position, ${N_{\rm HNCO}/N_{\rm HC_{3}N}}$ and dust temperature reach minima, while  ${N_{\rm HNCO}/N_{\rm SiO}}$ and ${N_{\rm HC_{3}N}/N_{\rm SiO}}$ are high.

\emph{G000.067$-$00.077}. As seen in the top panels of Fig.~\ref{fig:Figure_A4}, the emissions of the HNCO 4$_{04}$--3$_{03}$, SiO 2--1 and HC$_{3}$N 10--9 lines show a similar morphology with a centrally condensed structure. The emission peaks coincide with the 870 $\mu$m continuum emission. At this position, the dust temperature is relatively low. In addition, there are two higher dust temperature regions, in the north-east and south-west of the 870 $\mu$m emission peak. The HC$_{3}$N spectra have a simple Gaussian shape. However, the HNCO and SiO lines towards the yellow cross position show non-Gaussian wing emission, indicative of outflows/shocks. HNCO and SiO are each fitted with two Gaussians. ${N_{\rm HNCO}/N_{\rm SiO}}$ and ${N_{\rm HNCO}/N_{\rm HC_{3}N}}$ show a decreasing trend towards the central lower dust temperature region.

\emph{G000.104$-$00.080}. The integrated intensity maps of the HNCO 4$_{04}$--3$_{03}$, SiO 2--1 and HC$_{3}$N 10--9 lines show a similar morphology, and coincide well with the ATLASGAL 870 $\mu$m emission where the dust temperature is relatively low. As in G000.067-00.077, two higher dust temperature regions are located in the north-east and south-west of the source. The HNCO, SiO and HC$_{3}$N lines all show an asymmetry and possess wing emission. The ${N_{\rm HNCO}/N_{\rm HC_{3}N}}$ and dust temperature curves are similar towards the central region along Galactic longitude direction shown in the last panel of Fig.~\ref{fig:Figure_A4}.

\emph{G000.106$-$00.001}. All of the molecular line emissions have an offset of $\sim$30$^{\prime\prime}$ from the peak of the ATLASGAL 870 $\mu$m continuum emission and show a large gradient towards the north-western higher dust temperature region, which might be caused by external pressure from the arched filament H{\sc\,ii} complex \citep{2001AJ....121.2681L}. The SiO integrated intensity (white contours in the lower left panel of the upper half of Fig.~\ref{fig:Figure_A5}) and cold dust temperature structures have a similar morphology. HNCO and HC$_{3}$N show a line wing on the redshifted side, and are each fitted with two Gaussians. In the central part of G000.106$-$00.001, ${ N_{\rm HNCO}/N_{\rm HC_{3}N}}$ and dust temperature curves are similar.

\emph{G000.110$+$00.148}. From the lower panels of Fig.~\ref{fig:Figure_A5}, HC$_{3}$N and HNCO emission peaks towards the lower dust temperature region, but is offset to the north-east of the maximum of the 870 $\mu$m dust continuum emission ($\sim$36$^{\prime\prime}$). The structure is complex in the case of SiO 2--1, perhaps because of low signal-to-noise ratios. A high dust temperature region appears in the lower left corner. The HNCO 4$_{04}$--3$_{03}$ line shows a single Gaussian shape, while SiO 2--1 also shows wing emission.

\emph{G000.314$-$00.100}. The upper panels of Fig.~\ref{fig:Figure_A6} demonstrate that the HNCO 4$_{04}$--3$_{03}$ morphology shows a centrally condensed structure, with the emission peak coinciding with the 870 $\mu$m emission, which shows low dust temperatures. The SiO 2--1 and HC$_{3}$N 10--9 emissions also have a centrally condensed structure and show similar distributions, but the emission peaks are offset ($\sim$20$^{\prime\prime}$) from those of the HNCO 4$_{04}$--3$_{03}$ and 870 $\mu$m emission. The spectral profiles have low signal-to-noise ratios in the HNCO 4$_{04}$--3$_{03}$ and SiO 2-1 lines. HNCO, SiO and HC$_{3}$N are each fitted with two Gaussians. Abundance ratios of ${ N_{\rm HC_{3}N}/N_{\rm SiO}}$ show an increasing trend with respect to ${N_{\rm HNCO}/N_{\rm SiO}}$ and ${N_{\rm HNCO}/N_{\rm HC_{3}N}}$ towards the central low dust temperature region.

\emph{G000.497$+$00.021} and \emph{G000.645$+$00.027}. From the lower panels of Fig.~\ref{fig:Figure_A6} and the upper panels of Fig.~\ref{fig:Figure_A7}, the SiO emission shows a complex morphology in the observed area, while the other two molecular emission lines reveal compact cores in the integrated intensity map. In addition, extended SiO emission is seen particularly around the central core of HNCO emission in source G000.497$+$00.021. As in G000.314$-$00.100, the dust emission regions are cold, and the abundance ratio of ${N_{\rm HC_{3}N}/N_{\rm SiO}}$ shows an increasing trend towards the central region.

\emph{G000.892$+$00.143}. The lower panels of Fig.~\ref{fig:Figure_A7} show a pronounced core elongated from the south-east to the north-west in the three studied molecular lines. Compared to the south-eastern 870 $\mu$m emission peak, the north-western one shows a higher temperature and coincides with the local HNCO 4$_{04}$--3$_{03}$, SiO 2--1 and HC$_{3}$N 10--9 peak positions. The SiO and HC$_{3}$N spectra exhibit asymmetric profiles. Abundance ratios of ${N_{\rm HNCO}/N_{\rm SiO}}$ and ${N_{\rm HNCO}/N_{\rm HC_{3}N}}$ decrease along the Galactic longitude direction towards the central hot dust core, while ${N_{\rm HC_{3}N}/N_{\rm SiO}}$ does the opposite.

\emph{G000.908$+$00.116}. Towards G000.908$+$00.116, shown in the upper panels of Fig.~\ref{fig:Figure_A8}, HNCO 4$_{04}$--3$_{03}$ shows a dense clump in the central map, with emission peaking at a $\sim$29$^{\prime\prime}$ offset from the maximum of the dust continuum emission. However, the integrated intensity map of SiO 2--1 shows a more complex structure. HC$_{3}$N exhibits a simple morphology with a ridge elongated from the south-east to the north-west. The abundance ratios of ${N_{\rm HNCO}/N_{\rm SiO}}$, ${N_{\rm HNCO}/N_{\rm HC_{3}N}}$ and ${ N_{\rm HC_{3}N}/N_{\rm SiO}}$ show no obvious trends.

\emph{G001.226$+$00.095}. The HC$_{3}$N emission peak has an offset from the continuum and SiO emission peak by about a beam ($\sim$38$^{\prime\prime}$, Fig.~\ref{fig:Figure_A8}). In the dust temperature map the arc-like low dust temperature region coincides with the HNCO emission. The SiO 2-1 line shows obvious redshifts of 8.2 km s$^{-1}$ and 4.1 km s$^{-1}$ relative to the $v_{\rm LSR}$ of the HNCO 4$_{04}$--3$_{03}$ and HC$_{3}$N 10--9 lines. Abundance ratios of ${N_{\rm HC_{3}N}/N_{\rm SiO}}$  show an increasing and ${N_{\rm HNCO}/N_{\rm SiO}}$ and ${N_{\rm HNCO}/N_{\rm HC_{3}N}}$ a decreasing trend towards the yellow cross position, respectively.

\emph{G001.344$+$00.258}. A similarly centrally condensed structure is seen in HNCO, SiO and HC$_{3}$N emission in the top panels of Fig.~\ref{fig:Figure_A9}. All molecular line emissions peak towards the low temperature maximum of the 870 $\mu$m dust continuum emission. The spectra of HNCO 4$_{04}$--3$_{03}$, SiO 2--1 and HC$_{3}$N 10--9 show a single velocity component. The trend of ${N_{\rm HC_{3}N}/N_{\rm SiO}}$  is opposite to that of the dust emission temperature.

\emph{G001.381$+$00.201}. From the lower panels of Fig.~\ref{fig:Figure_A9}, it can be seen that both the integrated intensity maps of HNCO and HC$_{3}$N show two compact cores. In addition, the upper compact cores of SiO, HNCO and HC$_{3}$N do not coincide with each other. Moreover, compared to the south-eastern core of HNCO and HC$_{3}$N, SiO emission shows more extended and complex structure. Abundance ratios and dust temperature appear not to be clearly correlated.

\emph{G001.510$+$00.155}. In the top panels of Fig.~\ref{fig:Figure_A10}, the SiO, HNCO and HC$_{3}$N emissions clearly show three, two, and one compact core(s), respectively. The HC$_{3}$N core is elongated from the south-east to the north-west, and coincides well with the low dust temperature region. The SiO 2--1, HNCO 4$_{04}$--3$_{03}$ and HC$_{3}$N 10--9 lines can be fitted by a single Gaussian component. The abundance ratio curve of ${N_{\rm HC_{3}N}/N_{\rm SiO}}$ shows an obvious plateau near the yellow cross position along the Galactic latitude direction.

\emph{G001.610$-$00.172} and \emph{G001.655$-$00.062}. From Figs.~\ref{fig:Figure_A10} and \ref{fig:Figure_A11}, it can be seen that both sources present condensed structures in the HNCO 4$_{04}$--3$_{03}$ and HC$_{3}$N 10--9 emissions in the low dust temperature region. The peaks of HNCO 4$_{04}$--3$_{03}$ emission show offsets of $\sim$40$^{\prime\prime}$ and $\sim$55$^{\prime\prime}$ from the strongest 870 $\mu$m dust emission in G001.610$-$00.172 and G001.655$-$00.062, respectively. The SiO 2--1 emissions for these two sources all show complex structures. In addition, the SiO 2--1 line of G001.655$-$00.062 exhibits a double peak, while HNCO and HC$_3$N show a single velocity component at the dip of SiO 2--1. The abundance ratio curve of ${N_{\rm HNCO}/N_{\rm SiO}}$ and ${N_{\rm HC_{3}N}/N_{\rm SiO}}$ show a reverse trend to that of the dust emission temperature.

\emph{G001.694$-$00.385}. A north-south elongated and compact structure is shown in the HNCO 4$_{04}$--3$_{03}$ and HC$_{3}$N 10--9 emissions, which contains three cores (Fig.~\ref{fig:Figure_A11}). These coincide with the region of low temperature 870 $\mu$m dust emission. The SiO emission shows an extended complex structure also associated with the region of higher dust temperature in the south-east. The spectrum of SiO 2-1 is noisy. Near the yellow cross position, ${N_{\rm HC_{3}N}/N_{\rm SiO}}$ reaches maxima, while  ${N_{\rm HNCO}/N_{\rm SiO}}$ and ${N_{\rm HNCO}/N_{\rm HC_{3}N}}$ and the dust temperature are low.

\emph{G001.699$-$00.366}. Two compact cores can be seen in the integrated intensity maps of HNCO 4$_{04}$--3$_{03}$ and HC$_3$N 10--9 in the top panels of Fig.~\ref{fig:Figure_A12}. This is consistent with the distributions of low temperature dust emission. There are also two cores in the SiO 2--1 line map, but here the emission peaks are offset from the dust continuum emission peaks ($\sim$35$^{\prime\prime}$). Both of the HNCO and HC$_{3}$N lines show single Gaussian velocity component, but SiO shows two velocity components on either side of the systemic velocity of the source. ${N_{\rm HNCO}/N_{\rm HC_{3}N}}$ follows the trend of the dust temperature.

\emph{G001.734$-$00.410} and \emph{G001.883$-$00.062}. In Figs.~\ref{fig:Figure_A12} and Fig.~\ref{fig:Figure_A13}, our three main molecular tracers show similar morphologies and a centrally condensed structure. Moreover, obvious offsets of $\sim$15$^{\prime\prime}$ and $\sim$30$^{\prime\prime}$ between the molecular line emission peaks and the lower temperature 870 $\mu$m dust continuum emission peaks are found in G001.734$-$00.410 and G001.883$-$00.062, respectively. All spectra show one Gaussian component. Near the yellow cross position, ${N_{\rm HC_{3}N}/N_{\rm SiO}}$ reaches maxima, while  ${N_{\rm HNCO}/N_{\rm SiO}}$ and ${N_{\rm HNCO}/N_{\rm HC_{3}N}}$ and the dust temperature are low.

\emph{G003.240$+$00.635}. Extended emission can be seen in the integrated intensity maps of the SiO and HC$_{3}$N lines in the lower panels of Fig.~\ref{fig:Figure_A13}. In addition, HNCO has the most compact emission morphology, and coincides with the 870 $\mu$m emission peak, which shows the lowest dust temperature. The HNCO and SiO lines are Gaussian towards the yellow cross position. All three plotted abundance ratios follow the trend determined by the dust temperature.

\emph{G003.338$+$00.419}. The top panels of Fig.~\ref{fig:Figure_A14} show a compact core, which is seen in the 870 $\mu$m continuum and the HNCO and HC$_3$N emission. SiO shows an overall similar morphology as HC$_3$N but its distribution on smaller scales appear to be more complex. The SiO 2--1 line shows a skewed profile with a blue-shifted peak that is weaker than the red-shifted one. Near the yellow cross position, ${N_{\rm HC_{3}N}/N_{\rm SiO}}$ reaches maxima, while  ${N_{\rm HNCO}/N_{\rm SiO}}$ and ${N_{\rm HNCO}/N_{\rm HC_{3}N}}$ and the dust temperature are low.

\emph{G008.671$-$00.357}. Fig.~\ref{fig:Figure_A15} shows a bubble, CN129, nearby, in the north-west of the presented maps of integrated line emission. This corresponds to a hot dust core, which is close to a class II CH$_{3}$OH maser (red diamond shown in the lower left panel of Fig.~\ref{fig:Figure_A15}) \citep{1993MNRAS.260..425C}. The H$_{2}$O maser (red asterisk shown in the same panel) lies in between the two dust emission cores, which indicates a possible interaction. The three molecular lines all show a morphology extending from north-west to south-east. Two peaks of HC$_{3}$N emission are seen towards the dust continuum emission. The trend of ${N_{\rm HNCO}/N_{\rm SiO}}$ is opposite to that of the dust temperature along the Galactic latitude direction. Inflow and outflow activities are simultaneously detected in this source from analysing molecular line profiles of HCO$^{+}$ 1--0 (see Sect.~\ref{sec:outflow_inflow} and Fig.~\ref{fig:Figure_32} for details).

\emph{G010.473$+$00.028}. From the lower panels of Fig.~\ref{fig:Figure_A15}, it can be seen that the molecular line emissions exhibit centrally condensed structure. SiO and HNCO emissions peak towards the faint dust continuum emission core, which contains OH \citep{1998MNRAS.297..215C} and class II CH$_{3}$OH masers \citep{2005A&A...432..737P,1993MNRAS.260..425C}, while HC$_3$N emission peaks towards the maximum of hot dust continuum emission. There is a bubble, CN151, close to the peak of the HC$_{3}$N emission, and an OH maser between them. Because of the low signal-to-noise ratio of the SiO 2-1 data, the integrated intensity is calculated in the wide velocity range between 55.54 to 84.44 km s$^{-1}$. ${N_{\rm HNCO}/N_{\rm HC_{3}N}}$ increases with increasing Galactic longitude, while $N_{\rm HC_{3}N}$/$N_{\rm SiO}$ follows an opposite trend.

\emph{G322.159$+$00.635}. It is identified as an IRDC of SDC 322.149+0.639 by \citet{2009A&A...505..405P}. As Fig.~\ref{fig:Figure_A16} shows, the map consists of two hot molecular cores embedded in bubble MWP1G322154+006351 identified by \citet{2012MNRAS.424.2442S}. The north-western molecular core contains OH \citep{1987AuJPh..40..215C} and CH$_{3}$OH \citep{1993MNRAS.260..425C} masers, while the south-eastern molecular core is close to another OH maser, which lies in between two dust cores. All three molecular lines show a single velocity component. Abundance ratios of ${N_{\rm HNCO}/N_{\rm SiO}}$ and ${N_{\rm HNCO}/N_{\rm HC_{3}N}}$ show similarly varying tendencies.

\emph{G326.653$+$00.618}. In the lower panels of Fig.~\ref{fig:Figure_A16}, there is an obvious dust temperature gradient decreasing from the south-east to the north-west. Furthermore, there is a bubble, S79, to the south-east. The peaks of molecular line emission have a significant offset of $\sim$30$^{\prime\prime}$ from the peak of the south-eastern hot dust clump, which might be caused by the shock between the bubble and the molecular clouds. The SiO spectrum from the maximum of dust continuum emission presents an additional velocity component at $\sim-$30 km s$^{-1}$. All these lines present two cores located on two sides of the brighest dust peak also hosting an H$_{2}$O maser \citep{1980AuJPh..33..139B}. Abundance ratios and dust temperature appear not to be clearly correlated.

\emph{G327.293$-$00.579}. G327.293$-$00.579 is identified as an IRDC, named SDC 327.306$-$0.566 by \citet{2009A&A...505..405P}. As Fig.~\ref{fig:Figure_A17} shows, the spectral line maps reveal an isolated core. The peak molecular line emissions are slightly offset by $\sim$20$^{\prime\prime}$ from the maximum of the dust continuum emission, which shows a 12-GHz CH$_{3}$OH maser \citep{1995MNRAS.274.1126C} close to the bubble, MWP1G327281$-$005629. Because of many saturated pixels in the dust temperature map, we just show integrated intensity ratios of ${I_{\rm HNCO}/I_{\rm HC_{3}N}}$, ${I_{\rm HNCO}/I_{\rm SiO}}$ and ${I_{\rm HC_{3}N}/I_{\rm SiO}}$ and find that they exhibit similar trends.

\emph{G345.004$-$00.224}. G345.004$-$00.224 is identified as an IRDC, named SDC 345.000$-$0.232 by \citet{2009A&A...505..405P}. As shown in the lower panels of Fig.~\ref{fig:Figure_A17}, molecular line emissions show a similar, centrally condensed structure. The separation between the bubble centroid and dust emission peak is $\sim$160\arcsec, corresponding to a projected distance of $\sim$1 pc at a kinematic distance of 1.4 kpc \citep{2018MNRAS.473.1059U}. SiO and HC$_{3}$N emissions peak toward the hot dust clump which contains H$_{2}$O \citep{1983AuJPh..36..401C}, 12-GHz CH$_{3}$OH \citep{1995MNRAS.274.1126C} and OH \citep{1983AuJPh..36..361C} masers, while the HNCO emission peak has an offset of $\sim$20$^{\prime\prime}$ from the maximum of the dust continuum emission. Moreover, the SiO line shows non-Gaussian line-wing emission, which indicates outflow activities. Abundance ratios of ${N_{\rm HNCO}/N_{\rm SiO}}$ and ${N_{\rm HNCO}/N_{\rm HC_{3}N}}$ increase away from the central hot clump. Inflow and outflow activities are simultaneously detected in this source from analysing the molecular line profile of HCO$^{+}$ 1--0 and SiO 2-1, respectively (see Sect.~\ref{sec:outflow_inflow} and Fig.~\ref{fig:Figure_32} for details).

\emph{G350.101$+$00.083}. G350.101$+$00.083 is identified as an IRDC, named SDC 350.117+0.096 by \citet{2009A&A...505..405P}. In the top panels of Fig.~\ref{fig:Figure_A18}, the SiO, HNCO and HC$_{3}$N molecular emissions show similar morphologies, with extended structure associated well with the hot 870 $\mu$m, OH maser \citep{1983AuJPh..36..361C} and water vapor maser emission \citep{1983AuJPh..36..401C}. There are three bubbles, MWP1G350105$+$000838, MWP1G350107$+$000852, which is a smaller bubble inside the area of MWP1G350105$+$000838, and CS112 \citep{2012MNRAS.424.2442S} located south-west and south of the molecular emission peak. The SiO 2-1 spectrum shows a low signal-to-noise ratio. The abundance ratio curve of ${N_{\rm HNCO}/N_{\rm SiO}}$ and ${N_{\rm HNCO}/N_{\rm HC_{3}N}}$ show a reverse trend to that of the dust emission temperature.

\emph{G351.443$+$00.659}. In the lower panels of Fig.~\ref{fig:Figure_A18}, the integrated intensity maps of the SiO 2--1, HNCO 4$_{04}$--3$_{03}$ and HC$_{3}$N 10--9 lines all show a similar, centrally condensed structure. The peak molecular line emissions are obviously offset from the maximum of dust continuum emission by $\sim$15$^{\prime\prime}$, $\sim$30$^{\prime\prime}$ and $\sim$15$^{\prime\prime}$ for SiO, HNCO and HC$_{3}$N, respectively, which might be caused by the shocks from bubbles MWP1G351426$+$006596 and MWP1G351418$+$006431. The SiO 2--1 line shows broad wings. Abundance ratios of ${N_{\rm HNCO}/N_{\rm HC_{3}N}}$ and ${N_{\rm HNCO}/N_{\rm SiO}}$ show an opposite trend with respect to the dust temperature along Galactic latitude. Inflow and outflow activities are simultaneously detected in this source from analysing the molecular line profiles of HCO$^{+}$ 1--0 (see Sect.~\ref{sec:outflow_inflow} and Fig.~\ref{fig:Figure_32} for details).

\emph{G351.582$-$00.352}. In the top panels of Fig.~\ref{fig:Figure_A19}, the three molecular emissions show similar morphologies with a single emission peak towards the hot dust clump, which contains H$_{2}$O \citep{1983AuJPh..36..401C}, class II CH$_{3}$OH \citep{2010MNRAS.404.1029C} and OH \citep{1983AuJPh..36..361C} masers. A bubble, MWP1G351585$-$003560, identified by \citet{2012MNRAS.424.2442S} is close to the maximum of dust continuum emission. The velocity range of the SiO line is the largest. ${I_{\rm HC_{3}N}/I_{\rm SiO}}$ has an obvious maximum towards the north-east of the molecular emission peak.

\emph{G351.775$-$00.537}. In the lower panels of Fig.~\ref{fig:Figure_A19}, the three molecular lines show a similar morphology with emission peaks near the bubble MWP1G351780$-$005500S. The peak molecular line emission of HC$_{3}$N is offset by $\sim$30$^{\prime\prime}$ from the maximum of hot continuum emission. The SiO 2--1 line profile shows a particularly wide velocity range. The trend of the ${I_{\rm HNCO}/I_{\rm HC_{3}N}}$ ratio is opposite to that of ${I_{\rm HC_{3}N}/I_{\rm SiO}}$. Inflow and outflow activities are simultaneously detected in this source from analysing molecular line profiles of HCO$^{+}$ 1--0 (see Sect.~\ref{sec:outflow_inflow} and Fig.~\ref{fig:Figure_32} for details).

\emph{G329.030$-$00.202}. G329.030$-$00.202 is identified as an IRDC, named SDC 329.028$-$0.202 by \citet{2009A&A...505..405P}. As seen in Fig.~\ref{fig:Figure_A20}, the morphologies of our three molecular lines are consistent with that of the 870 $\mu$m hot dust emission. The source contains 12-GHz CH$_{3}$OH \citep{1995MNRAS.274.1126C} and H$_{2}$O \citep{1989A&A...221..105S} masers. The SiO 2--1 line may show a red-skewed profile with the red-shifted peak being stronger than the blue-shifted peak. HNCO and and HC$_{3}$N lines show a single velocity component peaking near the dip of the SiO 2--1 line. In addition, we note that the SiO line profile shows wing emission. For abundance ratios of ${N_{\rm HNCO}/N_{\rm SiO}}$ and ${N_{\rm HNCO}/N_{\rm HC_{3}N}}$, there is an obvious decrease towards the hot dense core, while ${N_{\rm HC_{3}N}/N_{\rm SiO}}$ does the opposite. Moreover, inflow and outflow activities are simultaneously detected in this source from analysing molecular line profiles of HCO$^{+}$ 1--0 (see Sect.~\ref{sec:outflow_inflow} and Fig.~\ref{fig:Figure_33} for details).

\emph{G331.709$+$00.602 \& G331.708$+$00.583}. From the lower panels of Fig.~\ref{fig:Figure_A20}, it can be seen that the three molecular line emissions show similar morphologies, and are consistent with two hot dust clumps, G331.709+00.602 and G331.708+00.583, which are indicated by a pink and a yellow cross, respectively. ${ N_{\rm HNCO}/N_{\rm SiO}}$ and ${N_{\rm HNCO}/N_{\rm HC_{3}N}}$ have a tendency to decrease with dust temperature towards the inner hot and dense region. Inflow and outflow activities are simultaneously detected in these two sources from analysing molecular line profiles of HCO$^{+}$ 1--0 (see Sect.~\ref{sec:outflow_inflow} and Fig.~\ref{fig:Figure_33} for details).

\emph{G335.586$-$00.289}. G335.586$-$00.289 is identified as an IRDC, named SDC 335.579$-$0.292 by \citet{2009A&A...505..405P}. As seen in the top panels of Fig.~\ref{fig:Figure_A21}, the emissions of the HNCO, SiO and HC$_{3}$N lines display similar morphologies with a single core peaking at the hot 870 $\mu$m dust emission clump, which contains CH$_{3}$OH \citep{1995MNRAS.272...96C} and OH \citep{1998MNRAS.297..215C} masers. The SiO line profile shows a peak and a blue shoulder, but the HNCO and HC$_3$N profiles exhibit single Gaussian components. Similar to sources G331.709+00.602 \& G331.708+00.583, ${N_{\rm HNCO}/N_{\rm SiO}}$ and ${N_{\rm HNCO}/N_{\rm HC_{3}N}}$ show a decreasing trend towards the warmer inner region. Inflow and outflow activities are simultaneously detected in this source from analysing molecular line profiles of HCO$^{+}$ 1--0 (see Sect.~\ref{sec:outflow_inflow} and Fig.~\ref{fig:Figure_33} for details).

\emph{G348.754$-$00.941}. G348.754$-$00.941 is identified as an IRDC, named SDC 348.765$-$0.924 by \citet{2009A&A...505..405P}. The peaks of the HNCO, SiO and HC$_3$N emission, showing a cometary distribution, are slightly offset from the maximum of low temperature filamentary dust continuum emission (Fig.~\ref{fig:Figure_A21}). SiO emission is relatively weak in G348.754$-$00.941. Moreover, the trends of ${N_{\rm HNCO}/N_{\rm SiO}}$ and ${N_{\rm HNCO}/N_{\rm HC_{3}N}}$ are opposite to that of the dust temperature.

\emph{G351.157$+$00.701}. In Fig.~\ref{fig:Figure_A22}, a north-west -- south-east elongated structure is shown in the three molecular emissions, which contain one core, with emission peaks obviously offset from the maximum of dust continuum emission by $\sim$30$^{\prime\prime}$, $\sim$45$^{\prime\prime}$ and $\sim$25$^{\prime\prime}$ for SiO, HNCO and HC$_{3}$N, respectively. SiO and HC$_{3}$N emission peaks towards H$_{2}$O \citep{1983AuJPh..36..401C} and OH \citep{1998MNRAS.297..215C} masers where the dust temperature is high. ${I_{\rm HNCO}/I_{\rm SiO}}$ and ${I_{\rm HC_{3}N}/I_{\rm SiO}}$ show a decreasing trend towards the yellow cross site.

\end{document}